\catcode`\@=11
\def\citen#1{\if@filesw \immediate\write \@auxout {\string\citation{#1}}\fi%
\@tempcntb\m@ne \let\@h@ld\relax \def\@citea{}%
\@for \@citeb:=#1\do {\@ifundefined {b@\@citeb}%
    {\@h@ld\@citea\@tempcntb\m@ne{\bf ?}%
    \@warning {Citation `\@citeb ' on page \thepage \space undefined}}%
    {\@tempcnta\@tempcntb \advance\@tempcnta\@ne
    \setbox\z@\hbox\bgroup\ifcat0\csname b@\@citeb \endcsname \relax
    \egroup \@tempcntb\number\csname b@\@citeb \endcsname \relax
    \else \egroup \@tempcntb\m@ne \fi \ifnum\@tempcnta=\@tempcntb
    \ifx\@h@ld\relax \edef \@h@ld{\@citea\csname b@\@citeb\endcsname}%
    \else \edef\@h@ld{\hbox{--}\penalty\@highpenalty
    \csname b@\@citeb\endcsname}\fi
    \else \@h@ld\@citea\csname b@\@citeb \endcsname \let\@h@ld\relax \fi}%
\def\@citea{,\penalty\@highpenalty\hskip.13em plus.13em minus.13em}}\@h@ld}
\def\@citex[#1]#2{\@cite{\citen{#2}}{#1}}%
\def\@cite#1#2{\leavevmode\unskip\ifnum\lastpenalty=\z@\penalty\@highpenalty\fi%
  \ [{\multiply\@highpenalty 3 #1%
  \if@tempswa,\penalty\@highpenalty\ #2\fi}]}   %
\makeatother 
 
\catcode`\@=12

\def\A             {{\rm A}}
\def\abihom        {alternating bi-homo\-mor\-phism}
\def\alfa          {f}
\def\alg           {algebra}

\newcommand\ao[1]  {a_{#1}^{}}
\newcommand\aO[1]  {a_{#1}}
\newcommand\Ao[3]  {\A_{#1\,#2}^{\!\!\circ\,#3}}
\def\AO            {\A\raisebox{.48em}{$\ssty\!\!\circ$}}
\newcommand\aor[3] {a_{#1#2;#3}^{}}
\newcommand\aOr[3] {a_{#1#2;#3}}
\def\atype         {automorphism type}
\def\auto          {automorphism}
\def\bashi         {boundary block}

\def\BB            {{\rm B}_\lambda}
\newcommand\bbb[1] {\rp\Beta_{\rp #1}}
\def\BBB           {\rp{\rm B}_\lambdab}

\def\BBJ           {\rp{\rm B}_{\J\lambdab}}

\def\bc            {boundary condition}
\def\Bc            {Boundary condition}
\def\be            {\begin{equation}}
\def\Be            {{\rm B}_\lambda^{(\bfe)}}
\def\bearl         {\begin{array}{l}}
\def\bearll        {\begin{array}{ll}}
\def\bearlll       {\begin{array}{lll}}
\def\Bet           {{\rm b}}
\def\Beta          {{\rm B}}

\def\Betab         {{\rp\Beta}}

\def\bfe           {{\bf1}}

\def\bihom         {bi-homo\-mor\-phism}
\def\br            {{\beta}}
\def\bS            {\rp S}

\def\cala          {{\mathfrak A}}

\def\calap         {{\rp\cala}}
\def\calb          {{\cal B}}
\def\calc          {{\cal C}}
\def\calh          {{\cal H}}
\def\calhb         {\bar{\cal H}}
\def\calhbj        {\plujl\calhb_{\J\lambdab}}
\def\calhbjp       {\plujl\calhb_{(\J\lambdab)^+_{\phantom I}}}
\def\calhbl        {\calhb_\lambdab}
\def\calhblp       {\calhb_\lambdabp}
\def\calhl         {{\cal H}_\lambda}
\def\calhlp        {{\cal H}_\lambdap}

\def\calo          {{\cal O}}

\def\calq          {\gg}
\def\cals          {{\cal S}}
\def\calso         {\cals^\oo}
\def\calsp         {\cals\oei}
\def\calu          {{\cal U}}

\newcommand\caluhp[1]{\calu_{#1}{\cap}\Hp}
\newcommand\caluhP[1]{|\calu_{#1}\cap\Hp|}
\def\caluo         {\calu^\oo}
\def\caluos        {\calu^{\oo*}}
\def\calup         {\calu\oei}

\def\calups        {\calu^{\prime*_{}}}

\def\cblock        {chiral block}
\def\cC            {C^\calb}
\def\cdoT          {\,{\cdot}\,}
\def\cft           {conformal field theory}
\def\cfts          {conformal field theories}
\def\chib          {\rp{\chii}}
\def\chii          {\raisebox{.15em}{$\chi$}}
\def\Check         {\check}
\def\chio          {{\cal X}^\oo}
\def\chip          {{\cal X}\oei}

\newcommand\Cl[1]  {{\rm C}^{B\,(\rp #1)}}
\def\cla           {classifying algebra}

\def\clAb          {\mbox{$\calc(\calap)$}}
\def\class         {classification}
\def\clo           {\calo }
\def\complex       {{\dl C}}
\def\corfu         {correlation function}
\newcommand\Cr[1]  {{\rm C}^{\calb\,(\rp #1)}}

\def\df            {\,{:=}\,}

\def\dl            {\mathbb }

\newcommand\dP[1]  {d_{#1}\oei}
\def\DPN           {\widehat{\rm N}}
\newcommand\DpNl[3]{\widehat{\rm N}_{#1,#2,#3}^{}}

\def\dsty          {\displaystyle}
\def\dud           {N}
\def\dudo          {N''}
\newcommand\dudu[1]{N_{#1}}

\def\dyd           {Dynkin diagram}

\def\ee            {\end{equation}}
\def\eE            {{\rm e}}
\def\eear          {\end{array}}

\def\End           {{\rm End}}

\def\eq            {\,{=}\,}
\newcommand\erf[1] {(\ref{#1})}
\newcommand\Erf[2] {(\ref{#1#2})}

\def\F             {{\cal F}}

\def\findim        {finite-dimensional}
\def\Fp            {{\F'}}
\newcommand\Frac[2]{\mbox{\large$\frac{#1}{#2}$}}

\def\fsi           {\epsilon}  
\def\furu          {fusion rule}
\def\futnote#1     {\footnote{~#1}\ }

\def\G             {{\rm G}}

\def\GF            {\G^F}

\def\gg            {{g}}

\def\Go            {\Gs^\oo}
\def\Gp            {{\G'}}
\def\Greg          {{\G^{\rm reg}}}
\def\GReg          {\G^{\rm reg}}
\def\Gregp         {{\Gp}^{\rm reg}}
\def\Gregps        {{{\Gp}^{\rm reg}}^*}
\def\Gregs         {{\Greg}^*}
\def\Gs            {{\cal G}}
\def\GS            {{G^*}}
\def\gt            {\succ}
\def\Hd            {\Gs\odr}
\def\gz            {generalized }

\def\HG            {{\rm H}}
\def\HK            {{\cal K}}
\newcommand\hsp[1] {\mbox{\hspace{#1 em}}}
\def\Hp            {\Gs\oei}
\def\hy            {$\mbox{-\hspace{-.66 mm}-}$}

\def\id            {\mbox{\sl id}}

\def\ii            {{\rm i}}
\def\II            {\cite{fuSc12}}

\def\imS           {{\cal I}}
\def\iN            {\,{\in}\,}
\def\infdim        {infinite-dimensional}
\def\inv           {invariance}

\def\irrep         {irreducible representation}
\def\J             {{\rm J}}
\def\JJ            {{}_{\J}}

\def\JJK           {{}_{\JK}}
\def\JJp           {{}_{{\J}'}}

\def\JK            {{\rm K}}
\def\Jp            {\J\oei}
\def\JP            {[\J]\oei}
\def\Kappa         {\kappa_\lambda}
\def\kappab        {\rp\kappa_\lambdab}
\def\kappabt       {\rp\kappa_{\rm tot}}
\def\kappai        {\kappa_{\Bet}}
\def\kappav        {\kappa_{\psu}}
\def\kappaw        {\kappa_{\psu^+}}
\renewcommand\ker[1]{{\rm ker}(#1)}
\def\KP            {[\JK]\oei}
\def\kzc           {Knizhnik\hy Zamolodchikov connection}

\long\def\labl#1   {\label{#1}\ee}
\long\def\Labl#1#2 {\label{#1#2}\ee}
\def\lambdab       {{\rp\lambda}}
\def\lambdaB       {{[\lambdab]}}
\def\Lambdab       {{[\lambdab,\psu]}}
\def\LambdaB       {{[\lambdab,\psu_\lambda]}}

\def\lambdaBo      {{[\lambdab]^\oo_{}}}
\def\LambdaBo      {{[\lambdab,\psu_\lambda]^\oo_{}}}
\def\lambdabp      {{\lambdab^{\!+}_{\phantom i}}}
\def\LambdabP      {{[\lambdab,\psu]^+_{\phantom i}}}
\def\lambdaBp      {{[\lambdab]\oei}}
\def\LambdaBp      {{[\lambdab,\psu_\lambda]\oei}}
\def\LambdaBP      {{[\lambdab,\psu_\lambda]^+_{\phantom i}}}
\def\lambdap       {{\lambda^{\!+}_{\phantom i}}}

\def\llb           {\mbox{\large(}}
\def\Llb           {\mbox{\Large(}}
\def\lrb           {\mbox{\large)}}
\def\Lrb           {\mbox{\Large)}}

\def\mimo          {minimal model}
\newcommand\Mo[3]  {{\rm M}^{\circ\,#1,#2}_{\ \ \ \ \ \ \ \ \ \ #3}}

\def\modinv        {modular invarian}
\def\Modinv        {Modular invarian}
\def\mub           {{\rp\mu}}
\def\muB           {{[\mub]}}
\def\MuB           {{[\mub,\psu_\mu]}}

\def\MuBo          {{[\mub,\psu_\mu]^\oo_{}}}
\def\mubp          {{\bar\mu^+_{\phantom i}}}
\def\MuBP          {{[\mu,\psu_\mu]^+_{\phantom i}}}

\def\n             {\beta_\circ}
\newcommand\N[3]   {\rp{\rm N}_{\rp #1,\rp #2}^{\;\ \ \rp #3}}

\def\nE            {\,{\ne}\,}
\newcommand\Ne[3]  {{}_{}^{\Hp\!}{\rm N}_{#1,#2}^{\,\ \ \ \ \ \ \ \ \ \ #3}}
\def\nj            {\beta_\circ^{(\psu;\J\lambdab)}}

\newcommand\norm[2]{{\cal N}_{#1,#2}}
\def\nub           {{\rp\nu}}
\newcommand\nxt[1] {\\\raisebox{.12em}{\rule{.35em}{.35em}}\hsp{.6}#1}

\def\oei           {'}
\def\odr           {''}

\def\omdual        {\gg^\star}

\def\one           {\mbox{\small $1\!\!$}1}
\def\onedim        {one-dimen\-sional}

\def\oo            {\circ}
\def\Opsu          {\clo}
\def\ot            {\raisebox{.07em}{$\scriptstyle\otimes$}}
\def\oT            {\,\ot\,}
\def\otimeS        {\,{\otimes}\,}
\def\ots           {\raisebox{.07em}{$\sss\otimes$}}
\def\parfu         {partition function}
\newcommand\pho[1] {\phi_{#1,\tilde{#1}}}

\def\phu           {{\hat\varphi}}
\def\phup          {{\hat\varphi\oei}}
\def\pj            {p^{\sss(\J\lambdab)}}
\def\plujl         {\bigoplus_{\J\in\Gs/\cals_\lambda}}

\newcommand\Plupsipsu[1]{\bigoplus_{\ssty\psi\in\cals_{#1}^* \atop\psi\gt\psu}}

\def\po            {p_\circ}
\def\pref          {N}
\def\psit          {\psu}
\def\psu           {{\hat\psi}}
\def\psuhp         {\Check\psi}
\def\psuo          {{\hat\psi^\oo}}
\def\psup          {{\hat\psi\oei}}
\def\q             {quantum }
\def\qj            {q^{\sss(\J\lambdab)}}

\def\qo            {q_\circ}
\def\raisa         {\!\!\raisebox{.54em}{$\phantom.$}}

\newcommand\Rc[3]  {{\rm R}^{#1}_{#2;#3}}
\def\rep           {representation}
\def\resp          {respectively}
\def\rhob          {{\rp\rho}}
\def\rhoB          {{[\rhob]}}
\def\RhoB          {{[\rhob,\psu_\rho]}}
\def\RHoB          {{[\rhob,\psu]}}

\def\RhoBd         {{[\rhob_3,\psu_3]}}

\def\rhobe         {{\rhob_1}}

\def\RhoBe         {{[\rhob_1,\psu_1]}}

\def\RhoBep        {{[\rhob_1,\psu_1]\oei}}

\def\RhoBp         {{[\rhob,\psu_\rho]\oei}}
\def\RhoBpp        {{[\rhob',\psu_\rho']}}

\def\RhoBv         {{[\rhob_4,\psu_4]}}
\def\rhobz         {{\rhob_2}}

\def\RhoBz         {{[\rhob_2,\psu_2]}}

\def\RhoBzp        {{[\rhob_2,\psu_2]\oei}}
\def\rhod          {{\rho_3}}
\def\rhoe          {{\rho_1}}
\def\rhov          {{\rho_4}}
\def\rhoz          {{\rho_2}}
\def\rhs           {right hand side}
\def\rp            {\bar}

\newcommand\s[1]   {{\rm s}_{#1}}
\newcommand\Sb[2]  {\rp S_{\rp{#1},\rp{#2}}}

\newcommand\sect[1]{\section{#1}\setcounter{equation}{0}}
\def\sigmab        {{\rp\sigma}}
\def\sigmaB        {{[\sigmab]}}
\def\SigmaB        {{[\sigmab,\psu_\sigma]}}

\def\SigmaBd       {{[\sigmab_3,\phu_3]}}
\def\SigmaBdp      {{[\sigmab_3,\phu_3]^+_{\phantom i}}}

\def\SigmaBe       {{[\sigmab_1,\phu_1]}}

\def\SigmaBep      {{[\sigmab_1,\phu_1]^+_{\phantom i}}}
\def\sigmaBp       {{[\sigmab]\oei}}
\def\SigmaBp       {{[\sigmab,\psu_\sigma]\oei}}
\def\sigmaBo       {{[\sigmab]^\oo_{}}}
\def\SigmaBo       {{[\sigmab,\psu_\sigma]^\oo_{}}}
\def\SigmaBO       {{[\sigmab,\phu_\sigma]}}

\def\SigmaBphu     {{[\sigmab,\phu]}}
\def\SigmaBphupp   {{[\sigmab',\phu']}}

\def\SigmaBz       {{[\sigmab_2,\phu_2]}}

\def\SJ            {S^\J}
\def\slz           {\mbox{SL$(2{,}\zet)$}}

\def\So            {S^\oo}
\newcommand\sO[1]  {{\rm s}_{#1}^\oo}

\renewcommand\sp[1]{{\rm s}_{#1}\oei}
\def\Sp            {S\oei}

\def\sss           {\scriptscriptstyle}
\def\ssty          {\scriptstyle}
\def\stts          {string theories}

\newcommand\sumbo[1]{\sum_{\ssty\rp #1 \atop Q_\Gs(#1)=0}}
\newcommand\sumBo[1]{\sum_{\ssty[\rp #1]\atop Q_\Gs(#1)=0}}

\newcommand\sumBoo[1]{\sum_{\ssty[\rp #1]^\oo\atop Q_\Gs(#1)=0}}
\newcommand\sumBop[1]{\sum_{\ssty[\rp #1]\oei\atop Q_{\Hp}(#1)=0}}
\newcommand\sumBoP[1]{\sum_{\ssty[\rp #1]\oei\atop Q_\Gs(#1)=0}}

\def\sumjl         {\sum_{\J\in\Gs/\cals_\lambda}}
\def\sumjm         {\sum_{\J'\in\Gs/\cals_\mu}}

\newcommand\Sumphiphu[1]{\sum_{\ssty\varphi\in\cals_{#1}^* \atop\varphi\gt\phu}}

\newcommand\Sumphipsu[1]{\sum_{\ssty\varphi\in\cals_{#1}^* \atop\varphi\gt\psu}}

\newcommand\Sumpsipsu[1]{\sum_{\ssty\psi\in\cals_{#1}^* \atop \psi\gt\psu}}
\newcommand\sumpsipsuo[1]{\sum_{\ssty\psi_{#1}\in\cals_{#1}^* \atop
                   \psi_{#1}\gt\psuo_{#1}}}
\newcommand\sumpsipsup[1]{\sum_{\ssty\psi_{#1}\in\cals_{#1}^* \atop 
                   \psi_{#1}\gt\psup_{#1}}}

\def\syms          {sym\-me\-tries}

\def\TauBp         {{[\tau,\psu_\tau]\oei}}
\def\tBeta         {{\Tilde\Beta}}
\def\tC            {C^B}

\def\Tilde         {\tilde}
\def\timeS         {\,{\times}\,}
\def\Times         {{\times}}
\def\tims          {\oT}

\def\tkappa        {\tilde\Kappa}
\newcommand\tN[3]  {\Tilde{\rm N}_{#1,#2}^{\ \ \ \ \ \ \ \ \ #3}}
\def\TN            {\Tilde{\rm N}}
\newcommand\tNl[3] {\Tilde{\rm N}_{#1,#2,#3}^{}}

\def\tPhi          {\Tilde\Phi}

\def\tr            {{\rm tr}}
\def\tR            {{\rm tr}\,}

\def\tS            {\Tilde S}

\def\twodim        {two-dimensional}
\renewcommand\u[1] {{\rm u}_{#1}}

\def\U             {{\rm U}}
\def\Uc            {{\U{\cap}\Gp}}
\newcommand\uo[1]  {{\rm u}_{#1}^\oo}

\newcommand\up[1]  {{\rm u}_{#1}\oei}
\def\Up            {{\U'}}
\def\Ups           {{{\U'}^*}}
\def\Us            {{\U^*}}
\def\ustab         {untwisted stabilizer}
\def\V             {{\cal V}}
\def\vac           {\Omega}
\def\vacb          {{\rp\vac}}
\def\vacB          {{[\vacb]}}
\def\vaco          {{\vac^\oo_{}}}
\def\vacp          {{\vac\oei}}

\def\vir           {\mbox{$\mathfrak V${\sl ir}}}
\def\voa           {vertex operator algebra}

\def\vop           {vertex operator}
\def\vphi          {{\varphi}}
\def\Vpsu          {\V_{\!\psu}}
\def\Vpsup         {\V_{\!\psu^+}}

\def\wrt           {with respect to }
\def\wrtt          {with respect to the }
\def\wzwm          {WZW model}
\def\wzwt          {WZW theory}
\def\wzwts         {WZW theories}
\newcommand\x[1]   {\xi_{#1}}
\def\X             {{\cal X}}

\def\zet           {{\dl Z}}

\def\zetay         {\zeta_Y}
\def\zetplus       {{\dl Z}_{>0}}
\def\zetpluso      {{\dl Z}_{\ge0}}

\documentclass[12pt]{article} \usepackage{amssymb,amsfonts,latexsym}

\begin{document}
 

\begin{flushright}  {~} \\[-1cm] {\sf hep-th/9902132} \\[1mm]
{\sf CERN-TH/99-35} \\[1 mm]
{\sf ETH-TH/99-03} \\[1 mm]
{\sf February 1999} \end{flushright}

\begin{center} \vskip 14mm 
{\Large\bf SYMMETRY BREAKING BOUNDARIES}\\[4mm] 
{\Large\bf I.\ GENERAL THEORY}\\[17mm]
{\large J\"urgen Fuchs}\\[3mm] CERN \\[.6mm] CH -- 1211~~Gen\`eve 23\\[7mm]
and\\[7mm]
{\large Christoph Schweigert}\\[3mm] Institut f\"ur Theoretische Physik \\
ETH H\"onggerberg \\[.2em] CH -- 8093~~Z\"urich
\end{center}
\vskip 17mm
\begin{quote}{\bf Abstract}\\[1mm]
We study conformally invariant boundary conditions that break part
of the bulk symmetries. A general theory is developped for those boundary
conditions for which the preserved subalgebra is the fixed algebra under an 
abelian orbifold group. We explicitly construct the boundary states and 
reflection coefficients as well as the annulus amplitudes. 
Integrality of the annulus coefficients is proven in full generality.
\end{quote}
\newpage


\sect{Introduction and summary}\label{s.1}

The space of conformally invariant boundary conditions
of \twodim\ \cfts\ is of interest in statistical mechanics, e.g.\
for the description of the Kondo effect and the theory of critical percolation,
as well as in open string theory, where particular attention followed the 
observation \cite{polc3} that string perturbation theory in solitonic sectors 
can be formulated in terms of world sheets with boundaries. In these 
applications it is crucial that the \bc s preserve conformal invariance;
in contrast, additional symmetries that the bulk theory may possess typically
need not be respected.

The special case of boundary conditions that preserve the full bulk symmetry
was already considered a long time ago \cite{card9}. In this case the 
consistent conformal boundary conditions are in one-to-one correspondence with 
the irreducible representations of the fusion rule algebra of the theory,
the so-called (generalized) quantum dimensions. To be precise, this result
holds when the torus \parfu\ is given by charge conjugation. More recently, it 
has been observed \cite{prss3,fuSc5} that also in the case when the torus 
\parfu\ corresponds to some simple current automorphism of the fusion rules, 
one can find a relative of the fusion algebra whose \irrep s precisely 
correspond to the boundary conditions that preserve the full bulk symmetry. 
This algebra has been dubbed the {\em classifying algebra}.\/

The consideration of Dirichlet boundary conditions for a free boson conformal
field theory brought yet another insight. Namely, for every conformal field 
theory, say with charge conjugation modular invariant, one should also study
boundary conditions that relate left and right movers by some
automorphism of the fusion rules \cite{fuSc6,reSC} that preserves conformal
weights. For a given fusion rule automorphism $\omdual$, \resp\ the 
corresponding automorphism $\gg$ of the chiral \alg, there will typically 
exist several distinct conformally invariant boundary conditions. They 
constitute the possible {\em Chan\hy Paton types\/} for the fixed 
{\em automorphism type\/} $\gg$. Again, it is natural to construct
these \bc s as \irrep s of some classifying \alg\ that generalizes
the fusion rule algebra \cite{fuSc6}.

One goal of this paper is to identify these \alg s; but to do so, it turns
out to be convenient to solve a more general problem. The boundary
conditions of automorphism type $\gg$ respect only a subset of the
bulk symmetries $\cala$, namely the subalgebra $\cala^{(\gg)}_{}$ of those 
elements that are fixed under $\gg$. More generally, one may therefore
address the following question. Given a subalgebra $\calap$ of the chiral \alg\
$\cala$, determine all those boundary conditions that preserve (at least)
$\calap$, but not necessarily all of $\cala$. It should be appreciated that even
when we ask this question for the subalgebra $\calap\eq\cala^{(\gg)}_{}$ 
associated to some automorphism $\gg$ of finite order, it is by no means 
clear that {\em all\/} the boundary conditions preserving $\calap$ possess 
a definite automorphism type. It will be a
non-trivial result of our analysis that this is indeed true.

As long as $\calap$ is
completely arbitrary, at present this problem is still too general to be
tractable. We will therefore restrict our attention to a particular subclass
of (consistent) subalgebras. Namely, we require that $\calap$ be the {\em fixed 
algebra\/} $\cala^G$ of some group $G$ of automorphisms
of the chiral algebra $\cala$. In other words, $\calap\eq\cala^G$ is the
chiral algebra of an {\em orbifold\/} of the theory that has chiral \alg\ 
$\cala$. The orbifold group $G$ need not necessarily be a finite group; one 
may even study orbifolds \wrt \findim\ Lie groups. But for the 
present purposes we assume that $G$ is indeed finite, and we still 
specialize further to the situation that $G$ is a finite abelian group. 

In this case the original chiral algebra $\cala$ can be reassembled
from its sub\alg\ $\calap$ by an {\em integer spin simple current extension\/}. 
This allows us to utilize simple current technology 
\cite{scya,intr,scya6,krSc,fusS3,fusS6}. This way we have several nice 
structures at our disposal, which have passed various rather non-trivial checks
in chiral \cft\ (see e.g.\ \cite{fusS6,fusS4,bant7,bant6}). They allow us 
to write down a natural candidate for a classifying algebra. We then take this
ansatz and compute reflection coefficients and annulus coefficients, and 
afterwards show that these quantities pass the usual consistency checks. 
In particular, the annulus coefficients are proven to be integral; it 
should be noticed that this property is an outcome of our
analysis rather than a requirement we impose.

For the convenience of the reader, we now present a brief summary of
our main results. We assume full reducibility (which is satisfied in all 
known examples), i.e.\ that we can decompose the \rep\ spaces 
$\calh_\lambda$ for all primary fields of the $\cala$-theory as
  \be  \calh_\lambda= \bigoplus_{\mub} V_\lambda^\mub
  \otimes \calhb_\mub  \labl1
into irreducible $\calap$-modules $\calhb_\mub$. The degeneracy spaces
$V_\lambda^\mub$ introduced this way are modules of suitable subgroups
$U_\lambda$ of the orbifold group $G$. We make the mild assumption that 
each of these $G$-modules is irreducible, so that $V_\lambda^\mub 
\,{\cong}\,V_\Psi$ where $\Psi\iN U_\lambda^*$. As a consequence, we can 
label the primary fields of the orbifold theory by pairs $(\lambda,\Psi)$ where 
$\lambda$ is an $\cala$-primary and $\Psi\iN U_\lambda^*$. Actually, at this
point we have somewhat oversimplified the story. Indeed, by assumption we have 
an action of $G$ on the chiral algebra, and thus on the vacuum primary field 
$\lambda\eq\vac$. While this does induce an action of $U_\lambda$ on the 
degeneracy spaces that arise in the decomposition for other primaries
as well, that action is in general only {\em projective\/}. Thus in general
we must allow for $V_\Psi$ to be only a projective 
module. Note that projective modules of an abelian group
do not necessarily have dimension one; accordingly, additional multiplicities 
will occur in our analysis. That this effect is indeed realized in concrete
models can already be seen for orbifolds of a free boson, compactified at 
self-dual radius; when one orbifoldizes by the dihedral group $D_2$, then
the dihedral group acts on the primary field of conformal dimension
$\Delta\eq1/4$ only projectively, and those projective \irrep s are, of
course, \irrep s of the universal central extension of
$D_2$, the quaternion group (for more details see \cite{dvvv}).

Technically, we will proceed in this work in a manner that is opposite to the
orbifold philosophy, i.e.\ we
express $\cala$-quantities in terms of quantities of the $\calap$-theory
instead of the other way round. It can be seen that under the above-mentioned 
non-degeneracy assumption the primaries $\J\eq(\vac,\Psi)$
of the $\calap$-theory that come from the vacuum sector $\vac$ of the original
theory form an abelian group $\Gs$ under fusion; in other words,
they are {\em simple currents\/}. This group is actually isomorphic to 
the character group of the orbifold group $G$, i.e.\ $\Gs\,{=}\,\GS$. 

Equipped with this information, we are then in a position to apply simple 
current technology. First, by its action through the fusion product, the
simple current group $\Gs$ organizes the $\calap$-primaries $\lambdab$
into orbits. Generically this action is not free, so one associates to 
every primary field 
$\lambdab$ its stabilizer, i.e.\ the subgroup $\cals_\lambda$ of $\Gs$ whose
elements leave $\lambdab$ fixed (as $\Gs$ is abelian, the stabilizer is the 
same for all fields on the same $\Gs$-orbit). Further, for every simple current
$\J\iN\Gs$ we associate to each primary field $\lambdab$ the rational number
  \be  Q_\J(\lambda):= \Delta_\lambdab+ \Delta_\J - \Delta_{\J\star\lambdab}
  \bmod\zet \,,  \Labl QJ
called the monodromy charge of $\lambdab$, which is constant on $\Gs$-orbits.
In orbifold terminology, the fields whose monodromy charge vanishes for
every $\J\iN\Gs$ are those in the untwisted sector of the orbifold. More
generally, the function $\calq_\lambda^{}\,{\equiv}\,\calq_\lambda^{(Q)}
{:}\ \Gs\,{\to}\,\complex$ with
  \be  \calq_\lambda^{}(\J):= \exp(2\pi\ii Q_\J(\lambda))  \ee
for all $\J\iN\Gs$ is an element of the character group $\Gs^*\eq(\GS)^*_{}
\,{\cong}\,G$ and can be identified with an element of the orbifold group; 
$\calq_\lambda$ characterizes the twist sector to which the field $\lambdab$
belongs.

Based on this description one might expect that it is possible to
express the $\cala$-primaries $\lambda$ in terms of $\calap$-quantities 
as follows. The label $\lambda$ is interpreted as a pair $(\lambdaB,\psi)$, 
consisting of a $\Gs$-orbit $\lambdaB$ and a character $\psi$
of the stabilizer $\cals_\lambda$. This would correspond to the decomposition 
  \be  \calh_\lambda \leadsto
  \bigoplus_{\J\in\Gs/\cals_\lambda} \calhb_{\J\star\lambdab}  \labl{s3.1}
of irreducible $\cala$-modules, with
the character $\psi\iN\cals_\lambda^*$ accounting for the fact 
that inequivalent $\cala$-modules can be equivalent as $\calap$-modules.
However, as established in \cite{fusS6}, this ansatz is too naive. 
The origin of the failure was actually already mentioned above; namely,
the decomposition \erf{s3.1} would exclude the possibility of having only
a projective action of the orbifold group on sectors other than the
vacuum. In contrast, the formalism developped in \cite{fusS6}, which is 
briefly reviewed in appendix \ref{s.a}, correctly takes this effect into 
account. What is required as an additional ingredient is to introduce
for each $\lambdab$ a certain subgroup $\calu_\lambda$ of $\cals_\lambda$,
called the {\em untwisted stabilizer\/} of $\lambdab$. This subgroup is
of quadratic index; the positive integer 
  \be  d_\lambda := \sqrt{|\cals_\lambda|\,/\,|\calu_\lambda|}  \Labl dl
is just the dimension of the relevant projective representation.
The analysis of \cite{fusS6} shows that the $\calap$-primaries are in fact
described by pairs $\Lambdab$, where $\psu$ is a character of the untwisted
stabilizer rather than of the full stabilizer. The action of 
$\Gs/\cals_\lambda$ is then implemented by an equivalence relation
that also involves the character $\psu$ (see formula \Erf eq).
In \cite{fuSc10}, where some of our results were announced, we have
concentrated on the case where for all fields $\lambdab$ the untwisted 
stabilizer coincides with the full stabilizer; in the present work, the 
whole structure is displayed for the most general situation.

We can now exhibit the boundary conditions that preserve
only the sub\alg\ $\calap$ of the bulk symmetries. Owing to factorization, 
boundary conditions are characterized \cite{card9,cale} by the one-point 
\corfu s of bulk fields on the disk. The corresponding \cblock s are
two-point blocks on the sphere. However, as only the symmetries in $\calap$
are preserved, these blocks are not the ordinary chiral two-point 
blocks of the $\cala$-theory; rather, we should take the \cblock s of the 
$\calap$-theory and combine them in a way compatible with
the decomposition of the spaces $\calhl$. Since states in
different $\calap$-modules that occur in such a decomposition 
are possibly reflected differently at the boundary, this way
we arrive at an independent chiral two-point block for each pair 
$(\lambdab,\psu_\lambda)$, where $\lambdab$ is a field in the untwisted 
sector of the orbifold theory and $\psu_\lambda$ is a character of the 
untwisted stabilizer of $\lambdab$. We must still be somewhat more careful, 
though. The chiral blocks of our interest are linear forms
  \be  \calh_\lambda^{}\otimes\calh_\lambdap \to \complex\,.  \ee
However, when the degeneracy space has dimension $d_\lambda\,{>}\,1$, then 
we cannot simply obtain boundary blocks for the $\cala$-theory by composing
the corresponding boundary blocks of the $\calap$-theory (which are linear
forms $\calhb_\lambdab^{}\,{\otimes}\,\calhb_\lambdabp\,{\to}\,\complex$);
rather, the construction of a boundary block then
requires in addition a linear form on the tensor product of the
$d_\lambda$-dimensional degeneracy spaces. There are
$d_\lambda^2\eq|\cals_\lambda|/|\calu_\lambda|$ such forms.

As a consequence, for each primary $\lambdab$ in the untwisted sector
of the $\calap$-theory we get 
  \be  N_{\rm block}(\lambdab) = d_\lambda^2\,|\calu_\lambda|
  = |\cals_\lambda|  \Labl Nb
many independent chiral two-point blocks. As we will demonstrate in section
\ref{s.4}, the labels characterizing these blocks naturally combine into a pair 
$(\lambdab,\psi_\lambda)$, where $\psi_\lambda$ is now a character of the 
{\em full\/} stabilizer.

Next we analyze also the way in which these blocks combine to \corfu s,
whereby we effectively characterize the boundary conditions. We first observe 
that in the case where the full bulk symmetry $\cala$ is conserved and the 
torus \parfu\ is given by charge conjugation, the \bc s correspond to the 
(generalized) quantum dimensions of the $\cala$-theory. Quantum dimensions, in 
turn, are related to primary fields via the modular S-matrix of the theory. 
(Actually in this simple case the structure is somewhat obscured by the fact 
that the modular S-matrix is symmetric so that there exists a natural 
identification between quantum dimensions and primary fields.) The fact that a 
modular transformation
relates boundary {\em blocks\/} to boundary {\em conditions\/}
has become even more apparent in the example considered in \cite{fuSc5}.

It is therefore not too surprising that also in the more general situation 
considered here, boundary blocks and boundary conditions are connected by a 
modular transformation. Let us further explore this idea heuristically. The 
labels $\lambdab$ of the boundary blocks are subject to $\calq_\lambda\,{\equiv}
\,1$. In orbifold language, this means that we are only dealing with the 
untwisted sector of the orbifold. Thus along the `space' direction of the
torus only the twist by the identity occurs. It follows that after
a modular S-transformation, only the identity appears as twist in the `time' 
direction of the torus, which in turn means that the usual orbifold projection 
does not take place. In simple current language, the corresponding
statement is that the boundary conditions are labelled by $\Gs$-{\em orbits\/}
$\rhoB$ of $\calap$-primaries rather than by individual primary 
fields. On the other hand, after the modular S-transformation arbitrary twists
in the `space' direction occur in the orbifold; this means that
in the labelling of the \bc s {\em all\/} $\Gs$-orbits $\rhoB$ appear, not just 
those with vanishing monodromy charges, i.e.\ not just the ones in the 
untwisted sector. In fact in \II\ we will show that the 
character $\gg_\rho\iN\Gs^*\cong G$ furnished by the monodromy charges
of $\rhob$ can be naturally identified with the
automorphism type of the \bc. This in turn allows us to {\em derive\/}
(rather than to assume ad hoc) that every boundary condition of the form 
considered here possesses a definite automorphism type. Finally, for the 
boundary conditions there is an additional degeneracy, too, this time governed 
by the {\em untwisted\/} stabilizer. As a matter of fact, in the structures
we are going to exhibit, consistency is achieved through a rather subtle (and 
beautiful) interplay between the untwisted stabilizer and the full stabilizer.

For the convenience of the reader, we now collect a few explicit formul\ae.
They are most conveniently presented in terms of a certain square matrix $\tS$.
This matrix diagonalizes the structure constants of the \cla; accordingly
its first index refers to a boundary block $(\lambdab,\psi_\lambda)$, while
the second index corresponds to a boundary condition $\RhoB$. The formula
for $\tS$ is
  \be  \tS_{(\lambdab,\psi_\lambda),\RhoB}
  = \Frac{|\Gs|}{(|\cals_\lambda|\,|\calu_\lambda|\,|\cals_\rho|\,|\calu_\rho|)
  ^{1/2}_{}} \sum_{\J\in\cals_\lambda\cap\calu_\rho} \psi_\lambda(\J)\,
  \psu_\rho(\J)^*\, S^\J_{\lambdab,\rhob} \,.  \ee
Roughly, one has to sandwich certain matrices $S^\J$ between the characters
$\psi_\lambda\iN\cals_\lambda^*$ and $\psu_\rho\iN\calu_\rho^*$; these
matrices are the modular transformation
matrices for one-point \cblock s with insertion of the simple current $\J$ on
the torus \cite{bant6} and appear naturally in the study of simple current
extensions \cite{fusS6}. In terms of the matrix $\tS$ the \onedim\ \irrep s 
of the \cla\ which provide the reflection coefficients read 
  \be  R^{}_\RhoB (\tPhi_{(\lambdab,\psi_\lambda)}) =
  \frac{\tS_{(\lambdab,\psi_\lambda),\RhoB}} {\tS_{\vacb,\RhoB}} \,.  \ee
We will also see that there is a natural conjugation on the boundary conditions,
a map of order two that implements the reversal of the orientation
of the boundary.

Finally, we display the annulus amplitude for an annulus with boundary
conditions $\RhoBe$ and $\RhoBz$. 
As we will see, it is natural to express the annulus amplitude as a linear
combination 
  \be  \A_{\RhoBe\,\RhoBz} = \sum_{\SigmaBp} 
  \A_{\RhoBe\,\RhoBz}^{\SigmaBp}\,\chip_{\SigmaBp}  \ee
of characters $\chip_{\SigmaBp}$ of the \cft\ that is obtained by extending the
$\calap$-theory by the simple currents in the subgroup
  \be  \Hp \equiv \Hp_{\rho_1\rho_2}
  :=\{ \J\iN \Gs \,|\, Q_\J(\rho_1)\eq0\eq Q_\J(\rho_2) \}  \Labl Hp
of $\Gs$. The annulus coefficients $\A_{\RhoBe\,\RhoBz}^{\SigmaBp}$ can then
be written, up to a prefactor, as a sum of the fusion rule coefficients 
$\Ne\RhoBp\SigmaBp\TauBp$ of the $\Hp$-extension:
  \be  \A_{\RhoBe\,\RhoBz}^{\SigmaBp}
  = \pref \sum_{\ssty\psup_1\in\calup_1 \atop\psup_1\gt\psuhp_1}
  \sum_{\ssty\psup_2\in\calup_2 \atop\psup_2\gt\psuhp_2}
  \sum_{\J\in\Gs/\Hd} \Ne{\RhoBzp}{\J\SigmaBp}{\RhoBep} \,. \ee
Here $\Hd$ is a certain subgroup of $\Gs$ which is intermediate between
$\Hp$ and $\Gs$, i.e.\ $\Hp\,{\subseteq}\,\Hd\,{\subseteq}\,\Gs$, and
$\psuhp_i\eq\psu_i|_{\caluhp i}^{}$. As an important consistency check we 
will present a general proof that the prefactor 
$\pref\,{\equiv}\,\pref_{\rho_1\rho_2}$, which is a quite complicated ratio
involving the sizes of various subgroups of $\Gs$ (see formula \erf{dud}), is 
always a non-negative integer, so that the annulus amplitude can 
be consistently interpreted as a partition function for open string states.

More precisely, the number $\pref$ can be written as a product of three
separate integral factors, each of which possesses a natural
group theoretic respectively representation theoretic interpretation
(see formul\ae\ \erf{dudo'} and \erf{dudu'}).
While this definitely implies that the annulus coefficients are non-negative 
integers (as befits the coefficients of a partition function), the 
interpretation of the prefactor $\pref$ should
also play a role in non-chiral field-state correspondence. One expects to be
able to associate to every open string state a field operator in the
full conformal field theory. Now the partition function for these states is the
annulus amplitude, and the presence of additional multiplicities in the latter 
means that several distinct operators in the {\em full\/} \cft\ must be built 
from one and the same {\em chiral\/} vertex operator. An explicit construction 
of these operators is not known for the moment, but in any case the information 
that the multiplicities all possess an interpretation in terms of fusion
rules and other representation theoretic objects seems to be highly relevant.

The rest of this paper is organized as follows.
We start in section \ref{s.2} with a description of our setup, i.e.\
\bc s which preserve a subalgebra of the bulk symmetries that is fixed under
a finite abelian group of automorphisms. The analysis
of such \bc s proceeds in two steps, where in the first step one works 
exclusively at the chiral level, while in the second non-chiral quantities
enter. The general features of the chiral part are collected in section 
\ref{s.3}, while in section \ref{s.4} a natural basis for the basic chiral
ingredients, the {\em boundary blocks\/}, is constructed. Section \ref{s.5}
is devoted to the non-chiral level. First, in subsection \ref{s.51}, we
show that the \bc s of interest to us are governed by a {\em \cla\/}; in
the rest of this section we establish the precise form of this \alg\ and 
investigate its properties. While we regard our arguments leading to these 
results as convincing, they are not mathematically rigorous. As further evidence
we therefore perform, in section \ref{s.6}, several additional consistency 
checks based on properties of the annulus amplitudes, the most important one 
(subsection \ref{s.64}) being a general proof of the 
integrality of the annulus coefficients that appear in the open string channel.
   
In a follow-up paper \II, we will address several complementary issues which 
concern the structure of the space of symmetry breaking \bc s and display
the \bc s for various classes of \cfts\ explicitly.
More concretely, we start by associating to each \bc\ its {\em \atype\/},
which arises as a direct consequence of the general structure.
Then we show that \bc s of definite \atype\ can be naturally
formulated with the help of certain twisted boundary blocks, obeying
twisted Ward identities, and that they carry their own individual \cla, which 
is a suitable quotient of the total \cla.
Further we study the realization of T-duality on the space of \bc s,
show that this space carries an action of the orbifold group (`boundary
homogeneity'), and introduce the concept of a {\em universal\/} classifying 
algebra for all conformally invariant \bc s.
Finally we will exhibit in a large number of examples the
concrete realization of various structures that we have uncovered.

\bigskip\noindent{\small {\bf Acknowledgement} \\
We are grateful to Peter Bantay and Bert Schellekens for very helpful
correspondence.}

\sect{Broken bulk symmetries}\label{s.2}

In this paper we analyze the following situation. We start with some
prescribed \cft\ that is consistently defined on all closed orientable
surfaces, and choose the charge conjugation modular invariant as the
partition function on the torus. This theory is, moreover, assumed to be
non-heterotic, i.e.\ for left and right movers we deal with one and
the same symmetry algebra $\cala$, the {\em chiral algebra\/}, which contains 
the Virasoro algebra $\vir$. (This condition in fact refers to the oriented 
Schottky cover of the surface, which for a closed orientable surface consists 
of two isomorphic disjoint sheets; the requirement is that we deal with one 
and the same chiral \cft\ on both sheets.) We call $\cala$ the algebra of 
{\em bulk symmetries\/}. For technical reasons we will assume that the theory
is rational, i.e.\ that it contains only finitely many $\cala$-primaries.

Boundary conditions that respect the full bulk symmetry have been studied
for quite a while \cite{card9,cale}.
In contrast, in the present work we are interested in boundary conditions 
that do not preserve all bulk symmetries, but only a subalgebra $\calap$ 
of $\cala$. This does not yet restrict at all the kind of boundary condition 
we consider, since to any arbitrary boundary condition one may associate
the subalgebra of $\cala$ that is preserved. Further, we are interested
in conformally invariant boundary conditions only, so that $\calap$ must
in particular contain the Virasoro subalgebra of $\cala$. Moreover, $\cala$
must be `consistent'; by this qualification we understand that the algebra 
is closed under charge conjugation and
allows for the definition of sheaves of \cblock s which come with a 
projectively flat \kzc\ and which obey consistent factorization rules.

A typical chiral algebra $\cala$ will, however, possess
very many, if not infinitely many, consistent subalgebras $\calap$. 
Accordingly, the first step towards a classification of all conformal boundary
conditions would be to classify all those subalgebras. This problem 
depends largely on the specific bulk \cft\ under consideration, and (except for 
a discussion of a possible limiting \alg\ of an inductive system of \cla s 
in \II) we will not have to say much about it.
On the other hand, for sufficiently simple theories, such as the
Virasoro minimal models or the free boson or its
$\zet_2$-orbifold, all consistent sub\alg s are known. More
generally, once the problem of classifying the consistent sub\alg s has
been solved for any single model, the methods presented below provide us
(possibly modulo the existence of so-called complex charges, compare
\cite{sasT2}) with all conformal boundary conditions of that model.

Here we rather concentrate on the task of classifying all boundary 
conditions that preserve some specified consistent subalgebra $\calap$.
As long as $\calap$ is a completely arbitrary subalgebra, this problem is 
still too general and cannot be solved with the methods that are available
at present. We will therefore restrict our attention 
to a particular subclass of consistent subalgebras. Namely, we require
that $\calap$ be the {\em fixed algebra\/} of some group $G$ of automorphisms
of the chiral algebra $\cala$. In other words, 
  \be  \calap = \cala^G \ee
is the chiral algebra of an {\em orbifold\/} 
of the original theory. In principle the orbifold group $G$ can be quite
arbitrary; for instance, it need not even be finite, but rather could be some 
\findim\ Lie group. Still, for the purpose of the present paper we restrict
our attention to the case when $G$ is finite, and when moreover it is abelian. 

This situation may seem rather special compared to the
general problem sketched above, but it nevertheless covers a variety of
cases of practical interest. Examples are provided by
the critical three-state Potts model and, more generally, by Virasoro
minimal models of $(A,D_{\rm even})$ type, by
Dirichlet \bc s for a free boson for which only the chiral algebra 
of the $\zet_2$-orbifold of the boson theory is preserved, by D-branes in 
toroidal compactifications at generic positions, by charge conjugation in
WZW theories, and by those \bc s for a free boson that correspond to a change 
in the compactification radius. (For a more extensive list, see the final
sections in the follow-up \II\ of this paper.) Moreover, already with this
restriction we can gain a number of additional physical insights,
e.g.\ concerning the relation between \bc s that preserve sub\alg s
$\calap_1$ and $\calap_2$ of $\cala$ that are contained in each other.

Let us briefly recall how to describe boundary conditions in \cft\ 
on surfaces $\cal C$ with boundaries. First \cite{fuSc6},
one must set up a chiral \cft\ on a closed oriented twofold covering surface
$\tilde{\cal C}$ of $\cal C$, the {\em Schottky cover\/} \cite{ales},
from which $\cal C$ is obtained by dividing out an anti-conformal 
involution. This amounts to specifying a system of {\em \cblock s\/}
that has a \kzc\ and obeys factorization rules;
as a consequence of factorization, the blocks most relevant to the \bc s
are the \cblock s for a single bulk field insertion on the disk, which
are two-point blocks on the projective line ${\dl P}^1$. In an independent 
second step we have to construct {\em \corfu s\/} as linear combinations 
of these blocks that satisfy \cite{card9,cale,prss3,bppz,runk}
locality and factorization constraints. As was emphasized in 
\cite{fuSc6}, these two conceptual levels should be carefully distinguished,
and accordingly we will divide our discussion in two parts. We start 
by analyzing, in the next two sections, the chiral \cft\ on the Schottky cover.

\sect{Chiral theory for symmetry breaking boundaries}\label{s.3}

As just pointed out, for the thorough investigation
of boundary conditions it is advisable
to distinguish clearly between the two conceptual levels of {\em chiral\/}
\cft\ and {\em full\/} \cft. In the chiral theory, boundaries are not yet
present explicitly; but as a prerequisite for analyzing the breaking of bulk 
symmetries by boundary conditions in the full theory various structures 
need to be understood already at this stage. These chiral concepts are the 
topic of the present and the next section.

\subsection{Simple currents versus orbifolds}

As already outlined above, our general situation is as follows. We are given 
a rational \cft\ with chiral algebra $\cala$, and we consider boundary 
conditions that preserve only a consistent subalgebra $\calap$ of $\cala$.
Let us assume that $\cala$ can be obtained from its subalgebra $\calap$ by the 
extension with a simple current group $\Gs$, which is some finite abelian group.
Each simple current $\J\iN\Gs$ corresponds to an irreducible \infdim\ \rep\
space $\calhb_\J$ of $\calap$; moreover, the vacuum module $\calhb_\vacb$ 
of $\calap$ and the vacuum module $\calh_\vac$ of $\cala$ are related as
  \be \calh_\vac \cong \bigoplus_{\J\in\Gs} \calhb_\J \,,  \ee
where the symbol `$\cong$' stands for isomorphism as $\calap$-modules;
the identity element in $\Gs$ corresponds to $\calhb_\vacb$.

In this situation the chiral algebra $\calap$ is necessarily
an orbifold subalgebra of $\cala$. Namely,
we can obtain an action of the dual group $G\eq\Gs^*$ on $\calh_\vac$
as follows. For every $\gg\iN G$, we define $R(\gg)$ to act on the subspace 
$\calhb_\J$ of $\calh_\vac$ as the multiple $\J(\gg)\,\id_{\calhb_\J}^{}$ of 
the identity map, where $\J$ is regarded 
as a character on $G$. Field-state correspondence relates the vectors in
$\calh_\vac$ to operators in the chiral algebra, and thus this prescription
provides us with a group of automorphisms of $\cala$ that is isomorphic to $G$. 

Conversely, suppose we are given an action of a finite abelian group $G$ on 
the chiral algebra $\cala$ that leaves the Virasoro sub\alg\ $\vir\,{\subseteq}
\,\cala$ pointwise fixed. Then $\cala$ contains as a subalgebra the algebra
$\cala^G$ of all elements that are left pointwise fixed under the action
of the orbifold group $G$, and $\cala^G$ contains the Virasoro sub\alg\ 
of $\cala$. Again by field-state correspondence, we then also
have an action of $G$ on the vacuum module $\calh_\vac$,\,%
 \futnote{Here we make the assumption that the action of $G$ on the vacuum 
module $\calh_\vac$ is a honest action rather than only a projective one. This 
condition should be regarded as part of the definition of the term orbifold.
If it were not satisfied, the structure to be divided out would no longer be
a group.}
 and this action commutes with the action of $\cala^G$. It can then be shown
\cite{dolm5,doMa2} that $\calh_\vac$ is completely reducible as an 
$\cala^G$-module, so that we can
decompose $\calh_\vac$ into irreducible submodules of\,\,%
 \futnote{Note that $G$ commutes with the Virasoro \alg\ so that it preserves
the grading of the \infdim\ space $\calh_\vac$. Thus each homogeneous
subspace of fixed conformal weight is a \findim\ $G$-module, which is
fully reducible. As $G$ even commutes with all of $\cala^G$, full reducibility 
\wrt $G\timeS\cala^G$ then follows from full reducibility \wrt $\cala^G$.
Incidentally, a vertex operator algebra for which
every graded representation is fully reducible possesses only finitely many
inequivalent \irrep s \cite{dolm3}, thus giving rise to a rational \cft.\\
A decomposition of the vacuum module of the form \erf{chi1} is expected to 
hold for arbitrary finite orbifold groups $G$, and has been proven for many 
non-abelian groups in \cite{dolm5,doMa2}. Analogous decompositions are valid 
\cite{doMa3} for other $\cala$-modules, including twisted modules.}
 $G\timeS\cala^G$ as
  \be  \calh_\vac \cong \bigoplus_\lambdab \bigoplus_{\Psi\in G^*_{}}
  V_{\Psi}\otimes \calhb_\lambdab \,,  \labl{chi1}
where $\calhb_\lambdab$ are irreducible $\cala^G$-modules and $V_\Psi$ are
irreducible modules of $G$ (and are thus \onedim).
It follows in particular that all $\cala^G$-modules $\calhb_\lambdab$ that
appear in \erf{chi1} are simple currents. This holds true because the fusion 
of these modules must be compatible with the decomposition of tensor products 
of irreducible $G$-\rep s. The latter decomposition, in turn, is just 
described by the dual group $\Gs\eq G^*_{}$, and hence we conclude that 
\wrtt fusion product the modules $\calhb_\lambdab$ 
appearing in \erf{chi1} form a simple current group isomorphic to $\Gs$.

\subsection{Simple current extensions}

That the boundary conditions of interest to us preserve only the subalgebra
$\calap$ implies that generically the fields corresponding to vectors in 
different $\calap$-submodules of a given $\cala$-module are reflected 
differently at the boundary. Accordingly we need to decompose every sector 
of the chiral \cft, i.e.\ every \irrep\ of $\cala$, into irreducible 
$\calap$-\rep s (again we impose full reducibility \wrt $\calap$). 

Fortunately, the decomposition of $\cala$-modules in terms of $\calap$-modules 
is a purely chiral issue, i.e.\ is in particular independent of any boundary 
effects, and this chiral issue is well understood in the simple
current framework. We summarize some relevant information here; for more 
details see appendix \ref{s.a} and \cite{scya6,fusS6}. The fusion product 
provides an action of the simple current group $\Gs$ on the fields $\lambdab$
of the $\calap$-theory. To each primary field $\lambdab$ one then associates 
a subgroup of $\Gs$, the stabilizer 
  \be  \cals_\lambda := \{\J\iN\Gs \,|\, \J\lambdab\eq\lambdab \} \,.  \ee
Stabilizer subgroups are constant on $\Gs$-orbits (which we already anticipated 
by writing $\cals_\lambda$ in place of $\cals_\lambdab$);
conjugate $\calap$-fields have identical stabilizers, too.

In the decomposition of a given $\cala$-module $\calh_\lambda$
only $\calap$-modules on a single $\Gs$-orbit appear. 
However, one and the same $\Gs$-orbit $\muB$ of primaries in the 
$\calap$-theory can give rise to several distinct primaries
of the $\cala$-theory. In other words, inequivalent $\cala$-modules can be
isomorphic as $\calap$-modules. This effect is controlled by 
a subgroup of the stabilizer, the so-called
\ustab, which in turn is obtained with the help of the following
structure. To each $\calap$-primary $\lambdab$ one can associate a \bihom\
  \be  F_\lambda:\quad \Gs\timeS\Gs \to \complex^\times \ee
that is alternating in the sense that $F_\lambda(\J,\J)\eq1$ for all $\J\iN\Gs$,
that again depends only on the orbit, and that is the same for any two 
conjugate orbits (for the precise definition see appendix \ref{s.b}). 
Every alternating \bihom\ is the commutator cocycle for some
two-cocycle $\F$ on $\Gs$, i.e.
  \be  F_\lambda(\J_1,\J_2) 
  = \F_\lambda(\J_1,\J_2) \,/\, \F_\lambda(\J_2,\J_1) \,,  \Labl ff
and the cohomology class of $\F_\lambda$ is uniquely determined by $F_\lambda$.

\subsection{The \ustab}

The commutator cocycle $F_\lambda$ allows us to single out the
{\em untwisted stabilizer\/} \cite{fusS6} as the subgroup 
  \be  \calu_\lambda := \{ \J\iN\cals_\lambda \,|\, F_\lambda(\J,\JK)\eq1 
  \mbox{ for all } \JK\iN\cals_\lambda \}  \ee
of the full stabilizer $\cals_\lambda$. As shown in \cite{fusS6}, those 
$\cala$-primaries that are isomorphic as $\calap$-modules 
are naturally labelled by characters of the group $\calu_\lambda$. As a 
consequence, the $\cala$-primaries can be denoted by $\Gs$-orbits $\LambdaB$ 
of pairs consisting of a primary label $\lambdab$ of the $\calap$-theory and 
a character $\psu_\lambda$ of the \ustab\ of $\lambdab$. (The equivalence 
relation that defines the classes $\LambdaB$ involves both constituents
of the pair, see formula \Erf eq.)

A second important piece of information that we can extract from the
commutator cocycle $F_\lambda$ is a collection of 
projective representations of $\cals_\lambda$. They are characterized by the
two-cocycle $\F_\lambda$, or rather its cohomology class. 
The theory of projective representations of finite abelian groups 
(for a brief summary see appendix \ref{s.b}) tells us that the projective
\irrep s are labelled by the characters $\psu$ of the {\em untwisted\/}
stabilizer $\calu_\lambda$ and, moreover, that they all have the same dimension 
$d_\lambda$ that was defined in \Erf dl, i.e.
  \be  d_\lambda = \sqrt{\s\lambda\,/\,\u\lambda}   \ee
with
  \be  \s\lambda := |\cals_\lambda| \,,\qquad \u\lambda := |\calu_\lambda| \,.
  \ee
Note that, even though not manifest from their definition,
the numbers $d_\lambda$ are indeed integral \cite{fusS6,bant7}; they constitute
the (additional) ground state
degeneracy of the resolved fixed point in the $\cala$-theory \cite{fusS6}.

Taking all this information together, we arrive at the decomposition
  \be  \calh_\lambda^{} \equiv \calh_\Lambdab = 
  \bigoplus_{\J\in\Gs/\cals_\lambda} \Vpsu \otimes \calhb_{\J\lambdab} \,, 
  \labl{deco}
where $\Vpsu$ is an irreducible projective $\cals_\lambda$-module.
In the special case of the vacuum $\lambda\eq\vac$ of the $\cala$-theory 
the stabilizer is trivial, $\cals_\vac\eq\{\vacb\}$; we then recover
\erf{chi1} with $\lambdab\eq\J\vacb$ and $\Psi\iN\GS\eq\Gs$ identified with
the simple current $\J$. It should also be kept in mind that on the \rhs\ of 
these decompositions only such
\irrep s\ $\mub$ of the $\calap$-theory arise for which the monodromy charges
$Q_\J(\lambda)$ \Erf QJ \wrt all currents $\J\iN\Gs$ vanish; this holds true
simply because it is satisfied \cite{scya6} for all \irrep s\ $\lambda$ in 
the extension and because monodromy charges are constant on $\Gs$-orbits.

\sect{Boundary blocks}\label{s.4}

We have now collected sufficient background material so as to be able
to address in more detail 
the basic ingredient needed for the analysis of boundaries at the chiral level. 
As already mentioned, as a consequence of factorization this ingredient is 
provided by the \cblock s for a single bulk insertion on the disk.
We will refer to these basic objects as the {\em boundary blocks\/} for broken 
bulk symmetries. (When all bulk symmetries are preserved, these blocks are 
also called Ishibashi states.) 

By definition, such boundary blocks are two-point chiral blocks on a world 
sheet ${\dl P}^1$ with the topology of the sphere. In more technical terms,
they are elements of 
$(\calh_\lambda^{}{\otimes}\calh_\lambdap)^\star_{\phantom I}$ -- the \alg ic 
dual of $\calh_\lambda^{}{\otimes}\calh_\lambdap$ -- i.e.\ linear forms
$\calhl^{}\otimeS\calhlp\,{\to}\,\complex$
on the tensor product spaces $\calh_\lambda\otimeS\calh_\lambdap$, which 
satisfy the Ward identities for $\calap$, i.e.\ are invariant under 
the symmetries in the chiral algebra that are preserved. 
In the special case when all bulk symmetries are respected,
every such two-point block is uniquely determined up to a scalar factor.

\subsection{Boundary blocks for broken symmetries}

We are interested in the situation where only the symmetries in the prescribed
subalgebra $\calap$ of the chiral \alg\ are preserved. {}From the decomposition
\erf{deco} it follows that the linear forms we are after are forms on
  \be  \calh_\lambda^{}\otimes\calh_\lambdap
  \equiv \calh_\Lambdab\otimes\calh_\LambdabP
  \cong \Vpsu^{}\otimeS\Vpsup\,\otimes\!\bigoplus_{\J,\JK\in\Gs/\cals_\lambda}\!
  \llb \calhb_{\J\lambdab}^{}\otimeS \calhb_{\JK\lambdabp} \lrb \,,  \Labl72
and hence they are sums of tensor products of linear forms on the tensor
product spaces $\Vpsu^{}{\otimes}\V_{\psu^+}$\, and 
$\,\bigoplus_{\J,\JK}\calhb_{\J\lambdab}^{}{\otimes}\calhb_{\JK\lambdabp}$. 
Moreover, since the $\calap$-symmetries are preserved, the latter forms satisfy
the Ward identities of the $\calap$-theory and hence are
precisely the two-point blocks for the relevant $\calap$-fields. These
blocks in turn can be non-vanishing only when one deals with tensor products of 
conjugate $\calap$-modules, i.e.\ effectively we are working with forms on the
subspace
  \be  \Vpsu^{}\otimeS\Vpsup\,\otimes\!\bigoplus_{\J\in\Gs/\cals_\lambda}\!
  \llb \calhb_{\J\lambdab}^{}\otimeS\calhb_{(\J\lambdab)^+_{\phantom i}} \lrb
  \ee
of the space \Erf72. Moreover, when non-vanishing, then these forms on the 
subspaces $\calh_\mub^{}\otimeS\calh_\mubp$ are uniquely fixed up to 
normalization, just as the two-point blocks for $\cala$ are. Thus, in short, 
they are just the ordinary boundary blocks
  \be  \Betab_\mub:\quad \calh_\mub^{}\otimes\calh_\mubp \to\complex  \Labl73
of the $\calap$-theory.

It follows that the \bashi s can be written as linear combinations
  \be  \sum_{\ssty i\in\{1,2,...,d_\lambda^2\} \atop
  \ssty \J\in\Gs/\cals_\lambda} \x{\J\lambdab;i}^{}\,
  \Bet_{\psu,(i)} \oT \Betab_{\J\lambdab} \,,  \Labl74
where the maps $\Bet_{\psu,(i)}$ constitute some basis of the linear forms
  \be  \beta_\psu:\quad \Vpsu\otimeS \Vpsup \to \complex \,.  \ee
The coefficients $\x{\J\lambdab;i}\iN\complex$ appearing in \Erf74 are 
undetermined at the level of chiral \cft, simply because the Ward identities
for the unbroken symmetries are satisfied independently of the values of
these coefficients. At the level of full \cft, however, we will be able to
determine them; each consistent set of values then corresponds to a 
boundary condition that preserves $\calap$.
  
Now for the $\calap$-part we are already given a natural basis of linear forms,
namely the ordinary boundary blocks \Erf73. On the other hand, at this point
we are still lacking a concrete basis for the linear forms $\beta_\psu$ 
on the degeneracy spaces. Therefore we now turn our attention to those forms 
$\beta_\psu$. As already mentioned, the degeneracy spaces are projective
modules of the stabilizer group $\cals_\lambda$ or, more precisely,
ordinary modules of that twisted group algebra $\complex_{\F_\lambda}
\cals_\lambda$ which corresponds to (the cohomology class of)
the two-cocycle $\F_\lambda$ that was introduced in formula \Erf ff. Thus to 
every $\J\iN\cals_\lambda$ is associated a linear map $R_\psu(\J)$ on $\Vpsu$, 
and these maps represent $\cals_\lambda$ projectively in the sense that
  \be  R_\psu(\J)\, R_\psu(\J') = \F_\lambda(\J,\J')\, R_\psu(\J\J')  \ee
for all $\J,\J'\iN\cals_\lambda$. When the simple current $\J$ is even 
contained in the untwisted stabilizer $\calu_\lambda\,{\subseteq}\,
\cals_\lambda$, whose group \alg\ coincides with the center 
of the twisted group algebra $\complex_{\F_\lambda}\cals_\lambda$,
then it is represented by a multiple of the unit matrix:
  \be  R_\psu(\J) = \psu(\J)\, \one_{d_\lambda} \quad {\rm for}\
  \J\iN\calu_\lambda \,.  \Labl1d
We also note that for any set $\{\JK\}$ of representatives of the quotient
$\cals_\lambda/\calu_\lambda$, the matrices $R_\psu(\JK)$ form a basis of the 
full matrix algebra $M_{d_\lambda}(\Vpsu)$ on $\Vpsu$.
(For further properties of twisted group algebras and their representations, 
see appendix \ref{s.b}.) 

By employing these maps $R_\psu(\J)$ we will now construct a natural basis for 
the linear forms $\beta_\psu$ and thereby for the boundary blocks. To this 
end we first establish an underlying basis $\{\clo_\psi\}$ for the 
endomorphisms of $\Vpsu$. 
Before doing so, however, we pause for a remark about the character $\psu^+_{}$
that first appeared in formula \Erf72 above. It arises via the formula
  \be  \LambdabP = [\lambdabp,\psu^+_{}]  \ee
for the conjugation of $\cala$-\rep s, and thus comes from the $\Gs$-orbit
that is conjugate to the $\Gs$-orbit of $\lambdab$. Now the commutator 
cocycles of conjugate orbits are just each others' complex conjugates (see 
relation \Erf Fs), so $\psu^+_{}$ is a character of the same group
$\calu_\lambda$ as $\psu$. However, it does not, in general, coincide with the 
complex conjugate character $\psu^*$ (see formula \Erf9n for the precise 
definition). Thus in particular the irreducible projective 
$\cals_\lambda$-module $\Vpsup$ in general neither coincides with $\Vpsu$ 
itself nor with the module $\Vpsu^\star$ that is dual to $\Vpsu$ in the sense 
that the \rep\ matrices are hermitian conjugate to those for $\Vpsu$.

\subsection{A natural basis for $\End(\Vpsu)$}

As an intermediate step towards constructing the desired basis for the
linear forms on $\Vpsu\otimeS \Vpsup$, we introduce in this subsection
a basis $\{\clo_\psi\}\,{\equiv}\,\{\clo_\psi^{(\psu)}\}$ 
for the linear maps on the degeneracy space $\Vpsu$.
To this end we first introduce the following concept. For every character 
$\psi$ of the full stabilizer, $\psi\iN\cals_\lambda^*$, the restriction 
of $\psi$ to $\calu_\lambda\,{\subseteq}\,\cals_\lambda$ is an element 
$\pi(\psi)\iN\calu_\lambda^*$. We write 
  \be  \psi\gt\psu \qquad{\rm or }\qquad \psu=\pi(\psi)
  \equiv \psi|_{\calu_\lambda}  \Labl49
to characterize this situation.
Each $\calu_\lambda$-character $\psu$ has $d_\lambda^2$ pre-images under the
projection $\pi$. We will show how these pre-images label the desired
basis of the endomorphisms of $\Vpsu$, according to
$\{\clo_\psi\,|\,\psi\,{\gt}\,\psu\}$. 

We start with the observation that
for $\psi\,{\gt}\,\psu$ the product $\psi^*(\J)R_\psu(\J)$
does not depend on the choice of representative $\J$ of a class in the quotient
$\cals_\lambda/\calu_\lambda$. For, if $\JK\iN\calu_\lambda$, then
  \be  \bearll  \psi^*(\J\JK)\, R_\psu(\J\JK) \!\!
  &= \psi^*(\J)\psi^*(\JK)\, R_\psu(\J) R_\psu(\JK) \\[.6em]
  &=\psi^*(\J)\psu^*(\JK)\,R_\psu(\J)\, \psu(\JK)\,\one_{d_\lambda}
  = \psi^*(\J)\, R_\psu(\J)  \eear \ee
owing to the identity \Erf1d.
Therefore for each $\psi\,{\gt}\,\psu$ we can introduce the endomorphism
  \be  \clo_\psi := \llb\Frac{\u\lambda}{\s\lambda}\lrb^{3/4}\!\!
  \sum_{\J\in\cals_\lambda/\calu_\lambda}\! \psi(\J)^* R_{\psu}(\J)
  \,\in \End(\Vpsu) \,, \labl{clo}
and these maps are well-defined. The following argument shows that the matrices 
$\clo_\psi$ form a basis of the full matrix algebra 
$M_{d_\lambda}\eq\End(\Vpsu)$.\,%
 \futnote{Naively one might also expect that these matrices are (proportional 
to) idempotents. But the non-triviality of the cocycle $\F_\lambda$ spoils 
this property.}
We use the fact, derived in the appendix after relation \Erf29,
that the partial sum over characters yields a non-zero result if and only
if $\JK\iN\calu_\lambda$ or, more precisely,
  \be  \Sumpsipsu\lambda \psi(\JK) = \Frac{\s\lambda}{\u\lambda}\,
  \delta_{\JK\in\calu_\lambda}\, \psu(\JK) \,.  \labl{lem}
The identity \erf{lem} implies that
  \be  \Sumpsipsu\lambda \psi(\JK)\, \clo_\psi
  = \llb\Frac{\s\lambda}{\u\lambda}\lrb^{1/4}\!\!
  \sum_{\J\in\cals_\lambda/\calu_\lambda} \delta^{}_{\JK\J^{-1}\in\calu_\lambda}
  \psu(\JK\J^{-1})\, R_{\psu}(\J)
  = d_\lambda^{1/2}\, \psu(\JK\J^{-1}_\JK)\, R_{\psu}(\J_\JK)  \ee
for all $\JK\iN\cals_\lambda$. Here $\J_\JK$ denotes the chosen representative 
in $\cals_\lambda/\calu_\lambda$ that is in the same class as $\JK$.
These sums, of course, depend on $\JK$, and not just on $\JK$ modulo
$\calu_\lambda$. However, for any set $\{\JK\}$ of representatives of
$\cals_\lambda/\calu_\lambda$, we recover all elements in a basis of the
space of endomorphisms of $\Vpsu$, because the operators 
$R_\psu(\J)$ span this space. 

It follows that the operators $\clo_\psi$ span the space $\End(\Vpsu)$ of 
endomorphisms; for dimensional reasons, they therefore constitute a basis
of $\End(\Vpsu)$, as claimed. Furthermore, since its construction is entirely 
specified in terms of the character $\psu$ and the simple currents in 
$\cals_\lambda$, this basis is indeed completely natural.

Later on we will also need the traces of the operators $\clo_\psi$ and of a
product of two of them, for two $\cals_\lambda$-characters $\psi,\varphi$ that 
restrict to the same $\calu_\lambda$-character $\psu$. To evaluate 
these traces we observe that the trace of $R_\psu(\J)$ is given by
  \be  \tR R_\psu(\J) = \delta^{}_{\J\in\calu_\lambda}\, \psu(\J)\,
  \tR\one_{d_\lambda} = d_\lambda\, \psu(\J)\,\delta^{}_{\J\in\calu_\lambda}
  \Labl oR
(compare the relations \Erf1p and \Erf0t). We then find
  \be  \bearll  \tR \Opsu_\psi \!\!
  &= d_\lambda^{-3/2} \dsty\sum_{\J\in\cals_\lambda/\calu_\lambda}\psi(\J)^*\,
  \tR R_\psu(\J)
  \\{}\\[-.8em]
  &= d_\lambda^{-1/2} \dsty\sum_{\J\in\cals_\lambda/\calu_\lambda} 
    \psi(\J)^*\psu(\J)\, \delta^{}_{\J\in\calu_\lambda}
   = d_\lambda^{-1/2}\, \psi(\J_1)^*\psu(\J_1) = d_\lambda^{-1/2} 
  \eear \ee
(here $\J_1$ is the chosen representative of the class of the unit element
of $\cals_\lambda/\calu_\lambda$)
as well as\,%
 \futnote{In the last line we assume that the cocycle $\F$ has been chosen to 
be standard, which means (see formula \Erf44) that for elements of the basis
of the twisted group algebra the operations of forming the inverse and of
conjugating with an element of the center look the same as in the untwisted 
case. This property of $\F$ can always be achieved by a suitable choice of
basis.}
  \be  \bearll  \tR \clo_\psi^\dagger \clo_\varphi \!\!
  &= d_\lambda^{-3}\! \dsty\sum_{\J_1,\J_2\in\cals_\lambda/\calu_\lambda}
  \psi(\J_1)\,\varphi(\J_2)^*\, \tR R_\psu^\dagger(\J_1)\,R_\psu(\J_2)
  \\{}\\[-.8em]
  &= d_\lambda^{-3}\! \dsty\sum_{\J_1,\J_2\in\cals_\lambda/\calu_\lambda}
  \psi(\J_1)\,\varphi(\J_2)^*\, \tR R_\psu(\J_1^{-1}\J_2) \cdot
  \F_\lambda(\J_1^{-1},\J_2) 
  \\{}\\[-.8em]
  &= d_\lambda^{-3+1}\! \dsty\sum_{\J_1,\J_2\in\cals_\lambda/\calu_\lambda}
  \psi(\J_1)\,\varphi(\J_2)^*\,\F_\lambda(\J_1^{-1},\J_2)\,
  \delta_{\J_1^{-1}\J_2\in\calu_\lambda}\, \psu(\J_1^{-1}\J_2)
  \\{}\\[-.8em]
  &= d_\lambda^{-2}\! \dsty \sum_{\J\in\cals_\lambda/\calu_\lambda}
  \psi(\J)\,\varphi(\J)^*\,\F_\lambda(\J^{-1}\!,\J)\,
  = \Frac{\u\lambda}{\s\lambda} \dsty \sum_{\J\in\cals_\lambda/\calu_\lambda}
  \psi(\J)\,\varphi(\J)^*
  = \delta_{\psi,\varphi}  \,.  \eear \Labl oo
Thus the basis $\{\clo_\psi\,|\,\psi\,{\gt}\,\psu\}$ is orthonormal.
Also, combining these results we learn that the endomorphisms
$d_\lambda^{-1/2}\Opsu_\psi$ form a partition of unity:
  \be  \Sumpsipsu\lambda\! \Opsu_\psi = d_\lambda^{1/2}\, \one_{d_\lambda}^{}
  \,. \Labl2d

\subsection{A natural basis for the boundary blocks}

Our next aim is to associate to each of the endomorphisms $\Opsu_\psi{:}\ 
\Vpsu\,{\to}\,\Vpsu$ with $\psi\,{\gt}\,\psu$ and to each 
$\J\iN\Gs/\cals_\lambda$ a linear form 
  \be  \Bet_\psi^{}\equiv\Bet_\psi^{(\psu;\J\lambdab)}:\quad
  \Vpsu \otimeS \Vpsup \to \complex \,,  \ee
in such a way that the collection of these forms constitutes a basis. (Thus
these maps are required to provide a concrete realization of the basis
elements $\Bet_{\psu,(i)}$ that were introduced in formula \Erf74; in 
particular the label $i$ appearing there turns out to be nothing but a character
$\psi\iN\cals_\lambda^*$ with $\psi\,{\gt}\,\psu$.)
This is achieved by first constructing a suitable non-degenerate linear form 
  \be  \n\equiv\nj:\quad \Vpsu \otimeS \Vpsup\to \complex  \Labl:n
and then defining 
  \be  \Bet_\psi\equiv\Bet_\psi^{(\psu;\J\lambdab)}
  := \n \circ (\Opsu_\psi\tims\id) \,,  \Labl nO
i.e.\ $\Bet_\psi(v\ot w)\df \n(\Opsu_\psi v\oT w)$
for all $v\iN\Vpsu$ and all $w\iN\Vpsup$.

To obtain $\n$, we observe that when restricted to an isotypic component in 
the decomposition \erf{deco}, the boundary block $\BB{:}\ \calhl^{}{\otimes}
\calhlp\,{\to}\,\complex$ of the $\cala$-theory satisfies the Ward identities of
the $\calap$-theory and is therefore proportional to the corresponding boundary
block $\BBB{:}\ \calhbl^{}{\otimes}\calhblp\,{\to}\,\complex$ of the 
$\calap$-theory. This implies that upon choosing any two fixed elements
$\po$ and $\qo$ of $\calhbl$ and $\calhblp$ for which $\BBB(\po\oT\qo)$
is non-zero, the prescription
  \be  \n(v\oT w)
  :=  \frac{\BB(v\ot\po\oT w\ot\qo)}{\BBB(\po\oT\qo)}  \labl n
yields a well-defined linear form on $\Vpsu{\otimes}\Vpsup$. Moreover,
by the non-degeneracy and uniqueness of $\BB$ and
$\BBB$, it is non-degenerate. Note that all forms in question are unique
up to a scalar. Of course, the scalar factor for $\n$ can be different for
different isotypic components of $\calhl$, so that for each 
$\J\iN\Gs/\cals_\lambda$ we obtain a different map $\nj$. Hence for
arbitrary elements $p,q$ of $\calhbj$ and $\calhbjp$, \resp, we have
  \be  \BB(v\ot\pj\oT w\ot\qj) = \nj(v\oT w) \cdot \BBJ(\pj\oT\qj)  \ee
for all $\J\iN\Gs/\cals_\lambda^*$, where $p\,{=:}\,\sumjl \pj$ with
$\pj\iN\calhb_{\J\lambdab}$ and analogously for $q$.

By the linear independence of the endomorphisms $\Opsu_\psi$, also the
forms \Erf nO are linearly independent; since there are $d_\lambda^2$ of
them, they therefore provide us indeed with a basis of the linear forms on
$\Vpsu{\otimes}\Vpsup$. We now combine this basis with the $\calap$-blocks 
\Erf73 with $\mub\eq\J\lambdab$. 
When doing so, we still have to allow for an arbitrary over-all
normalization of the blocks, which cannot be determined at the present
stage. We thus arrive at the linear forms
  \be  \tBeta^{}_{(\mub,\psi)}\equiv\tBeta_{(\mub,\psi)}^\lambda
  := \norm\mub\psi\,d_\lambda^{-1/2}\,\Bet_\psi \oT \bbb\mu  \labl{tBeta}
on $\calhl^{}{\otimes}\calhlp$ with some non-zero $\norm\mub\psi\iN\complex$,
acting as
  \be  \tBeta_{(\J\lambdab,\psi)}(v\ot p\oT w\ot q) 
  := \norm{\J\lambdab}\psi\,d_\lambda^{-1/2}\, \Bet_\psi(v\ot w)
  \cdot \BBJ(\pj\ot\qj)  \Labl bB
for all $v\iN\Vpsu$, $w\iN\Vpsup$, $p\iN\calhl$ and $q\iN\calhlp$.
When $\lambda\eq\Lambdab$, then for $\mub\eq\J\lambdab$ with $\J$ ranging over 
$\Gs/\cals_\lambda$ and $\psi$ over all $\psi\,{\gt}\,\psu$, the forms 
$\tBeta^{}_{(\mub,\psi)}$ constitute a natural basis for the boundary 
blocks on $\calhl^{}{\otimes}\calhlp$ that preserve $\calap$.
In short, the \bashi s are naturally labelled by pairs
$(\mub,\psi)$, one label referring to a primary field $\mub$ of the
$\calap$-theory with vanishing monodromy charge $Q_\Gs(\mu)\eq0$, the
other a character of the {\em full\/} stabilizer, $\psi\iN\cals_\mu^*$.

We remark that with the help of the identity \Erf2d\,%
 \futnote{The introduction of the explicit factor of $d_\lambda^{-1/2}$
in \erf{tBeta} was chosen with hindsight, so as to cancel the corresponding
factor in \Erf2d.}
one checks that the
ordinary boundary blocks of the $\cala$-theory can be expressed in terms of
the blocks $\tBeta_{(\mub,\psi)}$ as
  \be  \BB = \Plupsipsu\lambda \plujl (\norm{\J\lambdab}\psi)^{-1}\,
  \tBeta_{(\J\lambdab,\psi)} \,.  \Labl BJ
Moreover, it is easy to see that the form $\n$ satisfies a `degeneracy space 
Ward identity', i.e.
  \be  \n \circ (y\oT\bfe - \bfe\oT y) = 0  \Labl10
for every $y\iN\End(\Vpsu)$.\,%
 \futnote{Note that in the two terms $y$ acts on different spaces, i.e.\ in
more pedantic notation the identity reads $\n\,{\circ}\,[R_\psu^{}(y)\oT\bfe 
- \bfe\oT R_{\psu^+}(y)]\eq0$. Just like in the usual Ward identities,
in \Erf10 the \rep\ symbols are suppressed.}
This result, in turn, when combined with the 
Ward identities of the $\calap$-theory, immediately implies that the
linear combination \Erf BJ indeed satisfies the Ward identities of 
the $\cala$-theory, i.e.
  \be  \Beta_\lambda \circ \llb Y_n \oT\bfe + (-1)^{\Delta_Y-1}\, \bfe \oT 
  Y_{-n} \lrb = 0  \Labl WI
for all $Y{\in}\,\cala$ ($\Delta_Y$ denotes the conformal weight of $Y$).

\subsection{Scalar products}

For the computation of annulus amplitudes we need to deal with suitable
scalar products of the \bashi s. As a matter of fact,
the boundary blocks are not normalizable; but for every $t\,{>}\,0$ there
exists a modified inner product 
  \be  \langle \tBeta_{(\lambdab,\psi)} 
  |\, \eE^{-(2\pi/t)(L_0\ots\bfe+\bfe\ots L_0-c/12)} \,|
  \tBeta_{(\mub,\varphi)} \rangle  \Labl mf
\wrt which they become normalizable.\,%
 \futnote{While at this point this observation is a mere peculiarity without any
particular application, these modified inner products indeed appear 
in the computation of annulus amplitudes, see subsection \ref{s.62} below. 
In that context, $t$ is the modular parameter of the annulus.}
To substantiate this statement and perform the concrete calculation, we 
compare it to the analogous computation for the boundary blocks of the 
$\calap$-theory. As usual \cite{prss2,prss3}, we normalize the ordinary 
boundary blocks of the $\cala$-theory by prescribing 
the over-all factor in their modified inner product, according to
  \be \langle \Beta_\lambda
  |\, \eE^{-(2\pi/t)(L_0\ots\bfe+\bfe\ots L_0-c/12)} \,|
  \Beta_\mu \rangle = \Frac1{S_{\lambda,\vac}}\,
  \chii_\lambda(2\ii/ t)\, \delta_{\lambda,\mu}^{} \,,  \labl{skp1}
and analogously for the boundary blocks of the $\calap$-theory:
  \be \langle \Betab_\lambdab 
  |\, \eE^{-(2\pi/t)(L_0\ots\bfe+\bfe\ots L_0-c/12)} \,|
  \Betab_\mub \rangle = \Frac1{\Sb\lambda\vac\raisa}\,
  \chib_\lambdab(2\ii/ t)\, \delta_{\lambdab,\mub}^{} \,. \labl{skp2}

What we need as an additional new ingredient is to construct an inner product
on the space of linear maps \Erf nO; this is achieved as follows.
First we construct a scalar product on the degeneracy space $\Vpsu$, i.e.\ 
a sesquilinear map
  \be  
  \kappav:\quad \Vpsu \times \Vpsu \to \complex \, . \ee
The construction uses the invariant scalar products on the modules of
$\cala$ and $\calap$. Namely, on $\cala$ we have an antilinear conjugation map 
$c{:}\ Y\,{\mapsto}\,Y^\dagger$, and on each $\cala$-module $\calhl$ there is
a scalar product $\Kappa{:}\ \calhl{\times}\calhl\,{\to}\,\complex$
which is invariant in the sense that
  \be  \Kappa(Y_n p,p') + \Kappa(p,(Y_n)^\dagger p') = 0 \ee
for all $p,p'\iN\calhl$ and all $Y_n\iN\cala$.
Such a scalar product on an {\em irreducible\/} module is unique up to a scalar.
Moreover, since the subalgebra $\calap$ must be consistent, it is closed under 
$c$, and hence there is an analogous structure on $\calap$-modules.

The scalar product on $\Vpsu$ can now be constructed as follows. The subspace 
$\Vpsu\otimeS \calhbl$ of $\calhl$ inherits a scalar product from the scalar 
product $\Kappa$ of $\calhl$. For any two fixed vectors $v,v'$ in the degeneracy 
space $\Vpsu$, the mapping $(p,p')\,{\mapsto}\,\Kappa(v\ot p,v'\ot p')$
for $p,p'\iN\calhbl$ provides a sesquilinear form.
This sesquilinear form is still unitary \wrtt restriction of the
conjugation $c$ to the subalgebra $\calap$. It must thus be proportional
to the standard scalar product $\kappab$ on $\calhbl$. We call the
constant of proportionality $\kappav(v,v')$.
In formulae,
  \be  \kappav(v,v') = \frac{\Kappa(v\ot p,v'\ot p')}
  {\kappab(p,p')} \,,  \labl{kappav}
where any pair $p,p'\iN\calhbl$ of vectors can be chosen that obeys
$\kappab(p,p')\nE0$. One verifies that $\kappav$ constitutes a 
non-degenerate scalar product on the degeneracy space $\Vpsu$. 

The scalar product $\kappav$ possesses an invariance property as well. Consider
the elements of $\cala$ that commute with the subalgebra $\calap$. Since the 
conjugation $c$ is an automorphism of $\cala$, this commutant is mapped by 
$c$ to itself, so that the commutant, too, comes with its own conjugation. 
The scalar product is now invariant in the sense that
  \be  \kappav(yv,v') = \kappav(v,c(y) v')  \ee
for all $y\iN\End(\Vpsu)$. (In case the commutant should be smaller than
$\End(\Vpsu)$, one simply extends $c$ to the rest of $\End(\Vpsu)$.)

Now since the degeneracy spaces $\Vpsu$, and analogously also $\Vpsup$, carry 
an invariant scalar product, also the space of linear forms on 
$\Vpsu\otimes\Vpsup$ and the space of endomorphisms ${\rm End}(\Vpsu)\cong
(\Vpsu)^*_{}\otimeS\Vpsu$ have a scalar product. For the latter, it is given
by $\kappa_{\End(\Vpsu)}(y,y')\eq\tR(y^\dagger y')$. (Notice that the trace is 
independent
of the scalar product on $\Vpsu$; the latter does enter, however, through the 
hermitian conjugation.) On the space of linear forms, the scalar product reads
  \be  \kappai(\Bet_\psi,\Bet_\varphi) :=
  \sum_{i,j=1}^{d_\lambda} \Bet_\psi(v_i\ot w_j) \cdot \Bet_\varphi(v_i\ot w_j)
  \,,  \ee
where $\{v_i\}$ and $\{w_j\}$ are orthonormal bases of $\Vpsu$ and $\Vpsup$
\wrtt scalar products $\kappav$ and $\kappaw$, \resp.

The non-degenerate form $\n$ defined in \erf n provides us with an isomorphism 
  \be  y \,\mapsto\, \n \circ (y \oT \id)  \Labl35
between $\End(\Vpsu)$ and the space of linear forms on $\Vpsu\otimeS\Vpsup$. 
We would like to check that this isomorphism is a homothety, i.e.\ that it
preserves 
angles. With the orthonormal bases introduced above we need to show that
  \be \sum_{i,j=1}^{d_\lambda} \n(yv_i\oT w_j)^*_{} \n(y'v_i,w_j)
  = \xi\, \tR y^\dagger y'  \Labl36
for some non-zero number $\xi\,{\equiv}\,\xi_\Lambdab\iN\complex$, implying
in particular that 
  \be  \langle \Bet_\vphi | \Bet_\psi \rangle 
  \equiv \kappai(\Bet_\psi,\Bet_\varphi)  
  = \xi\, \tR \Opsu_\psi^\dagger \Opsu_\varphi
  = \xi\, \delta_{\psi,\varphi} \,.  \ee
In components with respect to the two chosen orthonormal bases, the relation 
\Erf36 amounts to $\sum_k (\n)_{ki}^*(\n)_{kj}^{}\eq\xi\delta_{ij}$, while
without reference to the basis of $\Vpsu$ it means
that for all $v,v'\iN\Vpsu$ one has
  \be  \sum_{j=1}^{d_\lambda} \n(v,w_j)^*_{}\, \n(v',w_j) =
  \xi\, \kappav(v,v') \,, \Labl y1
where the sum is over any arbitrary orthonormal basis $\{w_j\}$ of $\Vpsup$. 
The validity of this relation can be established by showing that
$\kappab(p,p')\sum_j\n(v\ot w_j)^*_{}\n(v'\ot w_j)$
provides a non-degenerate and invariant scalar product on $\calhl$.
This is indeed possible; the details are presented in appendix \ref{s.c}.

\subsection{The normalization of the boundary blocks}

We are now finally in a position to determine the value of the 
over-all normalization constant $\norm\lambdab\psi$
that was left undetermined in the definition \erf{tBeta} of the boundary blocks
$\tBeta_{(\lambdab,\psi)}$. To this end we have to prescribe some 
normalization of the modified inner product of these blocks,
much as was done in \erf{skp1} and \erf{skp2} for the ordinary 
boundary blocks. As it turns out, a convenient prescription is
  \be \langle \tBeta_{(\lambdab,\psi)}
  |\,\eE^{-(2\pi/t)(L_0\ots\bfe+\bfe\ots L_0-c/12)}\,| \tBeta_{(\mub,\varphi)}
  \rangle
  = \Frac1{(|\Gs|/\u\lambda)\, \Sb\lambda\vac\raisa}\, \chib_\lambdab(2\ii/ t)\,
  \delta_{\lambdab,\mub}\, \delta_{\psi,\varphi} \,.  \Labl ia
We also observe that relation \erf{oo} amounts to the statement
that the operators $\clo_\psi$ with $\psi\,{\gt}\,\psu$ form an orthonormal 
basis of the space of endomorphisms $\End(\Vpsu)$. It follows that
the constants $d_\lambda^{1/2}(\norm\lambdab\psi)^{-1}$ are precisely
the constants of proportionality between
the scalar product on $\End(\Vpsu)$ and on the space of linear
forms that appear in the relation \Erf36. This implies in particular that
$\norm\lambdab\psi$ actually depends only on the $\calu_\lambda$-character
$\psu$ and not on the particular $\psi$ that extends it to a
character of $\cals_\lambda$.

To proceed, we combine formula \Erf ia with the decomposition 
\Erf BJ of the ordinary boundary blocks $\Beta_\lambda$ of the 
$\cala$-theory. We then find 
  \be  \bearll \langle \Beta_\lambda
  |\, \eE^{-(2\pi/t)(L_0\ots\bfe+\bfe\ots L_0-c/12)} \,|
  \Beta_\mu \rangle \!\!
  &= \dsty \Sumpsipsu\lambda \Sumphiphu\mu \sumjl \sumjm
  (\norm{\J\lambdab}\psu^*)^{-1}_{} (\norm{\J'\mub}\phu^{})^{-1}_{}\,
  \\{}\\[-1.32em] & \hsp{4.9}
  \langle \tBeta_{(\J\lambdab,\psi)}
  |\, \eE^{-(2\pi/t)(L_0\ots\bfe+\bfe\ots L_0-c/12)} \,|
  \tBeta_{(\J'\mub,\varphi)} \rangle 
  \\{}\\[-.6em]
  &= \delta_{\psu,\phu}^{}\,\delta_{\lambdab,\mub}^{}\,
  {\dsty \Sumpsipsu\lambda \sumjl} |\norm{\J\lambdab}\psu|^{-2}_{}\,
  \Frac1{(|\Gs|/\u\lambda)\, \bS_{\J\lambdab,\vacb}\raisa}\,
  \chib_{\J\lambdab}(\frac{2\ii}t)
  \\{}\\[-.8em]
  &= \delta_{\psu,\phu}^{}\,\delta_{\lambdab,\mub}^{}\, d_\lambda^2\,
  \Frac1{(|\Gs|/\u\lambda)\, \Sb\lambda\vac \raisa}\,
  {\dsty\sumjl} |\norm{\J\lambdab}\psu|^{-2}_{}\,
  \chib_{\J\lambdab}(\frac{2\ii}t)
  \\{}\\[-.8em]
  &= \delta_{\lambda,\mu}^{}\, d_\lambda\,
  \Frac1{(|\Gs|/\u\lambda)\, \Sb\lambda\vac \raisa}\,
  |\norm\lambdab\psu|^{-2}_{}\, \chii_\lambda(\frac{2\ii}t)
  \,.  \eear \ee
Here in the last step we have used the information that according to
formula \erf{skp1} the result must 
be proportional to the $\cala$-character $\chii_\lambda$, so that the relation
\erf X between the $\cala$-characters and those of the $\calap$-theory tells 
us in particular that the normalizations $\norm{\J\lambdab}\psu$
in fact do not depend on $\J$, and hence only on $\lambda\eq\Lambdab$.
Moreover, by inspection of formula \erf S for the modular matrix $S$ of the
$\cala$-theory we have
  \be  S_{\Lambdab,\vac}
  = \Frac{|\Gs|}{\sqrt{\s\lambda\u\lambda}}\, \Sb\lambda\vac \,.  \Labl23
Thus the normalization condition \erf{skp1} also allows us to
determine the explicit value of the constants $\norm\lambdab\psu$, namely
$|\norm\lambdab\psu|^{-2}d_\lambda\u\lambda/|\Gs| \eq \sqrt{\s\lambda\u\lambda}
/|\Gs|$, and hence simply\,%
 \futnote{We admit that our conventions were chosen with some hindsight.}
  \be  |\norm\lambdab\psu| = 1 \,.  \ee
Note that we determine these constants only up to a phase. Manifestly, we 
cannot do better, because the relations \erf{skp1} and \erf{skp2} determine 
the boundary blocks also only up to a phase.

To conclude this section, we summarize our results about the natural
basis of boundary blocks for symmetry breaking boundary conditions. For
every primary $\lambda\eq\Lambdab$ of the $\cala$-theory we have 
$(|\Gs|/|\cals_\lambda|)\,{\cdot}\,d_\lambda^2$ basis elements 
$\tBeta_{(\mub,\psi)}$ which
are labelled by those $\calap$-primaries $\mub$ that are on the $\Gs$-orbit 
of $\lambdab$ and by the $\cals_\lambda$-characters $\psi$ that restrict to
$\psu$. These boundary blocks obey the normalization condition \Erf ia.

\sect{The classifying algebra} \label{s.5}

In this section we turn to the level of {\em full\/} \cft. We explain how 
the representation theory of a classifying algebra allows us to determine the
boundary conditions for a given \cft, and we explicitly construct the
classifying algebra for boundary conditions that preserve a prescribed 
sub\alg\ $\calap$ of the bulk symmetries.

\subsection{Boundary conditions and reflection coefficients}\label{s.51}

Because of factorization, a \bc\ should essentially be characterized by a 
consistent collection of one-point correlation functions for all bulk fields 
$\pho\lambda$ on the disk \cite{card9,cale,lewe3,prss3}. Thus in order to 
classify the \bc s, one needs to find all consistent one-point \corfu s 
$\langle\pho\lambda\rangle$ for those fields. As explained in \cite{fuSc6}, 
the correlation functions on a surface with boundaries are linear combinations 
of blocks on its Schottky cover; for the disk 
this oriented cover has the topology of the sphere, so that the relevant 
\cblock s are those studied in section \ref{s.4}. Our task is
now to determine the coefficients that give the correct physical correlators. 

For a more detailed discussion it is convenient to use the language of vertex 
operators and operator products.
For every vector $v\oT\tilde v\iN \calh_\lambda^{}\otimeS\calh_{\tilde\lambda}$
we have a vertex operator $\pho\lambda(v\ot\tilde v;z)$. Such vertex operators
are suitable linear combinations of pairs of chiral vertex operators, the
correlators of which are nothing but the boundary blocks discussed above.
In view of the description \Erf74 of the boundary blocks and their precise
definition \erf{tBeta}, we are thus looking 
for coefficients $\xi_{\mub,\psi}$ such that the value of the one-point 
correlator for $\pho\lambda(v\ot\tilde v;z)$ on the disk at $z\eq0$ is
  \be  \langle \pho\lambda(v\ot\tilde v;z{=}0) \rangle
  = \Sumpsipsu\lambda \sum_{\J\in\Gs/\cals_\lambda} \x{\J\lambdab,\psi}\,
  \tBeta_{(\J\lambdab,\psi)} (v\ot\tilde v) \,.  \labl{oben}
Moreover, it follows from the results at the chiral level that this correlator 
can be non-vanishing only for $\tilde\lambda\eq\lambdap$, which we therefore
assume from now on.

Note, however, that the index structure of the vertex operator in formula
\erf{oben} is tailored to the case of a closed orientable surface, where the 
bulk fields $\pho\lambda$ are the only fields present. In contrast, for 
surfaces with boundaries, where there are also boundary fields $\Psi(x)$,
and allowing for boundary conditions that break part of the bulk symmetries, 
the index structure can actually be more complicated. Accordingly
we have to be careful when interpreting formula \erf{oben}. 
What we have to implement correctly is the fact that, while chiral vertex 
operators can definitely be extracted from the three-point \cblock s on 
${\dl P}^1$, their concrete form does depend on which chiral symmetries are 
preserved. In the situation of interest to us we are not allowed to employ all
symmetries of the bulk, but rather we must take the three-point blocks of
the orbifold theory with symmetry $\calap$ for extracting the chiral vertex
operators. In other words, we must
take into account that states in different $\calap$-submodules of
$\calh_\lambda^{}\eq\bigoplus_{\J\in\Gs/\cals_\lambda}\Vpsu\otimeS
\calhb_{(\J\lambdab,\psu)}$ can cause different excitations on the boundary 
and can thus be reflected differently. Note that, unlike in formula \erf{deco},
here we have attached the label $\psu$ also to the $\calap$-modules $\calhb$, 
so as to indicate that the reflection at the boundary may also depend on the 
particular $\cala$-module into which a given $\calap$-module is embedded. 
In addition, we have to account for the dimensionality of the projective 
$\cals_\lambda$-module $\Vpsu$, which amounts to using characters 
$\psi\iN\cals_\lambda^*$ instead of $\psu\iN\calu_\lambda^*$.

Accordingly, when studying the behavior of bulk fields close to the boundary,
for vectors $v\oT\tilde v\iN(\Vpsu{\otimes}\calhb_{(\lambdab,\psu)})\otimeS
(\V_{\psu^+}{\otimes}\calhb_{(\lambdabp,\psu^+)})\,{\subset}\,
\calh_\lambda^{}\otimeS\calh_\lambdap$ 
we must work with vertex operators that are labelled as
  \be  \phi_{(\lambdab,\psi),(\lambdabp,\psi^+)}(v\ot\tilde v;z) \,.  \Labl52
As for the \corfu s, this means that in place of \erf{oben} we are interested
in the individual summands
  \be  \langle \phi_{(\lambdab,\psi),(\lambdabp,\psi^+)}(v\ot\tilde v;z{=}0)
  \rangle = \x{\lambdab,\psi}\,
  \tBeta_{(\lambdab,\psi)} (v\ot\tilde v) \,.  \labl{unten}
In order to determine the coefficients $\x{\mub,\psi}$ in this relation, 
we study the operator product expansion describing the excitation that a 
bulk field causes on the boundary when it approaches the boundary; this
operator product reads
  \be  \phi_{(\lambdab,\psi),(\lambdabp,\psi^+)}(r\eE^{\ii\sigma})
  = \sum_\mub (1{-}r^2)^{-2\Delta_\lambdab+\Delta_\mub}_{}\,
  \Rc a{(\lambdab,\psi)}\mub\,\Psi^{aa}_\mub(\eE^{\ii\sigma})
  + \mbox{descendants} \quad\; {\rm for}\;\ r\,{\to}\,1 \,.  \labl{oben2}
Comparing this expansion with relation \erf{unten} we learn that 
    \be  \x{\mub,\psi} = \Rc a{(\mub,\psi)}\vacb\, \langle\Psi^{aa}_\vacb\rangle
    \,.  \Labl xR
In words, up to a normalization given by the (constant) one-point correlator
of a boundary vacuum field $\Psi^{aa}_\vacb$, the coefficients $\x{\mub,\psi}$ 
are equal to the {\em reflection coefficients\/} $\Rc a{(\mub,\psi)}\vacb$.
   
We pause to comment on the index structure of the boundary fields 
$\Psi^{ab}_\mub(x)$. The underlying three-point blocks for the operator product 
\erf{oben2} are those of the orbifold theory, because boundary fields are
involved and the latter only need to preserve the symmetries in $\calap$.
As a consequence, the boundary field carries a chiral label $\mub$ of
the orbifold theory. In addition, there are two labels $a,b$ which account
for the fact that the insertion of a boundary field can change the \bc.
(And finally, in order to account for annulus coefficients that are bigger
than one -- which can appear for $\mub\nE\vacb$ -- one must allow for an 
additional degeneracy label, which we suppress.)
The presence of these boundary labels on the \rhs\ of \erf{oben2} tells us 
that, in contrast to \cft\ on surfaces that are closed and orientable,
on surfaces with boundaries the locality and factorization constraints for 
the \corfu s do not, in general, possess a unique solution.
Rather, there are several consistent collections of reflection coefficients
$\Rc a{(\lambdab,\psi)}\vacb$, and as a consequence there are several solutions 
  \be  \langle\phi_{(\lambdab,\psi),(\lambdabp,\psi^+)}\rangle
  = \langle\phi_{(\lambdab,\psi),(\lambdabp,\psi^+)}{\rangle}_a  \ee
which are indexed by the \bc s.

Note that up to this point it was not necessary to specify the values that the
boundary label $a$ can take.
To determine the possible boundary conditions, we analyze the factorization of
bulk-bulk-boundary correlators in much the same manner as
\cite{lewe3,sasT2,prss3} for \bc s that preserve all of $\cala$. This is
possible because, by the requirement that $\calap$ is a consistent chiral 
algebra, the orbifold chiral blocks obey the usual factorization rules.
Concretely, we consider two different factorization limits of the disk \corfu\
  \be  \langle \phi_{(\lambdab_1,\psi_1),(\lambdab_1^+,\psi_1^+)}(z_1)\,
  \phi_{(\lambdab_2,\psi_2),(\lambdab_2^+,\psi_2^+)}(z_2) {\rangle}_a  \ee
involving two bulk fields. On one hand we can use the operator product between
bulk fields (this is an operator product respecting the full $\cala$-symmetry,
although for fields that are only $\calap$-primaries but may be 
$\cala$-descendants)
and afterwards the operator product \erf{oben2}, so as to express the correlator
in terms of bulk operator product coefficients and a single reflection
coefficient $\Rc a{(\lambdab_3,\psi_3)}\vacb$. On the other hand, applying the
expansion \erf{oben2} twice expresses the correlator in terms of two 
bulk-boundary operator products, i.e.\ two reflection coefficients. The latter 
are to be understood as prefactors of a four-point block on the projective 
line ${\dl P}^1$, and since boundary insertions are involved, these are 
four-point blocks of the orbifold theory. The two different factorizations 
correspond to such blocks in different channels, so that for their comparison 
one must relate them through fusing matrices or, to be precise, through fusing 
matrices of the $\calap$-theory. Such matrices exist 
because by assumption the orbifold chiral blocks come with a \kzc.

Taking everything together we arrive at a relation of the form
  \be  \Rc a{(\lambdab_1,\psi_1)}\vacb\, \Rc a{(\lambdab_2,\psi_2)}\vacb
  = \sum_{\lambdab_3,\psi_3}
  \tN{(\lambdab_1,\psi_1)}{(\lambdab_2,\psi_2)}{(\lambdab_3,\psi_3)}\,
  \Rc a{(\lambdab_3,\psi_3)}\vacb \,,  \labl{res1}
where the numbers $\TN$ are combinations of bulk operator product coefficients 
and fusing matrices of the $\cala$- {\em and\/} of the $\calap$-theory. Notice 
that the fact that quantities of both the original
and the orbifold theory appear is in accordance with the index structure
of the vertex operators \Erf52, in which there appear individual fields 
$\lambdab$ of the orbifold theory rather than $\Gs$-orbits of such fields, 
but also a character $\psi$ that keeps track of the information
as submodule of which $\cala$-module a given $\calap$-module occurs.

None of the constituents of the numbers $\TN$ depends on the boundary label 
$a$. The result \erf{res1} can therefore be interpreted as follows. The 
conformally invariant boundary conditions that preserve $\calap$ correspond to
\onedim\ representations of some algebra, which we call the {\em classifying 
algebra\/} and denote by \clAb. It is expected \cite{fuSc6}
that the \alg\ \clAb\ shares most properties of fusion \alg s, i.e.\ it should
be a commutative associative semisimple algebra, so that in particular all
its \irrep s are \onedim. These properties imply
the existence of a diagonalizing matrix $\tS$ through which the structure 
constants of \clAb\ are expressible via an analogue of the Verlinde formula.

It is worth stressing that the two labels of the diagonalizing matrix $\tS$
are on a rather different footing; the row index labels the basis of the 
classifying algebra \clAb\ which is given by the allowed boundary blocks, while
the column index labels the irreducible representations of \clAb. In the case 
of boundary conditions that preserve the full bulk symmetry (and where
the pairing for the labels of the bulk fields is given by charge conjugation,
i.e.\ $\tilde\lambda\eq\lambdap$), it has already been argued long ago
\cite{card9} that $\tS$ is the modular matrix that implements the modular
transformation $\tau\,{\mapsto}\,{-}1/\tau$ on the characters. In this case
the \cla\ is just the fusion rule \alg\ and the reflection coefficients are
the generalized quantum dimensions; in particular there is a natural
correspondence between the two types of labels.

In the general case, this natural correspondence does not persist. But it has
been seen that even in more general situations (see \cite{prss3,fuSc5}
for an example) nevertheless the two sets of labels are still related by 
modular transformations. Moreover, it can be expected that the boundary 
conditions are labelled by {\em orbits\/} of fields rather than individual 
fields, as in \cite{fuSc5}. That this is indeed the case can be seen as follows.
As for the labels $\lambdab$ of the boundary blocks, only those occur 
which appear in the decomposition of some $\cala$-module, which means
that they satisfy $Q_\J(\lambda)\eq0$ for every $\J\iN\Gs$. In orbifold
terminology, we are only dealing with the untwisted sector of the orbifold 
or, in other words, along the `space' direction of the
torus only the trivial twist by the identity occurs. This implies that after
a modular S-transformation, only the identity appears as a twist
in the `time' direction of the torus, which in turn tells us that we must not
perform the usual orbifold projection in the twisted sector.
Translating this back into simple current terminology, we arrive at the
statement that the boundary conditions must not be labelled by
individual primary fields of the $\calap$-theory, but rather by
$\Gs$-{\em orbits\/} of $\calap$-primaries.
On the other hand, in the `time' direction we start with arbitrary 
twists, since the labelling is by individual primary fields; it follows that
after the S-transformation arbitrary twists
in the `space' direction occur in the orbifold. Thus in the labelling of the
\bc s {\em all\/} $\Gs$-orbits appear, not just those with vanishing monodromy
charges, i.e.\ not just the ones in the untwisted sector. 
Moreover, by comparison with the S-transformation of the $\cala$-characters
one is led to expect that these orbits are to be combined with the characters 
of the relevant \ustab; as we will see below, this provides us indeed with a
consistent ansatz for the \cla.

\subsection{The matrix $\tS$}

As advocated above, the boundary blocks are in one-to-one correspondence
with the elements of a basis of the \cla\ \clAb, while the $\calap$-preserving 
boundary conditions are in one-to-one correspondence with the (isomorphism 
classes of) \onedim\ \irrep s of \clAb. Thus 
a basis of \clAb\ is labelled by pairs $(\lambdab,\psi_\lambda)$
consisting of a primary label $\lambdab$ of the $\calap$-theory in the
untwisted sector (i.e.\ $Q_\J(\lambda)\eq0$ for all $\J\iN\Gs$) and a
character $\psi_\lambda\iN\cals^*_\lambda$ of the stabilizer 
of $\lambdab$, while the arguments at the end of the previous subsection
tell us that the \onedim\ irreducible \clAb-\rep s are labelled by 
$\Gs$-orbits $\RhoB$ of pairs consisting of an arbitrary primary label $\rhob$ 
of the $\calap$-theory and a character $\psu_\rho$ of the \ustab\ of $\rhob$.

According to our general expectations the
\cla\ \clAb\ should possess most properties of fusion \alg s,
in particular there should exist a diagonalizing square matrix $\tS$.
Note that according to the previous remarks the row and column labels of 
this matrix are on a rather different footing, so that at this point it is 
still far from obvious that the two sets of labels indeed have equal size.
Our strategy is now to start by making an educated ansatz for the matrix $\tS$
and then develop the \cla\ and its \rep\ theory along analogous lines as one
may study fusion \alg s by starting from the modular S-matrix $S$.
We stress that, unlike the considerations in the previous section, here we 
are indeed making an {\em ansatz\/}, and it will be necessary to support this 
ansatz by performing various consistency checks, the most basic one being that
$\tS$ is manifestly a square matrix. 
But once one accepts this ansatz, the arguments presented
in the previous subsection allow us to learn more about how the
fusing matrices in an integer spin simple current extension are related
to the fusing matrices of the original theory. 

As a matter of fact, by combining
the considerations that relate symmetry breaking, orbifolds and integer spin
simple current extensions with the results about simple current extensions
obtained in \cite{fusS6}, it follows that, up to normalizations, there is 
a natural candidate for the matrix $\tS$. It reads
  \be  \tS_{(\lambdab,\psi_\lambda),\RhoB}
  := \Frac{|\Gs|}{\sqrt{\s\lambda\u\lambda\s\rho\u\rho}}
  \sum_{\J\in\cals_\lambda\cap\calu_\rho} \psi^{}_\lambda(\J)\,
  \psu_\rho(\J)^*\, S^\J_{\lambdab,\rhob} \,.  \Labl tS
Here the matrices $S^\J$ are those which appear \cite{fusS6} in the modular 
S-matrix of the simple current extension; upon a canonical\,%
 \futnote{The choice of canonical basis still leaves some residual 
freedom in the normalization, which remains to be clarified.}
normalization of the one-point \cblock s with insertion $\J$ on the torus, the 
matrix $\SJ$ also represents the modular S-transformation on these blocks 
\cite{bant6}. (For the convenience of readers who are not familiar with the 
pertinent results of \cite{fusS6}, we summarize them in appendix \ref{s.a}. 
For a brief account, see also section 3 of \cite{fuSc10}.)

Note that at the chiral level, where one deals with \cft\ on a complex curve,
there is no direct influence of boundaries \cite{fuSc6}. The chiral \cft\ 
structures that are related to the matrices $S^\J$ which enter the discussion 
here are thus logically independent of any boundary data; they have passed 
independent tests \cite{fusS6,bant6} in the context of closed \cft. In the 
present situation, where the $\calap$-theory can be regarded as an orbifold 
of the $\cala$-theory, as compared to \cite{fusS6} there is actually even 
further evidence for the existence of the matrices $S^\J$. Namely, under a 
finiteness assumption on the codimension of a certain subspace of the vacuum 
module, it has been proven in \cite{zhu3} (see also \cite{dolm6,miya6}) that 
one can associate a modular S-transformation 
matrix to the chiral blocks on the torus for 
arbitrary descendants of the vacuum. In our case, we are precisely concerned 
with one-point blocks for descendants of the vacuum of the $\cala$-theory (which
are {\em not\/} descendants of the vacuum in the $\calap$-theory, though).
In connection with the reasoning of \cite{fusS6} one may say
that if an extension by integer spin simple currents is possible at all, then
such matrices $S^\J$ must necessarily exist in order to
comply with the general result of \cite{zhu3}.

As a first consistency check, we consider the special case where
$Q(\rho)\,{\equiv}\,0$. These boundary conditions correspond precisely to 
orbits that furnish primary fields in the extended theory. On the
other hand, the boundary conditions that respect the full bulk symmetry
should also be recovered from our ansatz, since a fortiori they preserve 
the subalgebra $\calap$. According to \cite{card9}, these boundary
conditions correspond to primary fields of the $\cala$-theory.
Indeed, the following consideration shows that for $Q(\rho)\,{\equiv}\,0$ we 
recover the modular S-matrix of the $\cala$-theory. The latter can
be expressed through the matrices $S^\J$ as in \erf S, i.e.
  \be  S_{\LambdaB,\RhoB}
  = \Frac{|\Gs|}{\sqrt{\s\lambda\u\lambda\s\rho\u\rho}}
  \sum_{\J\in\calu_\lambda\cap\calu_\rho} \psu_\lambda(\J)\,
  \psu_\rho(\J)^*\, S^\J_{\lambdab,\rhob} \,,  \Labl5S
where both $\lambdab$ and $\rhob$ have monodromy charge zero. Because of the 
latter property, we know that for every $\J\iN\cals_\lambda{\setminus}\,
\calu_\lambda$ there exists at least one $\JK\iN\cals_\lambda$ such that
  \be  S^\J_{\lambdab,\rhob} = S^\J_{\JK\lambdab,\rhob}
  = F_\lambda(\JK,\J)\cdot 1\cdot S^\J_{\lambdab,\rhob}  \Labl ss
with $F_\lambda(\JK,\J)\nE1$, from which we conclude that
$S^\J_{\lambdab,\rhob}\eq0$ for all $\J\iN\cals_\lambda{\setminus}\,
\calu_\lambda$. It follows that for $Q_\Gs(\rho)\eq0$ the $\J$-summations 
in the two expressions \Erf tS and \Erf5S actually extend over the 
same range; moreover, we then have $\psu_\lambda(\J)\eq\psi_\lambda(\J)$ for
all $\J$ that appear in the sum, so the two expressions indeed are equal, i.e.
  \be  \tS_{(\J\lambdab,\psi_\lambda),\RhoB} = S_{\LambdaB,\RhoB}  \ee
for all $\J\iN\Gs$ and all $\psi_\lambda\,{\gt}\,\psu_\lambda$.

\subsection{Properties of $\tS$}

Let us now establish further properties of the matrix $\tS$ that we defined in
\Erf tS. As a matter of fact, we first need to check that $\tS$ is well-defined,
i.e.\ does not depend on the choice of representative of the $\Gs$-orbit 
of the pair $(\rhob,\psu_\rho)$. To do so, we need 
the explicit form of the equivalence relation, which reads
$(\rhob,\psu_\rho)\sim \J'(\rhob,\psu_\rho)\eq(\J'\rhob,\JJp\psu_\rho)$
(see formulae \Erf eq and \Erf JJ). We observe that
  \be  \JJp\psu_\rho(\J)^* S^{\J}_{\lambdab,\J'\rhob}
  = [F_\rho(\J',\J)^*\,\psu_\rho(\J)]^* \cdot \eE^{2\pi\ii Q_{\J'}(\lambda)}_{}
  \, F_\rho(\J',\J)^*\, S^{\J}_{\lambdab,\rhob}
  = \psu_\rho(\J)^* S^{\J}_{\lambdab,\rhob} \,,  \Labl51
where we used the simple current property \Erf QF of $S^\J$ and the fact that
$\lambdab$ is in the untwisted sector.
This tells us that the corresponding part of formula \Erf tS, and hence the 
whole matrix $\tS$, is indeed independent of the choice of representative.

We may define an analogous transformation as in this equivalence relation
also when dealing with characters of full stabilizers, i.e.\ also for the row 
index of $\tS$, namely
  \be  \J'\,(\lambdab,\psi_\lambda):= (\J'\lambdab,\JJp\psi_\lambda)  \Labl eQ
for all $\J'\iN\Gs$, with
  \be  \JJp\psi_\lambda(\J):= F_\lambda(\J',\J)^*\,\psi_\lambda(\J)
  \,.  \Labl jJ
By the \bihom\ property of $F$, $\JJp\psi_\lambda$ is again a character of
$\cals_\lambda$. It then follows that $\tS$ satisfies the standard
simple current relation, too, i.e.\ we have
  \be  \tS_{\J'(\lambdab,\psi_\lambda),\RhoB}
  = \eE^{2\pi\ii Q_{\J'}(\rho)}_{} \cdot
  \tS_{(\lambdab,\psi_\lambda),\RhoB}  \Labl SQ
for every $\J'\iN\Gs$. This holds because in each term in the $\J$-summation
on the \rhs\ of \Erf tS the factor of $F_\lambda(\J',\J)^*$ that comes from 
the action of $\J'$ on $\psi_\lambda$ cancels against the factor
$F_\lambda(\J',\J)$ that accompanies the phase $\eE^{2\pi\ii Q_{\J'}(\rho)}$
in the simple current relation for $S^\J$.

Next we note that the matrix $\tS$ is (in general) not symmetric; in fact it 
does not even make sense to talk about symmetry, because the label sets for the
rows and columns are different. It does, however,
make sense to talk about invertibility and unitarity. A direct calculation
shows that $\tS$ is invertible, the inverse being given by
  \be  (\tS^{-1})^{\RHoB,(\lambdab,\varphi)}_{}
  = \Frac{\u\lambda}{|\Gs|}\, \tS^*_{(\lambdab,\varphi),\RHoB} \,.  \Labl3U
That \Erf3U is a right-inverse is seen by
  \be  \hsp{-.99} \begin{array}{ll}
  \dsty \sum_\rhoB \sum_{\psu_\rho\in\calu_\rho^*}
  \tS_{(\lambdab,\psi_\lambda),\RhoB}\,
  \llb \tS_{(\mub,\psi_\mu),\RhoB}\lrb_{}^* \!\!\!
  &= \Frac{|\Gs|^2}{\sqrt{\s\lambda\u\lambda\s\mu\u\mu}}
  \dsty\sum_\rhoB \sum_{\J\in\cals_\lambda\cap\cals_\mu\cap\calu_\rho}
  \!\Frac1{\s\rho}\, \psi_\lambda(\J) \psi_\mu(\J)^* \,
  S^{\J}_{\lambdab,\rhob}\, (S^{\J}_{\mub,\rhob})^*
  \\{}\\[-.8em]
  &= \Frac{|\Gs|}{\sqrt{\s\lambda\u\lambda\s\mu\u\mu}}
  \dsty\sum_{\J\in\cals_\lambda\cap\cals_\mu} \psi_\lambda(\J) \psi_\mu(\J)^*
  \sum_\rhob S^{\J}_{\lambdab,\rhob}\, (S^{\J}_{\mub,\rhob})^*
  \\{}\\[-.8em]
  &= \Frac{|\Gs|}{\s\lambda\u\lambda}\, \delta_{\lambdab,\mub}^{}
  \dsty\sum_{\J\in\cals_\lambda} \psi_\lambda(\J) \psi_\mu(\J)^*
   = \Frac{|\Gs|}{\u\lambda}\cdot \delta_{(\lambdab,\psi_\lambda),
  (\mub,\psi_\mu)}^{}
  \,.  \eear  \Labl1U
Here in the first step we inserted the definition \Erf tS and performed the
$\psu_\rho$-summation,
while in the second step we replaced
$\sum_\rhoB$ by $\sum_\rhob\,(\s\rho/|\Gs|)$, which is possible owing 
to the fact that $Q_J(\lambda)\eq0\eq Q_J(\mu)$ and that $\J\iN\calu_\rho$,
and furthermore we dropped taking the intersection with $\calu_\rho$ in the 
$\J$-summation. To see that this change in the summation range is allowed,\,%
 \futnote{Compare also the remarks before eq.\ (C.2) in \cite{fusS6}.}
let first $\J\,{\not\in}\,\cals_\rho$; then according to \Erf00 we simply have
$S^\J_{\lambdab,\rhob}\eq 0\eq S^\J_{\mub,\rhob}$. Otherwise, i.e.\ when
$\J\iN\cals_\rho{\setminus}\,\calu_\rho$, there must exist a
$\J'\iN\cals_\rho$ with $F_\rho(\J',\J)\ne1$, and we have
  \be  S^\J_{\lambdab,\rhob} = S^\J_{\lambdab,\J'\rhob}
  = F_\rho(\J',\J)^*\, \eE^{2\pi\ii Q_{\J'}(\lambda)}\, S^\J_{\lambdab,\rhob}
  = F_\rho(\J',\J)^*\, S^\J_{\lambdab,\rhob} \,,  \Labl FQ
where again we use that $Q(\lambda)\eq0$; thus in this case we have 
$S^\J_{\lambdab,\rhob}\eq0$ as well.

To check that the matrix \Erf3U is also a left-inverse of $\tS$, we start by
calculating
  \be \bearll
  \dsty \sumbo\lambda \sum_{\psi_\lambda\in\cals_\lambda^*}
  \tS_{(\lambdab,\psi_\lambda),\RhoB} \cdot \Frac{\u\lambda}{|\Gs|}\,
  \tS_{(\lambdab,\psi_\lambda),\SigmaB}^* \!\!\!
  &= \Frac{|\Gs|}{\sqrt{\s\rho\u\rho\s\sigma\u\sigma}}
  \dsty\sumbo\lambda \sum_{\J\in\calu_\rho\cap\calu_\sigma\cap\cals_\lambda}\!\!
  \psu_\rho(\J)^*\psu_\sigma(\J)\,
  S^{\J}_{\lambdab,\rhob}\, (S^{\J}_{\lambdab,\sigmab})^*_{}
  \\{}\\[-.8em]
  &= \Frac{|\Gs|}{\sqrt{\s\rho\u\rho\s\sigma\u\sigma}}
  \dsty\sum_{\J\in\calu_\rho\cap\calu_\sigma}
  \psu_\rho(\J)^*\psu_\sigma(\J)\, \!\!\!
  \sumbo\lambda S^{\J}_{\lambdab,\rhob}\, (S^{\J}_{\lambdab,\sigmab})^*_{}
  \,.  \eear \ee
Here after first performing the $\psi_\lambda$-summation,
we dropped taking the intersection with $\cals_\lambda$ in the $\J$-summation, 
which is allowed for the same reason as above.\,%
 \futnote{Note, however, that in the present case we would {\em not\/} be
allowed to drop the {\em untwisted\/} stabilizer $\calu_\lambda$ if it were
present, because for currents in $\cals_\lambda{\setminus}\,\calu_\lambda$ the
above reasoning would not go through: in \Erf FQ one would now have a factor of
$\eE^{2\pi\ii Q_{\J'}(\rho)}$, which unlike $\eE^{2\pi\ii Q_{\J'}(\lambda)}$
is not necessarily equal to one, since $\rhob$ can be in any twist sector.}
Next we extend the $\lambdab$-summation to all sectors by inserting a projector
and use the unitarity of $S^\J$ (see formula \Erf ip), so as to obtain
  \be \hsp{-.8} \bearll
  \dsty \sumbo\lambda \sum_{\psi_\lambda\in\cals_\lambda^*}
  \tS_{(\lambdab,\psi_\lambda),\RhoB} \cdot \Frac{\u\lambda}{|\Gs|}\,
  \tS_{(\lambdab,\psi_\lambda),\SigmaB}^* \!\!
  &= \Frac1{\s\rho\u\rho}\, \dsty\sum_{\J'\in\Gs} \delta^{}_{\rhob,\J'\sigmab}
  \sum_{\J\in\calu_\rho} \psu_\rho(\J)^*\, F_\sigma(\J',\J)^*\, \psu_\sigma(\J)
  \\{}\\[-1.1em]
  & \hsp{-6.3}
   = \Frac1{\s\rho}\, \dsty\sum_{\J'\in\Gs} \delta^{}_{\rhob,\J'\sigmab}\,
  \delta^{}_{\psu_\rho,\JJp\psu_\sigma}
  = \Frac1{\s\rho}\, \dsty\sum_{\J'\in\Gs}
  \delta^{}_{(\rhob,\psu_\rho),\J'(\sigmab,\psu_\sigma)}
  = \delta^{}_{\RhoB,\SigmaB}
  \,.  \eear \ee
(Here we also used the fact that $\rhob\eq\J'\sigmab$ already implies 
$\calu_\rho\eq \calu_\sigma$ and $\cals_\rho\eq\cals_\sigma$.)

The fact that $\tS$ has a two-sided inverse means
in particular that $\tS$ is a square matrix. This implies the sum rule
  \be  \sumbo\lambda |\cals_\lambda| = \sum_\rhoB |\calu_\rho| \,.  \Labl sr
In words: the number of primary {\em fields\/} in the untwisted (i.e., charge 
zero) sector, counted with their (full) stabilizer, is the same as the number 
of {\em orbits\/} in all sectors, counted with their untwisted stabilizer.

It is also worth pointing out that $\tS$ is (in general) not unitary.
Of course, we could redefine the matrix $\tS$ so as to make it unitary; however,
this would spoil some other nice properties of $\tS$
and hence we refrain from doing so.

\subsection{Conjugation}

Since the row and column labels of $\tS$ are on different footings,
there are two distinct matrices which are candidates for conjugations,
namely
  \be  \tC := \tS\, \tS^{\rm t}  \labl tC
and
  \be  \cC := \tS^{\rm t}\,U\,\tS  \,;  \labl cC
here the superscripts $B$ and $\calb$ indicate that the entries of these two
matrices are labelled by boundary blocks and boundary states (i.e.\ \bc s),
\resp. The presence of the diagonal matrix $U$, defined as
  \be  U_{(\lambdab,\psi_\lambda),(\mub,\psi_\mu)} 
  := \Frac{\u\lambda}{|\Gs|}\,\delta_{(\lambdab,\psi_{\lambda}),(\mub,\psi_\mu)}
  \,,  \labl U
in \Erf cC accounts for the natural weight of the boundary blocks, cf.\ 
for instance formula \Erf3U.

Both matrices are manifestly symmetric. To establish further properties, we
write them in the form
  \be \begin{array}{ll}
  \tC_{(\lambdab,\psi_{\lambda}),(\mub,\psi_\mu)} \!\!\!
  &= \Frac{|\Gs|}{\sqrt{\s\lambda\u\lambda\s\mu\u\mu}}
  \dsty\sum_\rhob \sum_{\J\in\cals_\lambda\cap\cals_\mu}
  \psi_\lambda(\J) \psi_\mu(\J^{-1}) \,
  S^\J_{\lambdab,\rhob}\, S^{\J^{-1}}_{\mub,\rhob}
  \\{}\\[-.8em]
  &= \Frac{|\Gs|}{\sqrt{\s\lambda\u\lambda\s\mu\u\mu}}
  \dsty\sum_{\J\in\cals_\lambda\cap\cals_\mu}
  \psi_\lambda(\J) \psi_\mu(\J^{-1})\, \delta_{\lambdab,\mub^+} \,
  \eta^\J_\lambdab
   = \Frac{|\Gs|}{\u\lambda} \,
  \Cl\lambda_{\psi_\lambda,\psi_\mu}\, \delta_{\lambdab,\mub^+}
  \,, \\{}\\[-.5em]
  \cC_{\RhoB,\SigmaB} \!\!
  &= \Frac{|\Gs|}{\sqrt{\s\rho\u\rho\s\sigma\u\sigma}}
  \dsty\sumbo\lambda \sum_{\J\in\calu_\rho\cap\calu_\sigma}
  \psu_\rho(\J^{-1})^*\,\psu_\sigma(\J)^*\,
  S^{\J^{-1}}_{\lambdab,\rhob}\,S^\J_{\lambdab,\sigmab}
  \\{}\\[-.8em]
  &= \Frac1{\sqrt{\s\rho\u\rho\s\sigma\u\sigma}}
  \dsty\sum_\lambdab \sum_{\JK\in\Gs} \eE^{2\pi\ii Q_\JK(\lambda)}
  \sum_{\J\in\calu_\rho\cap\calu_\sigma} \psu_\rho(\J)\psu_\sigma(\J^{-1})\,
  S^\J_{\rhob,\lambdab}\,S^\J_{\lambdab,\sigmab}
  \\{}\\[-.8em]
  &= \Frac1{\sqrt{\s\rho\u\rho\s\sigma\u\sigma}}
  \dsty\sum_{\J\in\calu_\rho\cap\calu_\sigma}\psu_\rho(\J)\psu_\sigma(\J^{-1})\,
  \sum_{\JK\in\Gs} F_\sigma(\JK,\J)\,
  \eta^\J_\rhob\, \delta_{\rhob,(\JK\sigmab)^+_{\phantom I}}
  \\{}\\[-.8em]
  &= \Frac1{\s\rho} \dsty\sum_{\JK\in\Gs} \Cr\rho_{\psu_\rho,\JJK\psu_\sigma}\,
  \delta_{\rhob,(\JK\sigmab)^+_{\phantom I}}
  \,. \eear \ee
Here we used the identities \Erf-J and \Erf eJ, and introduced
  \be  \Cl\lambda_{\psi,\psi'}
  := \Frac1{\s\lambda} \sum_{\J\in\cals_\lambda} \psi(\J)\, \eta^\J_\lambdab\,
  \psi'(\J)^*  \Labl Cl
as well as
 \futnote{Compare formula (C.3) of \cite{fusS6}.}
  \be  \Cr\rho_{\psu,\psu'}
  := \Frac1{\u\rho} \sum_{\J\in\calu_\rho} \psu(\J)\, \eta^\J_\rhob\,
  \psu'(\J)^*  \Labl Cr
for any two characters $\psi,\psi'\iN\cals_\lambda^*$, \resp\
$\psu,\psu'\iN\calu_\rho^*$.

To proceed, we need several properties of the matrices $\Cl\lambda$ and
$\Cr\rho$. First, with the help of the identity \Erf es we have
  \be  \llb \Cl\lambda_{\psi,\psi'} \lrb^* 
  = \Frac1{\s\lambda} \dsty\sum_{\J\in\cals_\lambda}
  \psi(\J)^*\,(\eta^\J_\lambdab)^*\,\psi'(\J)
  = \Frac1{\s\lambda} \dsty\sum_{\J\in\cals_\lambda}
  \psi(\J^{-1})\,\eta^{\J^{-1}}_\lambdab\,\psi'(\J^{-1})^*
  = \Cl\lambda_{\psi,\psi'}  \,,  \Labl*C
i.e.\ $\Cl\lambda$ is real.
Second, combining this reality property with the identity \Erf ep we see that
  \be  \Cl{\lambda^+}_{\psi,\psi'} 
  = \Frac1{\s\lambda} \dsty\sum_{\J\in\cals_\lambda}
  \psi(\J)\, \eta^\J_{\lambdab^+}\, \psi'(\J)^*
  = \Frac1{\s\lambda} \dsty\sum_{\J\in\cals_\lambda}
  \psi'(\J)^*\,(\eta^\J_\lambdab)^*\,\psi(\J)
  = \llb \Cl\lambda_{\psi',\psi} \lrb^* = \Cl\lambda_{\psi',\psi}  \,,  \ee
i.e.
  \be  \Cl{\lambda^+} = \llb \Cl\lambda \lrb^{\rm t}_{} \,.  \Labl Ct
Analogous computations yield
  \be  \llb \Cr\rho \lrb^* = \Cr\rho \qquad{\rm and}\qquad
  \Cr{\rho^+} = \llb \Cr\rho \lrb^{\rm t}_{} \,.  \Labl CT
Finally, implementing the identity \Erf er, i.e.\ the fact that 
$\eta^\J_\rhob$ is a character of $\calu_\rho$, we have
  \be  \sum_{\psu'\in\calu_\rho^*} \Cr\rho_{\psu,\psu'}\, \psu'(\J')
  = \Frac1{\u\rho} \dsty\sum_{\J\in\calu_\rho}
  \psu(\J)\, \eta^\J_\rhob \sum_{\psu'\in\calu_\rho^*} \psu'(\J)^*\,\psu'(\J')
  = \psu(\J')\, \eta^{\J'}_\rhob = \psu^+(\J')  \Labl8n
with the character $\psu^+\iN\calu_\rho^*$ (not to be mixed up with the 
complex conjugate character $\psu^*\iN\calu_\rho^*$) as defined by \Erf9n, which
means that the map $\Cr\rho$ on the \bc s is a permutation. We can thus write
  \be  \Cr\rho_{\psu,\psu'} \equiv \delta^{}_{\psu',\psu^+}
  = \delta^{}_{\psu',\pi_\rhob(\psu)}  \ee
with some permutation $\pi_\rhob$. Because of \Erf CT we also have
  \be  \pi_{\rhob^+}^{} = (\pi_\rhob)^{-1} \,.  \Labl2n

Having obtained these properties of $\Cr\rho$, we can finally
conclude that $\cC$ is a {\em conjugation\/}, i.e.\ it is symmetric and
each column and row contains just a single non-zero element:
  \be  \cC_{\RhoB,\SigmaB} = \Frac1{\s\rho} \sum_{\JK\in\Gs}
  \, \delta^{}_{\rhob,(\JK\sigmab)^+_{\phantom I}}\,
  \Cr\rho_{\psu_\rho,\JJK\psu_\sigma} = \delta_{\RhoB,\SigmaB^+}  \ee
with $\SigmaB^+\eq[\sigmab^+,\psu_\sigma^+]$
(and $\psu_\sigma^+(\J)\,{\equiv}\,\psu_\sigma(\J)\eta^{\J}_\sigmab$).
In particular $\cC$ is an involution, i.e.\ we have
  \be  (\cC)_{\phantom I}^2 = \one \,.  \ee
(The crucial property of $\pi_\rhob$ entering here is \Erf2n;
that relation is not tied to the order of $\pi_\rhob$, so that in particular
$\pi_\rhob$ need not have order two itself.)

For the map $\tC$ on boundary blocks,
the conclusions are somewhat different. We still have 
  \be  \sum_{\psi'\in\cals_\lambda^*} \Cl\lambda_{\psi,\psi'}\, \psi'(\J')
  = \psi(\J')\, \eta^{\J'}_\lambdab  \ee
similarly to \Erf8n; but since $\eta_\lambdab$ is only a character of
$\calu_\lambda$, but not necessarily of $\cals_\lambda$, the expression on the
\rhs, seen as a function on $\cals_\lambda$, is not a character any more. 
Therefore $\tC$ in general no longer constitutes a permutation.
(Of course, when $\calu_\lambda$ coincides with $\cals_\lambda$, it still does.
In this case the arguments are completely parallel to those above, leading to 
the conclusion that $U\tC$ is a conjugation as well.) As a consequence, the 
matrix $\tC$ is not, in general, a (weighted) conjugation. Nevertheless we can 
again conclude that $\tC$ is (weighted) involutive. Namely, we find
  \be  \bearll  
  \dsty\sum_{\psi'\in\cals_\lambda^*} \Llb \sum_{\psi''\in\cals_\lambda^*} 
  \Cl\lambda_{\psi,\psi''}\Cl\lambda_{\psi',\psi''} \Lrb \, \psi'(\J) \!\!
  &= \Frac1{\s\lambda} \dsty\sum_{\J'\in\cals_\lambda} \eta^{\J'}_\lambdab
  \eta^{\J^{-1}}_\lambdab \sum_{\psi''\in\cals_\lambda^*}
  \psi(\J')\psi''(\J')^*\psi''(\J)
  \\{}\\[-.8em]
  &= \eta^\J_\lambdab\eta^{\J^{-1}}_\lambdab\, \psi(\J) 
   = \psi(\J)
  \,,  \eear \ee
where in the last step we used the identity \Erf et, and hence
  \be  \bearll  
  ((\tC)^2)_{(\lambdab,\psi_\lambda),(\nub,\psi_\nu)} \!\!\!
  &= \Frac{|\Gs|^2}{\u\lambda^2} \, \delta^{}_{\lambdab,\nub}\,
  \dsty\sum_{\psi_\mu\in\cals_\mu^*}
  \Cl\lambda_{\psi_\lambda,\psi_\mu}\Cl{\lambda^+}_{\psi_\mu,\psi_\nu}
   = \Frac{|\Gs|^2}{\u\lambda^2} \, \delta^{}_{\lambdab,\nub}\,
  \delta^{}_{\psi_\lambda,\psi_\nu}
   = \Frac{|\Gs|^2}{\u\lambda^2} \,
  \delta^{}_{(\lambdab,\psi_\lambda),(\nub,\psi_\nu)}
  \,.  \eear \ee
We can also deduce that
the inverse of $\tC$ is given in terms of the inverse of $\tS$ as
  \be  (\tC)^{-1}_{}
  = (\tS^{-1})^{\rm t}\, \tS^{-1}  \,.  \labl{icon}
 
Let us point out that the existence of a conjugation $\cC$ on the boundary
{\em conditions\/} does not come as a big surprise. Indeed, it precisely
implements what one heuristically expects as the result of changing the 
orientation of the boundary. The latter manipulation is required e.g.\ when 
one wants to glue surfaces along boundaries.
In contrast, for the boundary {\em blocks\/} such a manipulation would not
make any sense; accordingly it is not too surprising that in the most
general case a genuine conjugation on the boundary blocks does not exist.

\subsection{Structure constants}
 
According to our general expectations, the structure constants of the 
classifying algebra are to be defined through a Verlinde-like formula
via the matrix $\tS$. We first introduce a corresponding quantity 
  \be \tNl{(\lambdab_1,\psi_{\lambda_1})}
  {(\lambdab_2,\psi_{\lambda_2})} {(\lambdab_3,\psi_{\lambda_3})}
  := \sum_\RhoB \frac{
  \tS_{(\lambdab_1,\psi_{\lambda_1}),\RhoB}
  \tS_{(\lambdab_2,\psi_{\lambda_2}),\RhoB}
  \tS_{(\lambdab_3,\psi_{\lambda_3}),\RhoB}} {\tS_{\vacb,\RhoB}}  \Labl4m
with only lower indices
and then raise the third index with the inverse \erf{icon} of $\tC$, so as
to arrive at the expression
  \be  \bearll
  \tN{(\lambdab_1,\psi_1)}{(\lambdab_2,\psi_2)}{(\lambdab_3,\psi_3)} \!\!
  &= \dsty\sum_\rhoB\sum_{\psu_\rho\in\calu^*_\rho}
  \Frac{ \tS_{(\lambdab_1,\psi_1),\RhoB}\,
  \tS_{(\lambdab_2,\psi_2),\RhoB}\,
  (\tS^{-1})^{\RhoB,(\lambdab_3,\psi_3)} } {\tS_{\vacb,\RhoB}}
  \\{}\\[-.95em]
  &= \Frac{\u{\lambda_3}}{|\Gs|}\,\dsty\sum_\rhoB\sum_{\psu_\rho\in\calu^*_\rho}
  \Frac{ \tS_{(\lambdab_1,\psi_1),\RhoB}\,
  \tS_{(\lambdab_2,\psi_2),\RhoB}\,
  \tS_{(\lambdab_3,\psi_3),\RhoB}^* } {\tS_{\vacb,\RhoB}}
  \eear \Labl4n
for the structure constants. More explicitly, we find
  \be \hsp{-.8} \bearll
  \tN{(\lambdab_1,\psi_1)}{(\lambdab_2,\psi_2)}{(\lambdab_3,\psi_3)} \!\!\!
  &= \Frac{|\Gs|\,\sqrt{\u3}}{\sqrt{\s1\u1\s2\u2\s3}}\,
  \dsty\sum_{\ssty\J_i\in\cals_i\atop \J_3=\J_1\J_2}\!
  \psi_1(\J_1)\psi_2(\J_2)\psi_3(\J_3)^* \,
  \sum_\rhoB \Frac1{\s\rho}\, S^{\J_1}_{\lambdab_1,\rhob}
  S^{\J_2}_{\lambdab_2,\rhob} (S^{\J_3}_{\lambdab_3,\rhob})^*
  \,/\, \bS_{\vacb,\rhob}
  \\{}\\[-.8em]
  &= \Frac{\sqrt{\u3}}{\sqrt{\s1\u1\s2\u2\s3}}\,
  \dsty\sum_{\ssty\J_i\in\cals_i\atop \J_3=\J_1\J_2}\!
  \psi_1(\J_1)^*\psi_2(\J_2)^*\psi_3(\J_3) \,
  \sum_{\rhob} S^{\J_1}_{\rhob,\lambdab_1,}
  S^{\J_2}_{\rhob,\lambdab_2} (S^{\J_3}_{\rhob,\lambdab_3})^*
  \,/\, \bS_{\vacb,\rhob}
  \,.  \eear \Labl5n

\subsection{Semisimplicity and \irrep s}

The following properties of the structure constants and of the \cla\ \clAb\
now follow directly:
\nxt
The structure constants \Erf4n are manifestly 
symmetric in the first two indices. Thus \clAb\ is {\em commutative\/}.
\nxt
We have
  \be  \hsp{-1.36} \bearl  \left(
  \tPhi_{(\lambdab_1,\psi_1)} \star \tPhi_{(\lambdab_2,\psi_2)}\right)\star
  \tPhi_{(\lambdab_3,\psi_3)}    \\{}\\[-.8em] \hsp{.8}
  = \dsty\sum_\rhoB \sum_{\psu_\rho\in\calu^*_\rho}\!
  \sumbo{\lambda_4}\sum_{\psi_4\in\cals^*_\lambda}\!
  \frac{ \tS_{(\lambdab_1,\psi_1),\RhoB}
  \tS_{(\lambdab_2,\psi_2),\RhoB} \tS_{(\lambdab_3,\psi_3),\RhoB}}
  {\tS_{\vacb,\RhoB}^2}\, (\tS^{-1})^{\RhoB,(\lambdab_4,\psi_4)}\,
  \tPhi_{(\lambdab_4,\psi_4)} \,. \eear \ee
This is totally symmetric in the three labels $(\lambdab_i,\psi_i)$ for
$i\eq1,2,3$. It follows that \clAb\ is {\em associative\/}.
\nxt
It is also immediately verified that \clAb\ is {\em unital\/}. The
unit element is $\lambdab\eq\vacb$.
\nxt
By construction, the matrix $\tS$ simultaneously diagonalizes all
matrices $\TN_{(\lambdab_1,\psi_{\lambda_1})}$. Explicitly,
  \be  \bearl
  \dsty\sum_{\lambdab_2,\psi_2} \sum_{\lambdab_3,\psi_3}
  (\tS^{-1})^{\RhoB,(\lambdab_2,\psi_2)}\,
  \tN{(\lambdab_1,\psi_1)}{(\lambdab_2,\psi_2)}{(\lambdab_3,\psi_3)}\,
  \tS_{(\lambdab_3,\psi_3),\RhoBpp}\,
  \\{}\\[-.9em] \hsp{3.32}
  = \dsty\sum_{\lambdab_2,\psi_2} \sum_{\lambdab_3,\psi_3}
  \sum_{\SigmaBphu} \sum_{\SigmaBphupp} \sum_{\lambdab_4,\psi_4}
  (\tS_{\vacb,\SigmaBphu})^{-1}_{} (\tS^{-1})^{\RhoB,(\lambdab_2,\psi_2)}\,
  \tS_{(\lambdab_3,\psi_3),\RhoBpp}\,
  \\{}\\[-.69em] \hsp{6.7}
  \tS_{(\lambdab_1,\psi_1),\SigmaBphu} \tS_{(\lambdab_2,\psi_2),\SigmaBphu}
  \tS_{(\lambdab_4,\psi_4),\SigmaBphu}
  (\tS^{-1})^{\SigmaBphupp,(\lambdab_3,\psi_3)}
  (\tS^{-1})^{\SigmaBphupp,(\lambdab_4,\psi_4)}
  \\{}\\[-.5em] \hsp{3.32}
  = \delta^{\RhoB}_{\ \RhoBpp}\,
  \tS_{(\lambdab_1,\psi_1),\RhoB}\,/\, \tS_{\vacb,\RhoB}
  \,.  \eear \Labl-n
Thus the regular \rep\ of \clAb\ is fully reducible.
\nxt
Together these properties imply in particular that the associative \alg\
\clAb\ is {\em semisimple\/}.
\nxt
The matrix $\tC$ can be expressed through the structure constants as
  \be \tC_{(\lambdab_1,\psi_{\lambda_1}),(\lambdab_2,\psi_{\lambda_2})}
  = \tNl{(\lambdab_1,\psi_{\lambda_1})}{(\lambdab_2,\psi_{\lambda_2})}\vacb
  = |\Gs|\, \tN{(\lambdab_1,\psi_{\lambda_1})}{(\lambdab_2,\psi_{\lambda_2})}
  {\ \ \ \ \ \ \ \vacb} \,.  \ee
Note, however, that generically this is not a conjugation.
(Recall that while the matrix $\cC$ provides a conjugation on the \bc s,
$\tC$ is in general only an involution, but not a conjugation on the
boundary blocks; for a given pair $(\lambdab_1,\psi_1)$ the matrix element
$\tC_{(\lambdab_1,\psi_{\lambda_1}),(\lambdab_2,\psi_{\lambda_2})}$
can be non-vanishing for several pairs $(\lambdab_2,\psi_{\lambda_2})$.)

The calculation in \Erf-n implies that each equivalence class $\RhoB$
furnishes a \onedim\ \irrep\ $R_\RhoB$ of \clAb. According to the result
\erf{res1} these 
\irrep s are, in turn, precisely the reflection coefficients. We thus have
  \be  \Rc\RhoB{(\lambdab,\varphi)}\vacb
  = R^{}_\RhoB (\tPhi_{(\lambdab,\varphi)})
  = \frac{\tS_{(\lambdab,\varphi),\RhoB}}{\tS_{\vacb,\RhoB}} \,.  \labl R
Moreover, due to the sum rule \Erf sr and the result that the algebra is
semisimple we can conclude that in fact the reflection coefficients
\erf R even provide {\em all\/} inequivalent \irrep s and that these
are all \onedim.

\subsection{Relation with \cblock s}

Our next aim is to write the structure constants of the classifying algebra
in terms of quantities related to chiral blocks. In connection with blocks,
the natural quantities are the structure constants with only lower indices.
We find that
  \be  \bearll
  \tNl{(\lambdab_1,\psi_1)}{(\lambdab_2,\psi_2)}{(\lambdab_3,\psi_3)} \!\!
  &= d_1 d_2 d_3\,|\Gs|
  \dsty\sum_{(\J_1,\J_2,\J_3)\in\cals_1\times\cals_2\times\cals_3
  \atop \J_1\J_2\J_3=1} \prod_{i=1}^3 \Frac{\psi_i(\J_i)}{\s i} \sum_\rhob
  \Frac{ S^{\J_1}_{\lambdab_1,\rhob}
  S^{\J_2}_{\lambdab_2,\rhob} S^{\J_3}_{\lambdab_3,\rhob}} {\bS_{\vacb,\rhob}}
  \\{}\\[-.8em]
  &=: \Frac{d_1 d_2 d_3\,|\Gs|}{ |\cals_1\cdoT\cals_2\cdoT\cals_3|} \cdot
  \DpNl{(\lambdab_1,\psi_1)}{(\lambdab_2,\psi_2)}{(\lambdab_3,\psi_3)} \,,
  \end{array} \Labl3n
where $d_i\,{\equiv}\,d_{\lambda_i}$ etc., and where 
in the second step we introduced {\em projected fusion coefficients\/} $\DPN$;
also, $\cals_1\cdoT\cals_2\cdoT\cals_3$ is by definition the subgroup of $\Gs$
that is generated by the three subgroups $\cals_i$.

To rewrite the coefficients $\DPN$ in a more convenient form, we
consider the group homomorphism $p{:}\ \cals_1\timeS\cals_2\timeS\cals_3\to\Gs$
that is defined by taking the product in $\Gs$ of three elements,
  \be  p: \quad \cals_1\timeS\cals_2\timeS\cals_3\ni\,
  (s_1,s_2,s_3) \,\mapsto\, p(s_1,s_2,s_3):=s_1s_2s_3 \,.  \ee
By the homomorphy theorem, the number of elements in the kernel of $p$ is
  \be  |\ker p| = \s1\,\s2\,\s3 \,/\, |\cals_1\cdoT\cals_2\cdoT\cals_3| \,.
  \labl{kern}
Thus we can replace the product over the $\s i$ by the product of $|\ker p|$
and the number of elements in the group generated by all $\cals_i$, so that
  \be
  \DpNl{(\lambdab_1,\psi_1)}{(\lambdab_2,\psi_2)}{(\lambdab_3,\psi_3)}
  = \Frac1{|\ker p|}\,\sum_{(\J_1,\J_2,\J_3)
  \in\cals_1\times\cals_2\times\cals_3 \atop \J_1\J_2\J_3=1}\,
  \prod_{i=1}^3 \psi_i(\J_i)\, \sum_{\rhob} \Frac{ S^{\J_1}_{\lambdab_1,\rhob}
  S^{\J_2}_{\lambdab_2,\rhob} S^{\J_3}_{\lambdab_3,\rhob}} {S_{\vacb,\rhob}}\,.
  \labl{prfus}
We remark that $|\ker p|$ is precisely the number of elements
of the group over which we have to perform a
Fourier transformation in order to attain a projection on the \cblock s.
Accordingly $|\ker p|$ indeed needs to be absorbed into the definition of the
coefficients
$\DpNl{(\lambdab_1,\psi_1)}{(\lambdab_2,\psi_2)}{(\lambdab_3,\psi_3)}$.

These relations suggest that the structure constants of the classifying
algebra should be related to an appropriate action of the simple current
group $\Gs$ on the space of \cblock s. This feature is familiar from
the situation studied in \cite{fuSc5}. However, owing to the fact that 
the action of $\Gs$ is only projective, here 
its precise realization is more involved and remains to be uncovered.

\sect{Annulus coefficients}\label{s.6}

\subsection{The $\Hp$-extension}\label{s.61}
 
In this section we compute the annulus amplitude $\A_{\rho_1\,\rho_2}$ for
two arbitrary boundary conditions
  \be  \rho_i \equiv [\rhob_i,\psu_i]  \ee
($i\eq1,2$) and study its properties.
One way to obtain the annulus amplitude is to evaluate it in the closed string
channel, where it can be regarded as factorizing into two disk one-point 
functions and a sphere two-point function, so that
it corresponds to propagation between the boundary 
states $\langle\calb_{[\rhob_i,\psu_i]}|$, according to
  \be  \A_{\RhoBe\,\RhoBz}(t)
  = \langle \calb_\RhoBz|\, \eE^{-(2\pi/t)(L_0\ots\bfe+\bfe\ots L_0-c/12)} \,|
  \calb_\RhoBe \rangle \,.  \labl A
Here by a {\em boundary state\/}
\cite{card9,poca,clny3,dpfhls,ishi,grwa,fuSc6,reSC} one means a linear form 
$\calb_\RHoB$ on the space $\bigoplus_\mub\calhb_\mub^{}\otimeS\calhb_\mubp$ of 
all closed string states (which correspond to bulk fields) that is characterized
by the property that when applied to an element $v\oT\tilde v\iN\V_\phu{\otimes}
\calhb_{(\lambdab,\phu)}\otimeS\V_{\phu^+}{\otimes}\calhb_{(\lambdabp,\phu^+)}$
it yields the one-point correlator of the corresponding bulk field on the disk,
i.e.
  \be  \calb_\RHoB(v\ot\tilde v) =
  \langle\phi_{(\lambdab,\varphi),(\lambdabp,\varphi^+)}(v\ot\tilde v;z{=}0)
  {\rangle}_\RHoB \,.  \Labl88
It follows that $\calb_\RHoB$ can be written as the specific linear combination
  \be  \calb_\RHoB
  = \bigoplus_{\ssty\lambdab \atop Q_\Gs(\lambda)=0}
  \bigoplus_{\varphi\in\cals_\lambda^*}
  \Rc\RHoB{(\lambdab,\varphi)}\vacb\, \langle\Psi^{\RHoB\,\RHoB}_\vacb\rangle\,
  \tBeta_{(\lambdab,\varphi)}  \Labl89
of boundary blocks $\tBeta_{(\lambdab,\varphi)}$.

To each boundary condition $\rho_i$ we associate the character
  \be  \gg_i \equiv \gg_{\rho_i}^{(Q)}
  \in \Gs^* = (\GS)^*_{} \cong G  \ee
of $\Gs$ that maps every simple current to the value of the corresponding
monodromy charge \Erf QJ, i.e.
  \be  \calq_i(\J) := \exp(2\pi\ii Q_\J(\rho_i))  \ee
for all $\J\iN\Gs$. As already mentioned in the introduction, this quantity 
can be regarded as an element of the orbifold group $G$, and indeed it 
coincides with the so-called automorphism type of the \bc\ (for details, see 
\II). Inspection shows that in the amplitude \erf A one deals with linear 
combinations of characters of that orbifold theory in which the cyclic subgroup 
$\langle\calq_1^{-1}\calq_2\rangle$ of $G$ that is
generated by $\calq_1^{-1}\calq_2\iN G$ is broken. (These combinations are
known as {\em twining characters\/} \cite{fusS3,fusS6} of the $\cala$-theory.)
This orbifold theory can equivalently be described as an integer spin simple
current extension of the $\calap$-theory by a subgroup of $\Gs$ which is
a proper subgroup when $\calq_1\nE\calq_2$.

In more precise terms the situation is described as follows. The exact sequence
  \be  0 \to \langle \calq_1^{-1}\calq_2\rangle \to G \to
  G / \langle \calq_1^{-1}\calq_2\rangle \to 0  \ee
of finite abelian groups implies the exact sequence
  \be 0 \to (\Gs/ \langle \calq_1^{-1}\calq_2\rangle)^* \to \Gs \to
  \langle \calq_1^{-1}\calq_2\rangle^* \to 0 \ee
of their character groups. Therefore we can extend the $\calap$-theory by
${(\Gs/\langle\calq_1^{-1}\calq_2\rangle)}^*_{\phantom I}$. This is the subgroup
of those characters of $G$ which descend to characters of the quotient, and 
these are precisely those which are the identity on $\langle\calq_1^{-1}
\calq_2\rangle$, i.e.\ those simple currents $\J$ which obey $\calq_1^{-1}
\calq_2(\J)\eq1$, which in turn is the same as $Q_\J(\rho_1)\eq Q_\J(\rho_2)$. 

Accordingly, one expects that the annulus amplitude can be expressed as a 
linear combination of characters of the extension of the $\calap$-theory by 
the subgroup
  \be  \Go \equiv \Go_{\rho_1\rho_2}
  :=\{ \J\iN \Gs \,|\, Q_\J(\rho_1)\eq Q_\J(\rho_2) \} \Labl Go
of $\Gs$. As we will see, this is indeed possible; still, as it turns out, 
this is not the most natural choice for the following reason. Ultimately, 
our goal is to express the annulus amplitude as a linear combination
  \be  \A_{\rho_1\,\rho_2} = \sum_\sigma\A_{\rho_1\,\rho_2}^\sigma\,
  \X^{(\HK)}_\sigma \ee
of irreducible characters $\X^{(\HK)}_\sigma$ in some extension $\HK$ 
of the $\calap$-theory. Now the interpretation of the annulus amplitude
as an open string {\em partition function\/} imposes the requirement that when
we expand $\A_{\rho_1\rho_2}(t)$ as a function of $q\eq\exp(2\pi\ii(\ii t/2))$,
then the coefficients in this expansion are non-negative integers.
While this does not necessarily imply that already all the numbers
$\A_{\rho_1\rho_2}^\sigma$ are
integers, it has been observed in many situations \cite{fuSc5,fuSc9} that the
multiplicities $\A_{\rho_1\rho_2}^\sigma$ possess a natural interpretation as
the rank of (a subsheaf of) a sheaf of \cblock s. (This interpretation also
enables one to establish the integrality property in full generality.)
In order to relate $\A_{\rho_1\,\rho_2}^\sigma$ to such a rank of a
\cblock, we need to work with an extended theory in which both $\rho_1$ and 
$\rho_2$ are allowed fields, which is the case if their monodromy charges
both vanish. The $\Go$-extension does not meet this condition in general;
rather, we need to consider the extension by the subgroup
  \be  \Hp \equiv \Hp_{\rho_1\rho_2}
  :=\{ \J\iN \Gs \,|\, Q_\J(\rho_1)\eq0\eq Q_\J(\rho_2) \} \ee
of $\Gs$, which is the largest subgroup of $\Go$ that has the desired
property.

Thus we define the {\em annulus coefficients\/} $\A_{\rho_1\,\rho_2}^\sigma$ 
as the multiplicities of characters in the extension of the $\calap$-theory 
by $\Hp$; in more precise notation,
  \be  \A_{\RhoBe\,\RhoBz}(t) = \sum_\sigmaBp \A_{\RhoBe\,\RhoBz}^\SigmaBp
  \, \chip_\SigmaBp(\Frac{\ii t}2) \,,  \ee
where $\SigmaBp$ is the $\Hp$-orbit of $(\sigmab,\psu_\sigma)$. We will 
demonstrate that
these quantities can be expressed as a sum of fusion rule coefficients in the
$\Hp$-extension, and check that various consistency requirements are satisfied.
Our first task is to make sure that the characters of the $\Hp$-extension
indeed appear in the expression of the annulus partition function in the 
closed string channel. These characters read (compare formula \erf X)
  \be  \chip_\LambdaBp = \Frac1{\sqrt{\sp\lambda\up\lambda}}
  \sum_{\J\in\Hp} \chib_{(\J\lambdab,\psup_\lambda)}
  \equiv \Frac1{\sqrt{\sp\lambda\up\lambda}}\,
  \llb \sum_{\J\in\Hp} \chib_{\J\lambdab} \lrb^{}_{\psup_\lambda} \,.  \Labl cp
Here $\sp\lambda\eq|\calsp_\lambda|$ and $\up\lambda\eq|\calup_\lambda|$ are
the cardinalities of the full and untwisted stabilizer $\calsp_\lambda$ and of
  \be  \calup_\lambda = \{ \J\iN\calsp_\lambda  \,|\,
  F_\lambda(\J,\JK)\eq1\;{\rm for\; all}\; \JK\iN\calsp_\lambda\} \,,  \ee
\resp, which are relevant in the $\Hp$-extension, and $\psup_\lambda$ is a
character of $\calup_\lambda$. Note that while one has
  \be  \calsp_\lambda = \cals_\lambda \cap\Hp \,,  \Labl90
there is no simple relation between $\calup_\lambda$ and $\calu_\lambda$. In 
particular, $\calup_\lambda$ differs, in general, from the intersection
$\caluhp\lambda$, though it always contains it as a subgroup:
  \be  \caluhp\lambda \,\subseteq\, \calup_\lambda \,;  \ee
in fact, already in simple examples it happens that
$\calup_\lambda$ is larger than $\calu_\lambda$.\,%
 \futnote{An example occurs for the case of the $D_4$ level 2 \wzwt. In this
case for the fixed point with stabilizer $\zet_2\times\zet_2$ the untwisted
stabilizer is trivial, but when the second \bc\ is taken to be in a
twisted sector, then both $\calsp$ and $\calup$ are equal to the
corresponding $\zet_2$ under which the twisted sector is fixed.}

\subsection{Expressions for the annulus coefficients}\label{s.62}

To proceed, we need more explicit expressions for the annulus coefficients.
To this end we insert the result \Erf ia for the regulated inner product 
\Erf mf of the boundary blocks and the relation \erf{unten} 
between one-point correlators and boundary blocks into formula \erf A. We
also substitute for the coefficients in the latter relation the explicit
expressions \Erf xR and \erf R as well as the normalization 
  \be  \langle\Psi^{\RhoB\,\RhoB}_\vacb\rangle = \tS_{\vacb,\RhoB}  \Labl xr
for the vacuum boundary fields. This yields
  \be  \bearll  \A_{\RhoBe\,\RhoBz}(t) \!\!
  &= \dsty\sum_{\ssty\lambdab;\,\psi_\lambda\in\cals_\lambda^* \atop\ssty 
  \mub;\,\psi_\mu\in\cals_\mu^*}
  \llb \Rc\RhoBe{(\lambdab,\psi_\lambda)}\vacb\,
  \langle\Psi^{\RhoBe\,\RhoBe}_\vacb\rangle {\lrb}^* \,
  \llb \Rc\RhoBz{(\mub,\psi_\mu)}\vacb\,
  \langle\Psi^{\RhoBz\,\RhoBz}_\vacb\rangle \lrb \,
  \\{}\\[-1.82em]& \hsp{10.9}
  \langle \tBeta_{(\mub,\psi_\mu)}
  |\, \eE^{-(2\pi/t)(L_0\ots\bfe+\bfe\ots L_0-c/12)} \,|
  \tBeta_{(\lambdab,\psi_\lambda)} \rangle
  \\{}\\[-.04em]
  &= \dsty\sumbo\lambda\sum_{\psi_\lambda\in\cals_\lambda^*}
  \tS_{(\lambdab,\psi_\lambda),\RhoBe}^* \tS_{(\lambdab,\psi_\lambda),\RhoBz}\,
  \Frac1{(|\Gs|/\u\lambda)\, \Sb\lambda\vac \raisa}\,
  \chib_{(\lambdab,\psi_\lambda)}(\Frac{2\ii}t) \,. \eear  \Labl1+
This result is in agreement with the requirement that in the closed channel
only those fields $(\lambdab,\psi_\lambda)$ are exchanged whose monodromy 
charges with respect to the currents in $\Hp$ vanish; indeed, the summation 
even extends only over those fields
for which all monodromy charges of currents in the larger group $\Gs$ are zero.
Our next goal is to rewrite \Erf1+ entirely in terms of $\Hp$-quantities;
first we arrive at a sum over $\Hp$-orbits $\lambdaBp$:\,%
 \futnote{In the product of the two $\tilde S$-elements, one is supposed to
choose a representative of the orbit $\lambdaBp$; it does not matter which one,
because the monodromy charges vanish.
If one worked with the larger group $\Go$ instead of $\Hp$,
one would again be able to show that only characters of the $\Go$-extension
appear, as the monodromy charges of $\rho_1$ and $\rho_2$ are
equal and hence cancel due to the complex conjugation. However, the
two factors $\tS$ separately would depend on the choice of representative 
on the $\Go$-orbits, and for having a well-defined expression one would
have to choose one and the same representative in both matrix elements.}
  \be  \bearll  \A_{\RhoBe\,\RhoBz}(t) \!\!
  &= \dsty\sumBoP\lambda \sum_{\J\in\Hp/\calsp_\lambda}
  \sum_{\psi_\lambda\in\cals_\lambda^*}
  \tS_{(\J\lambdab,\psi_\lambda),\RhoBe}^*
  \tS_{(\J\lambdab,\psi_\lambda),\RhoBz}\,
  \Frac1{(|\Gs|/\u\lambda)\, \Sb\lambda\vac \raisa}\,
  \chib_{(\J\lambdab,\psi_\lambda)}(\Frac{2\ii}t)
  \\{}\\[-.8em]
  &= \dsty\sumBoP\lambda \sum_{\psi_\lambda\in\cals_\lambda^*}
  \tS_{(\lambdab,\psi_\lambda),\RhoBe}^*
  \tS_{(\lambdab,\psi_\lambda),\RhoBz}\,
  \Frac1{(|\Gs|/\u\lambda)\, \Sb\lambda\vac \raisa}\,
  \Llb \Frac1{\s\lambda\oei} \sum_{\J\in \Hp}\chib_{(\J\lambdab,\psi_\lambda)}
  (\Frac{2\ii}t) \Lrb \,.  \eear \Labl1;
In \Erf1; we are still dealing with the characters $\psi_\lambda\iN
\cals_\lambda^*$ and $\psu_i\iN\calu_i^*$. To express the amplitude through the
correct quantities $\psup_\lambda\iN\calups_\lambda$ and $\psup_i\iN\calups_i$, 
analogously to \Erf49 we write
  \be \psi_\lambda \gt \psi\oei_\lambda \ee 
when the restriction of the 
$\Gs$-character $\psi_\lambda$ to the subgroup $\Hp$ of $\Gs$ is equal
to the $\Hp$-character $\psi\oei_\lambda$, and similarly when we deal with
other embedded pairs of groups, e.g.\ stabilizers and untwisted stabilizers
(however, for the $\psu_i$ we will have to be careful because
in general $\calup_i$ is not a subgroup of $\calu_i$). We then arrive at 
  \be  \bearll  \A_{\RhoBe\,\RhoBz}(t) \!\!\!
  &= \!\dsty\sumBoP\lambda \sum_{\psup_\lambda\in\calups_\lambda}
  \llb \sumpsipsup\lambda
  \tS_{(\lambdab,\psi_\lambda),\RhoBe}^*
  \tS_{(\lambdab,\psi_\lambda),\RhoBz} \lrb\,
  \Frac1{(\dP\lambda^{}\,|\Gs|/\u\lambda)\, \Sb\lambda\vac \raisa}\,
  \chip_\LambdaBp(\Frac{2\ii}t)
  \,.  \eear \ee
We are now in a position to perform a modular transformation that involves
the S-matrix $\Sp$ of the $\Hp$-extension; we get
  \be  \bearll  \A_{\RhoBe\,\RhoBz}(t) \!\!\!
  &= \dsty\sumBoP\lambda \sum_{\psup_\lambda\in\calups_\lambda}
  \Llb \sumpsipsup\lambda
  \tS_{(\lambdab,\psi_\lambda),\RhoBe}^*
  \tS_{(\lambdab,\psi_\lambda),\RhoBz}^{} \Lrb\,
  \Frac1{(\dP\lambda^{}\,|\Gs|/\u\lambda)\, \Sb\lambda\vac \raisa}\,
  \\{}\\[-1.54em]
  & \hsp{12.9} \dsty\sum_\sigmaBp\sum_{\psup_\sigma\in\calups_\sigma}
  \Sp_{\LambdaBp,\SigmaBp} \,
  \chip_\SigmaBp(\Frac{\ii t}2) \,,  \eear \Labl5r
from which we finally can read off the annulus coefficients as the 
coefficients of the $\Hp$-characters $\chip_\SigmaBp$.
Thus we finally see that the annulus coefficients are given by
  \be  \bearll
  \A_{\RhoBe\,\RhoBz}^\SigmaBp \!\!
  &= \dsty\sumBoP\lambda \Frac{|\Hp|}{|\Gs|}\,\Frac{\u\lambda}{\sp\lambda}
  \sum_{\psup_\lambda\in\calups_\lambda} \sumpsipsup\lambda
  \tS_{(\lambdab,\psi_\lambda),\RhoBe}^* \tS_{(\lambdab,\psi_\lambda),\RhoBz}\,
  \Sp_{\LambdaBp,\SigmaBp} \,/\, \Sp_{\LambdaBp,\vacp}
  \,.  \eear \Labl6r

As a check of the normalization of the annulus coefficients, let us specialize
to the case of \bc s that preserve the full bulk symmetry, in which case
$\Hp\eq\Gs$ and the annulus coefficients coincide with the structure constants 
of the fusion rule \alg. As seen after \Erf ss, in this case the matrix 
elements of $\tS$
coincide with those of $\Sp\eq S$ where one takes the orbit corresponding
to $\lambdab$. In addition we then have $\psup\eq\psu$, so that the summation 
over $\psi\,{\gt}\,\psup$ just amounts to a factor of $\s\lambda/\u\lambda$, and
we can use (compare \Erf23)
  \be  \Sp_{\LambdaBp,\vacp}
  = \Frac{|\Hp|}{\sqrt{\s\lambda\oei\u\lambda\oei}}\, \bS_{\lambdab,\vacb}
  \,.  \ee
Thus in the case $\Hp\eq\Gs$ the result \Erf5r reduces to
  \be  \bearll  \A_{\RhoBe\,\RhoBz}^\SigmaB \!\!
  &= \dsty\sumBo\lambda \sum_{\psu_\lambda\in\calu_\lambda^*}
  S_{\LambdaB,\RhoBe}^*\, S_{\LambdaB,\RhoBz} \, S_{\LambdaB,\SigmaB}
  \,/\, S_{\LambdaB,\vac}
  \,,  \eear \ee
from which by comparison with the Verlinde formula for the $\cala$-theory we
learn that the annulus coefficients indeed coincide with the structure constants
of the fusion \alg.
  
Let us mention one immediate consequence of the result \Erf6r. Relabelling the
summation variable $(\lambdab,\psi_\lambda)$ to $\J(\lambdab,\psi_\lambda)$
for an arbitrary current $\J\iN\Gs$ and inserting the simple current
symmetries \Erf SQ and \Erf QQ of $\tS$ and $\Sp$, \resp, one finds that
  \be  \A_{\RhoBe\,\RhoBz}^\SigmaBp
  = \eE^{2\pi\ii[Q_\J(\rho_2)-Q_\J(\rho_1)+Q_\J(\sigma)]}_{} \cdot
  \A_{\RhoBe\,\RhoBz}^\SigmaBp \,.  \Labl AQ
It follows that $\A_{\RhoBe\,\RhoBz}^\SigmaBp$ vanishes unless
$Q_\J(\sigma)\eq Q_\J(\rho_1)-Q_\J(\rho_2)$ for all $\J\iN\Gs$. In short, the
annulus coefficients are graded by the monodromy charge.
 
\subsection{Relation with fusion coefficients of the $\Hp$-extension}

To proceed we insert the formul\ae\ \Erf tS for $\tS$ and (the analogue for
the $\Hp$-extension of) \erf S for $\Sp$ into \Erf6r, leading to
  \be  \bearll
  \A_{\RhoBe\,\RhoBz}^{\SigmaBp} \!\!\!
  &=\dsty\sumBoP\lambda \Frac{\u\lambda^{}}{|\Gs|\,\dP\lambda}\,
  [\bS_{\lambdab,\vacb}]^{-1}_{}
  \sum_{\psup_\lambda\in\calups_\lambda}\sumpsipsup\lambda
  \Frac{|\Gs|^2}
  {\s\lambda^{}\u\lambda^{}\,\sqrt{\s{\rho_1}\u{\rho_1}\s{\rho_2}\u{\rho_2}}} \,
  \Frac{|\Hp|}{\sqrt{\sp\lambda\up\lambda\sp\sigma\up\sigma}}
  \\{}\\[-1.0em]
  &  \hsp{-3.3} 
  \dsty\sum_{\ssty \J_1\in\cals_\lambda\cap\calu_{\rho_1} \atop 
             \ssty \J_2\in\cals_\lambda\cap\calu_{\rho_2}}
  \dsty\sum_{\J_3\in\calup_\lambda\cap\calup_\sigma}
  \psi_\lambda(\J_1)^*\psu_1(\J_1)\, (S^{\J_1}_{\lambdab,\rhobe})^*_{} \,
  \psi_\lambda(\J_2)\psu_2(\J_2)^*\, S^{\J_2}_{\lambdab,\rhobz} \,
  \psup_\lambda(\J_3)\psup_\sigma(\J_3)^*\,S^{\J_3}_{\lambdab,\sigmab}
  \,. \eear \ee
This somewhat unwieldy expression simplifies a lot when one performs
the $\psi_\lambda$-summation (obtained by combining the $\psup_\lambda$- and
$\psi_\lambda\,{\gt}\,\psup_\lambda$-summation) and implements the fact that
$S^{\J_i}_{\lambdab,\rhob}$ is non-zero only if $Q_{\J_i}(\rho)\eq0$ (which 
follows from the identities \Erf00 and \Erf0Q), i.e.\ only if $\J_i\iN\Hp$:
  \be  \bearll
  \A_{\RhoBe\,\RhoBz}^{\SigmaBp} \!\!\!
  &= \Frac{|\Gs|\,|\Hp|}{\sqrt{\s{\rho_1}\u{\rho_1}\s{\rho_2}\u{\rho_2}
  \sp\sigma\up\sigma}}
  \dsty\sumBoP\lambda \Frac1{\sp\lambda}
  \\{}\\[-1.0em]
  &\quad \dsty\sum_{\ssty \Jp_1\in\caluhp1,\,\Jp_2\in\caluhp2,\,\Jp_3\in
  \calup_\sigma \atop \Jp_1=\Jp_2\Jp_3}
  \psu_1(\Jp_1)\psu_2(\Jp_2)^*\psup_\sigma(\Jp_3)^*\,
  (S^{\Jp_1}_{\lambdab,\rhobe})^*_{}
  S^{\Jp_2}_{\lambdab,\rhobz}S^{\Jp_3}_{\lambdab,\sigmab} \,
  [\bS_{\lambdab,\vacb}]^{-1}_{}
  \,. \eear \ee
Next we insert the analogue of \erf{lem} for the embedding
$\caluhp i\,{\subseteq}\,\calup_i$ to arrive at an expression with 
$\calup_i$-characters:
  \be  \bearll
  \A_{\RhoBe\,\RhoBz}^{\SigmaBp} \!\!\!
  &= \Frac{|\Gs|\,|\Hp|}{\sqrt{\s{\rho_1}\u{\rho_1}\s{\rho_2}\u{\rho_2}
  \sp\sigma\up\sigma}}
  \dsty\sumBoP\lambda \Frac1{\sp\lambda}
  \cdot \Frac{\caluhP1}{\up{\rho_1}}\!\! \sum_{\ssty\psup_1\in\calup_1 \atop
  \psup_1\gt\psuhp_1}
  \Frac{\caluhP2}{\up{\rho_2}}\!\! \sum_{\ssty\psup_2\in\calup_2 \atop
  \psup_2\gt\psuhp_2}
  \\{}\\[-.8em]
  &\quad
  \dsty\sum_{\ssty \Jp_1\in\calup_1,\,\Jp_2\in\calup_2,\,\Jp_3\in\calup_\sigma
  \atop \ssty \Jp_1=\Jp_2\Jp_3}
  \psup_1(\Jp_1)\psup_2(\Jp_2)^*\psup_\sigma(\Jp_3)^*\,
  (S^{\Jp_1}_{\lambdab,\rhobe})^*_{}
  S^{\Jp_2}_{\lambdab,\rhobz}S^{\Jp_3}_{\lambdab,\sigmab} \,
  [\bS_{\lambdab,\vacb}]^{-1}_{}
  \,. \eear \Labl8s
Here $\psuhp_i$ denotes the character 
  \be  \psuhp_i\,{:=}\,\psu_i|_{\caluhp i}^{}  \ee
of $\caluhp i$.

The $\lambdab$-summation in formula \Erf8s is still over all $\Hp$-orbits that
are even $\Gs$-allowed. We now rewrite it such that we sum over all
$\Hp$-orbits that are just $\Hp$-allowed; we first convert the summation 
to a sum over {\em all\/} orbits by inserting the projector \erf P, and then
restrict again to $\Hp$-allowed orbits, which means that
the factor $\eE^{2\pi\ii Q_\J(\lambda)}$ in \erf P is equal to 1 for $\J\iN\Hp$
and constant on the cosets of $\Gs$ \wrt $\Hp$. This amounts to using
the projector
  \be  \Frac{|\Hp|}{|\Gs|}\sum_{\JP\in\Gs/\Hp} \eE^{2\pi\ii Q_\J(\lambda)} 
  \,.  \ee
Afterwards we get rid of the phase factor $\eE^{2\pi\ii Q_\J(\lambda)}$ by
exploiting  the simple current symmetry \Erf QF of the $S^\J$-matrices;
this yields
  \be  \bearll
  \A_{\RhoBe\,\RhoBz}^{\SigmaBp} \!\!\!
  &= \Frac{\sqrt{\sp{\rho_1}\up{\rho_1}\sp{\rho_2}\up{\rho_2}}
  \,\caluhP1\,\caluhP2}{\sqrt{\s{\rho_1}\u{\rho_1}\s{\rho_2}
  \u{\rho_2}}\,\up{\rho_1}\up{\rho_2}}
  \dsty\sum_{\ssty\psup_1\in\calup_1 \atop\psup_1\gt\psuhp_1}
  \sum_{\ssty\psup_2\in\calup_2 \atop\psup_2\gt\psuhp_2} \sum_{\J\in\Gs/\Hp}
  \Ne\RhoBzp{\JP\star\SigmaBp}{\ \ \RhoBep}
  \,,  \eear \Labl8r
where $\JP$ denotes the surviving simple current of the $\Hp$-extension that 
comes from the simple current $\J$ of the $\calap$-theory, and where
  \be  \bearll
  \Ne\RhoBzp\SigmaBp\RhoBep \!\!\!
  &= \Frac{|\Hp|^2}{\sqrt{\sp{\rho_1}\up{\rho_1}\sp{\rho_2}\up{\rho_2}
  \sp\sigma\up\sigma}} \dsty\sumBop\lambda \Frac1{\sp\lambda}
  \\{}\\[-1.0em]
  &\quad \dsty\sum_{\ssty \Jp_1\in\calup_1,\,\Jp_2\in\calup_2,\,\Jp_\sigma\in
  \calup_\sigma \atop \Jp_1=\Jp_2\Jp_\sigma}
  \psup_1(\Jp_1)\psup_2(\Jp_2)^*\psup_\sigma(\Jp_\sigma)^*\,
  (S^{\Jp_1}_{\lambdab,\rhobe})^*_{}
  S^{\Jp_2}_{\lambdab,\rhobz}S^{\Jp_3}_{\lambdab,\sigmab} \,
  /\, \bS_{\lambdab,\vacb}
  \,. \eear \Labl7r
The numbers \Erf7r are precisely the fusion coefficients of the $\Hp$-extension,
as can be seen by
inserting (the analogue for $\Sp$ of) \erf S into the Verlinde formula.

We thus have succeeded in writing the annulus coefficients as a linear
combination of fusion coefficients of the $\Hp$-extension. Still we would like
to manipulate our result further. To this end we observe that in \Erf8r we 
are free to let $\JP$ act on the label of the $\Hp$-fusion coefficients where 
we like it most. In particular for suitable $\JP$ the action will 
then be trivial. To determine these currents, consider first the requirement
  \be  \JP\star \SigmaBp \equiv [\J\sigmab,\JJ\psup_\sigma]\oei
  \,\stackrel!=\, \SigmaBp \,.  \ee
This implies, first, that we need $\J\sigmab\eq\Jp\sigmab$ for some $\Jp\iN\Hp$,
which is solved by $\J\iN\cals_\sigma\cdoT\Hp$. In addition we then need
$\JJ\psup_\sigma\eq\JJp\psup_\sigma$, which is equivalent to
  \be  F_\sigma(\J(\Jp)^{-1},\J_3') = 1\quad{\rm for\; all}\;
  \J_3'\iN\calup_\sigma \,;  \Labl8u
because of $\calup_\sigma\subseteq\cals_\sigma$, for the latter equality it is
sufficient (though not necessary, in general)\,%
 \futnote{It is of course also sufficient that $\J$ is in $\calsp_\sigma$, which
in general is not a subgroup of $\calu_\sigma$. However, we have $\calsp_\sigma
\subseteq\Hp$, and hence because of the explicit appearance of $\Hp$
on the \rhs\ of \Erf Hd this is in fact not relevant.}
that
  \be  \J\in\calu_\sigma\cdoT\Hp \,.  \ee
Similar arguments apply to $\rho_1$ or
$\rho_2$, but now we can also take into account the summations over the
$\psup_i$ which satisfy $\psup_i\gt\psuhp_i$; accordingly, while the first
part of the argument is identical, leading to the requirement that
  \be  \J\in\cals_i\cdoT\Hp \,,  \ee
in the second part the equality between characters only needs to hold for the
restriction of the $\calup_i$-characters to $\caluhp i$, so that the analogue
of \Erf8u gets relaxed to
  \be  F_{\rho_i}(\J(\Jp)^{-1},\J_i') = 1\quad{\rm for\; all}\;
  \J_i'\iN\caluhp i \,,  \ee
which in turn is satisfied for every $\J\iN\cals_i\cdoT\Hp$.
Thus we conclude that whenever $\J$ is in the group
  \be  \Hd \equiv \Hd_{\rho_1\rho_2\sigma} := \cals_{\rho_1}\cdot
  \cals_{\rho_2}\cdot \calu_\sigma \cdot \Hp_{\rho_1\rho_2} \,, \Labl Hd
then $\JP$ acts trivially. It follows that we can rewrite \Erf8r as
  \be  \A_{\RhoBe\,\RhoBz}^{\SigmaBp}
  = \dud \sum_{\ssty\psup_1\in\calup_1 \atop\psup_1\gt\psuhp_1}
  \sum_{\ssty\psup_2\in\calup_2 \atop\psup_2\gt\psuhp_2}
  \sum_{\J\in\Gs/\Hd} \Ne{\RhoBzp}{\J\SigmaBp}{\RhoBep}  \Labl9r
with
  \be  \dud := \Frac{|\Hd|}{|\Hp|}\,
  \Frac{\sqrt{\sp{\rho_1}\up{\rho_1}\sp{\rho_2}\up{\rho_2}}}
  {\sqrt{\s{\rho_1}^{}\u{\rho_1}^{}\s{\rho_2}^{}\u{\rho_2}^{}}}\,
  \Frac{\caluhP1\,\caluhP2}{\up{\rho_1}\up{\rho_2}} \,.  \labl{dud}

Note that both the fusion coefficients and the prefactor $\dud$ are
manifestly non-negative, and hence the result \Erf9r shows that
the annulus coefficients are {\em non-negative\/}. For the interpretation of
the annulus amplitude as a \parfu\ they must even be non-negative
{\em integers\/}. To establish this stronger property will require some more
work. As the fusion coefficients are manifestly integral, we only have to show 
integrality for the prefactor $\dud$.
As a preparation we rewrite this number as a product
  \be  \dud = \dudo \cdot \dudu{\rho_1} \cdot \dudu{\rho_2}  \ee
of three factors
  \be  \dudo := \Frac{|\Hd|\, \sp{\rho_1}\sp{\rho_2}}
  {|\Hp|\,\s{\rho_1}\s{\rho_2}}  \labl{dudo}
and $\dudu{\rho_i}$, where
  \be  \dudu\rho :=
  \Frac{\sqrt{\s\rho}\;\caluhP\rho} {\sqrt{\u\rho^{}\,\sp\rho\up\rho}}
  = \Frac{d_\rho}{\dP\rho}\, \Frac{\caluhP\rho}{\up\rho} \,.  \labl{dudu}
As we will see in the next subsection, actually each of the three factors
is already integral individually; furthermore, those integers possess a natural
\rep\ theoretic interpretation.

\subsection{Integrality}\label{s.64}

We first show the integrality of $\dudo$ \erf{dudo}. Consider the map
$p{:}\ \cals_{\rho_1}\Times\cals_{\rho_2}\Times(\caluhp\sigma)\,{\to}\,\Hd$
that is defined by
  \be  p:\quad (\J_1,\J_2,\J_3) \,\mapsto\, \J:=\J_1^{-1}\J_2\J_3 \,, \Labl1h
which of course we can also interpret as a map to the subgroup
  \be  \imS:=p(\cals_{\rho_1}\timeS\cals_{\rho_2}\timeS\caluhp\sigma)
  \subseteq\Hd \ee
on which \Erf1h is a surjection. We would like to determine when the image $\J$
is already in $\Hp\,{\subseteq}\,\imS$. Let us look at the monodromy charges
for $\J$. Using the fact that the monodromy charge of a fixed point vanishes
(see \Erf0Q) and using the gradation property \Erf AQ of the annulus 
coefficients,
we conclude that
  \be \bearll
  Q_{\J_1}(\rhoe) = 0\,,\quad & Q_{\J_1}(\rhoz)=-Q_{\J_1}(\sigma)\,, \\[.5em]
  Q_{\J_2}(\rhoz) = 0\,, \ & Q_{\J_2}(\rhoe)   = Q_{\J_2}(\sigma) \,, \\[.5em]
  Q_{\J_3}(\sigma)= 0\,, \ & Q_{\J_3}(\rhoe) = 0 = Q_{\J_3}(\rhoz) \,.
  \eear \ee
Additivity of monodromy charges then implies
  \be  Q_\J(\rhoe) =  Q_{\J_2}(\rhoe) \,, \qquad
  Q_\J(\rhoz) = -Q_{\J_1}(\rhoz) \,,  \qquad
  Q_\J(\sigma)=  Q_{\J_1}(\rhoz) + Q_{\J_2}(\rhoe) \,, \ee
which tells us that in order to have $\J\iN\Hp$, i.e.\
$Q_\J(\rhoe)\eq0\eq Q_\J(\rhoz)$, it is necessary and sufficient that
$Q_{\J_1}(\rhoz)\eq0\eq Q_{\J_2}(\rhoe)$, which in turn is equivalent to
$\J_1,\J_2\iN\Hp$. We conclude that the kernel of the map
$(\J_1,\J_2,\J_3) \,{\mapsto}\, [\J_1^{-1}\J_2\J_3] \iN\imS/\Hp$
is the subgroup $\calsp_{\rho_1}\Times\calsp_{\rho_2}\Times(\caluhp\sigma)$
of $\cals_\rhoe\Times\cals_\rhoz\Times(\caluhp\sigma)$.
By the homomorphism theorem this in turn implies that
  \be  |\imS|\, \sp{\rho_1}\sp{\rho_2} = |\Hp|\,\s{\rho_1}^{}\s{\rho_2}^{} \,.
  \ee
Moreover, $\imS$ is a subgroup (not just a subset) of $\Hd$, so
$|\Hd|/|\imS|$ is integral, and hence also
  \be  \dudo \equiv \Frac{|\Hd|\, \sp{\rho_1}\sp{\rho_2}}
  {|\Hp|\,\s{\rho_1}\s{\rho_2}} = [\Hd\,{:}\,\imS] \,\in\, \zetplus \,. 
  \labl{dudo'}
This proves the integrality of the number $\dudo$; it also provides us with a
simple reason for the integrality property, namely that $\dudo$ is the index of
the subgroup $\imS\eq p(\cals_{\rho_1}\Times\cals_{\rho_2}\Times\,
\caluhp\sigma)$ in the subgroup $\Hd$ of $\Gs$.

Note that for the case where all untwisted stabilizers are equal to the full
stabilizers (which immediately implies $\dudu\rho\eq1$), this already settles
the integrality problem. To establish integrality of $\dudu\rho$ as defined in 
\erf{dudu} in the general case, we use information about the \rep\ theory 
of twisted group \alg s (see appendix \ref{s.b} for an 
introduction to twisted group \alg s). Concretely, we just need to observe
that the twisted group algebra $\complex_\Fp\calsp_\rho$ is a semisimple
subalgebra of $\complex_\F\cals_\rho$, where $\F$ is the two-cocycle
(determined uniquely up to a coboundary) whose commutator cocycle is
$F_\rho|_{\cals_\rho\times\cals_\rho}$, while $\Fp$ is the analogous two-cocycle
whose commutator cocycle is $F_\rho|_{\calsp_\rho\times\calsp_\rho}$.

The results \Erf2c and \Erf4c about the decomposition of 
$\complex_\F\cals_\rho$-\rep s into irreducible 
$\complex_\Fp\calsp_\rho$-\rep s then tell us that the 
number $\dudu\rho$ has the natural interpretation as the multiplicity $\br$
\Erf4c that occurs in those branching rules, and hence in particular that
  \be  \dudu\rho = \br_{}^{(\calsp_\rho\subseteq\cals_\rho)}\in \zetpluso \,,
  \labl{dudu'}
as announced.

\subsection{Further consistency checks}

We would like to stress once more that the natural annulus coefficients are 
the quantities $\A_{\RhoBe\,\RhoBz}^\SigmaBp$ that were defined in subsection
\ref{s.61}, for which the upper and lower indices are, in general, of different 
type. (In fact, they differ in a rather subtle way, as even the very meaning of 
the upper index depends, via the definition of the group 
$\Hp\,{\equiv}\,\Hp_{\rho_1\rho_2}$, on the value of the two lower indices.)
In particular, it is the integrality of these numbers that guarantees
that the coefficients in an expansion of $A_{\RhoBe\,\RhoBz}(t)$ in powers of 
$q\eq\eE^{-\pi t}$ are integral and therefore allows for the 
interpretation of the annulus amplitude as a partition function.
On the other hand, for certain purposes it is also desirable to have at one's
disposal some closely related numbers $\AO$ for which all three labels are on
an equal footing, which means that the upper index should be of the same
form, i.e.\ $\SigmaB$, as the labels for the \bc s. In this subsection we
show that numbers of the latter form can indeed be introduced, and that they
satisfy two interesting systems of relations, see \Erf MA and \Erf AA below. 
(Further inspection shows that in many cases the $\AO$-coefficients are
just multiples of the annulus coefficients, although the constants of 
proportionality are generically non-integral.)
 
Let us start by inspecting the formula \Erf6r for the annulus coefficients. It 
may be noticed that to derive that result, no other property of $\Hp$ was used 
than that it is a subgroup of $\Gs$. Accordingly, analogous expressions are
obtained when any other subgroup of $\Gs$ is used. In particular, let us
introduce the quantities $\Ao\RhoBe\RhoBz\SigmaBo$ as the coefficients of the
annulus amplitude in the expansion
  \be  \A_{\RhoBe\,\RhoBz}(t) = \sum_\sigmaBo \Ao\RhoBe\RhoBz\SigmaBo
  \, \chio_\SigmaBo(\Frac{\ii t}2)  \Labl5t
\wrtt characters $\chio$ of the extension of the $\calap$-theory by the simple 
currents in the group $\Go\,{\subseteq}\,\Gs$ that was defined in \Erf Go.
We then have
  \be  \Ao\RhoBe\RhoBz\SigmaBo
  = \sumBoo\lambda \Frac{|\Go|}{|\Gs|}\,\Frac{\u\lambda}{\sO\lambda}
  \sum_{\psuo_\lambda\in\caluos_\lambda} \sumpsipsuo\lambda
  \tS_{(\lambdab,\psi_\lambda),\RhoBe}^* \tS_{(\lambdab,\psi_\lambda),\RhoBz}\,
  \So_{\LambdaBo,\SigmaBo} \,/\, \So_{\LambdaBo,\vaco} \,,  \Labl6s
where
  \be  \So_{\LambdaBo,\MuBo}
  := \Frac{|\So|}{\sqrt{\sO\lambda\uo\lambda\sO\mu\uo\mu}}
  \sum_{\J\in\caluo_\lambda\cap\caluo_\mu} \psuo_\lambda(\J)\,
  \psuo_\mu(\J)^*\, S^\J_{\lambdab,\mub}  \labl So
is the modular S-matrix of the $\Go$-extension.
While by construction the upper index of the numbers $\AO$ is a priori again
of a type different from the two lower ones, we will now show that actually 
their values only depend on full $\Gs$-orbits $\SigmaBO$.

To this end we compare the expressions \Erf So for $\So$ and \Erf tS for
$\tS$ and take into account the specific way in which $\So$ appears in \Erf6s.
Our aim is then to show that up to numerical factors, we are allowed to
replace $\So$ by $\tS$. To this end we observe that apart from the 
different prefactors, we have to deal with the presence
of different group characters and with the different summation range for the
simple currents. As for the characters, we simply need to implement their
restriction properties. Concerning the simple current summation, the following
reasoning shows that in the expression \Erf6s only terms with $\J$ in the
intersection of the two groups $\caluo_\lambda{\cap}\caluo_\sigma$ and
$\cals_\lambda{\cap}\calu_\sigma$ give non-vanishing contributions.
\nxt
For $\J\iN\cals_\lambda{\setminus}\,\caluo_\lambda$, we distinguish between
two cases. First, when $\J\iN\cals_\lambda{\setminus}\,\calso_\lambda$, we
deduce from the definition \Erf Go of $\Go$ that $Q_\J(\rhoe)\nE Q_\J(\rhoz)$,
so that by the gradation property of the numbers \Erf6s (which follows by a
consideration analogous to that for the annulus coefficients) we can assume that
$Q_\J(\sigma)\nE0$; but then $\sigmab$ cannot be a fixed point of $\J$, and
hence $S^\J_{\lambdab,\sigmab}\eq0$, so that the corresponding contribution to 
the annulus tensor vanishes.
Second, when $\J\iN\calso_\lambda{\setminus}\,\caluo_\lambda$, then there exists
a $\JK\iN\calso_\lambda$ such that $F_\lambda(\J,\JK)\nE0$, and hence from
  \be  S^\J_{\lambdab,\sigmab} = S^\J_{\JK\lambdab,\sigmab}
  = \eE^{2\pi\ii Q_\JK(\sigma)} F_\lambda(\J,\JK)\,S^\J_{\lambdab,\sigmab}
  = F_\lambda(\J,\JK)\,S^\J_{\lambdab,\sigmab}  \ee
we can again conclude that $S^\J_{\lambdab,\sigmab}$ vanishes.
Here in the last equality we have also used the fact that $\JK\iN\Go$, so that
we can again invoke the grading property so as to set $Q_\JK(\sigma)\eq0$.
\nxt
For $\J\iN\calu_\sigma{\setminus}\,\caluo_\sigma$ the same reasoning as in
the first part of the previous case applies.
\nxt
For $\J\iN\caluo_\sigma{\setminus}\,\calu_\sigma$, there exists 
a $\JK\iN\cals_\sigma{\setminus}\,\calso_\sigma$ with $F_\lambda(\J,\JK)\nE0$,
so that now
  \be  S^\J_{\lambdab,\sigmab} = S^\J_{\lambdab,\JK\sigmab}
  = \eE^{2\pi\ii Q_\JK(\lambda)} F_\sigma(\J,\JK)\,S^\J_{\lambdab,\sigmab}
  = F_\sigma(\J,\JK)\,S^\J_{\lambdab,\sigmab}  \ee
tells us that $S^\J_{\lambdab,\sigmab}$ must be zero.

Furthermore, we can combine the summations over $\psuo_\lambda$ and 
$\psi_\lambda{\gt}\psuo_\lambda$ to a summation over all $\psi_\lambda$,
while the $\lambdaBo$-summation can be converted to a sum over all $\lambdab$ 
times a factor of $\sO\lambda/|\Go|$. It follows that
  \be  \Ao\RhoBe\RhoBz\SigmaBo
  = \ao\sigma \!
  \sumbo\lambda\sum_{\psi_\lambda\in\cals_\lambda^*} \Frac{\u\lambda}{|\Gs|}\,
  \Frac{ \tS_{(\lambdab,\psi_\lambda),\RhoBe}^*
  \tS^{}_{(\lambdab,\psi_\lambda),\RhoBz}\,
  \tS^{}_{(\lambdab,\psi_\lambda),\SigmaB} }
  {\tS_{(\lambdab,\psi_\lambda),\vacB}} \,,  \Labl7t
with
  \be  \ao\sigma \equiv \aor\rhoe\rhoz\sigma
  := \frac{\sqrt{\s\sigma\u\sigma}}{\sqrt{\sO\sigma\uo\sigma}} \,.  \Labl ao
Note that, as indicated by the notation $\aor\rhoe\rhoz\sigma$, this prefactor
not only depends on the upper label $\sigma$ of the coefficient \Erf7t, but
implicitly on the values of the two lower labels $\rhoe$ and $\rhoz$ as
well, namely through the relevant subgroup $\Go\,{\equiv}\,\Go_{\rho_1\rho_2}$
of $\Gs$. What is more interesting, however, is that
the prefactor is constant on the $\Gs$-orbit of $\sigmab$; this implies
that we can replace the upper label according to
  \be  \Ao\RhoBe\RhoBz\SigmaBo \equiv \Ao\RhoBe\RhoBz\SigmaBO  \,,  \ee
where $\phu_\sigma$ is any $\calu_\sigma$-character satisfying
  \be  \phu_\sigma|^{}_{\calu_\sigma\cap\caluo_\sigma}
  = \psu_\sigma|^{}_{\calu_\sigma\cap\caluo_\sigma} \,.  \ee
Thus, as announced, we are dealing with quantities where all three labels are
on the same footing. By comparison with formula \Erf4n for the structure
constants of \clAb, the coefficients $\AO$ are, up to the prefactor \Erf ao,
just the `opposite structure constants', i.e.\ those obtained when summing
over the other index of the non-symmetric $\tS$-matrix.

To see how the numbers $\Ao\RhoBe\RhoBz\SigmaB$ are related to the
annulus coefficients $\A_{\RhoBe\,\RhoBz}^\SigmaBp$, we recall the definition
\erf X of extended characters. Consider first the situation where the relevant
untwisted stabilizer groups are related as $\calup_\sigma\,{\subseteq}\,\caluo
_\sigma$; then we can immediately conclude that
  \be  \bearll  \chio_\SigmaBo \!\!
  &= \Frac1{\sqrt{\sO\sigma\uo\sigma}}
  \dsty\sum_{\J\in\Go} \chib_{(\J\sigmab,\psu_\sigma)}
   = \Frac{\sqrt{\sp\sigma\up\sigma}}{\sqrt{\sO\sigma\uo\sigma}}
  \dsty\sum_{\KP\in\Go/\Hp} \chip_{\KP\star[\sigmab,\psu_\sigma|^{}_
  {\calup_\sigma}]\oei}
  \,,  \eear \ee
where $\psu_\sigma|^{}_{\calup_\sigma}$ is the $\calup_\sigma$-character that
is obtained by restricting the $\caluo_\sigma$-character $\psu_\sigma$ to the
subgroup $\calup_\sigma$. This implies that the corresponding coefficients of
the annulus amplitude are linearly related as well,
  \be  \Ao\RhoBe\RhoBz\SigmaBO
  = \Frac{\sqrt{\sO\sigma\uo\sigma}}{\sqrt{\sp\sigma\up\sigma}}
  \cdot \A_{\RhoBe\,\RhoBz}^{[\sigmab,\phu_\sigma|^{}_{\calup_\sigma}]\oei}
  \,.  \ee
In contrast, in the case where $\calup_\sigma$ is not a subgroup of
$\caluo_\sigma$ (which, in spite of $\calsp_\sigma\,{\subseteq}\,\calso_\sigma$,
can happen for the same reasons as in the case of $\calup_\sigma$ versus
$\calu_\sigma$, see the remarks
after formula \Erf90), more complicated linear combinations 
arise that mix those characters of the $\Hp$- and of the $\Go$-extensions for 
which the group characters $\psuo_\sigma\iN\caluos_\sigma$ and
$\psup_\sigma\iN\calups_\sigma$ have common restrictions to the intersection
$\caluo_\sigma\,{\cap}\,\calup_\sigma$. As the precise form of this relation
does not seem to play any particular role, we refrain from writing it out here.

Having arrived at sensible coefficients $\AO$ with three labels of equal type,
we are now in a position to perform a few additional consistency checks.
We first compute the product of two of these quantities, regarding them as
matrices in their lower indices. By direct computation we find
  \be \bearl
  \dsty\sum_\rhoB\sum_{\psu_\rho\in\calu_\rho^*}\!
  \Ao\RhoBe\RhoB\SigmaBe\, \aOr\rhoe\rho{\sigma_1}^{-1}
  \aOr\rho\rhoz{\sigma_2}^{-1}\, \Ao\RhoB\RhoBz\SigmaBz
  \\{}\\[-.8em] \hsp{2.7}
  = \dsty\sumbo\lambda\!\sum_{\psi_\lambda\in\cals_\lambda^*}
  \Frac{\u\lambda}{|\Gs|}\, [\tS_{(\lambdab,\psi_\lambda),\vac}]^{-2}_{}\,
  \tS^{}_{(\lambdab,\psi_\lambda),\SigmaBe}
  \tS^{}_{(\lambdab,\psi_\lambda),\SigmaBz}\,
  \tS^*_ {(\lambdab,\psi_\lambda),\RhoBe}
  \tS^{}_{(\lambdab,\psi_\lambda),\RhoBz}
  \\{}\\[-.8em] \hsp{2.7}
  = \dsty\sum_\SigmaBd
  \sumbo\lambda\sum_{\psi_\lambda\in\cals_\lambda^*} \Frac{\u\lambda}{|\Gs|}\,
  \tS^{}_{(\lambdab,\psi_\lambda),\SigmaBe}
  \tS^{}_{(\lambdab,\psi_\lambda),\SigmaBz}
  \tS^*_ {(\lambdab,\psi_\lambda),\SigmaBd}
  [\tS_{(\lambdab,\psi_\lambda),\vac}]^{-1}_{}
  \\{}\\[-1.0em] \hsp{6.26}
  \dsty\sumbo\mu\sum_{\psi_\mu\in\cals_\mu^*} \Frac{\u\mu}{|\Gs|}\,
  \tS^{}_{(\mub,\psi_\mu),\SigmaBd} \tS^*_ {(\mub,\psi_\mu),\RhoBe}
  \tS^{}_{(\mub,\psi_\mu),\RhoBz} [\tS_{(\mub,\psi_\mu),\vac}]^{-1}_{}
  \,.  \eear \Labl56
Here we have introduced a weight factor $\aO{\sigma_1}^{-1}\aO{\sigma_2}^{-1}$
into the summation, which correctly accounts for the number of chiral
boundary labels (i.e.\ boundary blocks) that is subsumed in the upper index
$\sigma$; because of the dependence of $\ao\sigma$ on the lower labels,
this weight factor need not be constant. (Of course one could avoid the
presence of a weight factor by simply considering the quantities
$\aO\sigma^{-1}\Ao\RhoBe\RhoB\SigmaB$ instead, but this appears to be
less natural.) {}From formula \Erf56 we can read off that
  \be  \sum_\RhoB \Ao\RhoBe\RhoB\SigmaBe\, \aOr\rhoe\rho{\sigma_1}^{-1}
  \aOr\rho\rhoz{\sigma_2}^{-1}\, \Ao\RhoB\RhoBz\SigmaBz
  = \sum_\SigmaBd \Mo\SigmaBe\SigmaBz\SigmaBd\, \Ao\RhoBe\RhoBz\SigmaBd
  \Labl MA
with
  \be  \Mo\SigmaBe\SigmaBz\SigmaBd
  := \frac{\aor\rhoe\rho{\sigma_1}\,\aor\rho\rhoz{\sigma_2}}
  {\aor{\sigma_1}{\sigma_2}{\sigma_3}\,\aor\rhoe\rhoz{\sigma_3}}\,
  \Ao\SigmaBep\SigmaBz\SigmaBdp \,.  \Labl Mo
Thus the coefficients $\AO$ can be regarded as the basis elements of a
\findim\ \alg\ with structure constants \Erf Mo. The presence of such an
\alg ic structure is often interpreted as a `completeness relation' for the
\bc s.

It is already apparent from the fact that the structure constants \Erf Mo
are essentially equal to suitable numbers $\AO$ that in addition some kind
of `associativity relation'  holds, where one sums over the upper index of
these objects. Indeed, by the same kind of calculation as above one checks
that
  \be  \bearl  \dsty\sum_\sigmaB\sum_{\psu_\sigma\in\calu_\sigma^*}
  \Ao\RhoBe\RhoBz\SigmaB\, \aOr\rhoe\rhoz\sigma^{-1}
  \aOr\rhod\rhov{\sigma^+}^{-1}\, \Ao\RhoBd\RhoBv{\SigmaB^+_{}}
  \\{}\\[-.98em]\hsp{6.1}
  = \dsty\sum_\sigmaB\sum_{\psu_\sigma\in\calu_\sigma^*}
  \Ao\RhoBe{\RhoBd^+_{}}\SigmaB\, \aOr\rhoe\rhod\sigma^{-1}
  \aOr\rhoz\rhov{\sigma^+}^{-1}\, \Ao{\RhoBz^+_{}}\RhoBv{\SigmaB^+_{}}
  \,.  \eear \Labl AA

Relations of the form \Erf MA and \Erf AA are expected on the basis of 
factorization arguments \cite{lewe3,sasT2,prss3}. But a rigorous derivation of
these identities from factorization, in particular for \bc s that do not 
preserve the full bulk symmetry, still remains to be established.
Moreover, such relations are technically rather difficult to exploit
in non-trivial theories. In our opinion, they do not constitute
an optimal starting point for the classification of \bc s.

\newpage
\appendix

\sect{Simple current extensions}\label{s.a}

\subsection{The spectrum of primary fields}

This appendix summarizes the results about integer spin simple current 
extensions \cite{scya6,fusS6} that are needed in the main text. When some
\cft\ with chiral \alg\ $\cala$ is obtained as an extension of a theory
with chiral \alg\ $\calap$ by a group $\Gs$ of integer spin simple currents,
then its primary fields are labelled by pairs $\LambdaB$, where $\lambdaB$ is a
$\Gs$-orbit with vanishing monodromy charge $Q_\J(\lambda)$ for all $\J\iN\Gs$,
$\psu_\lambda$ is a certain group character (see below), and the square brackets
refer to classes \wrtt equivalence relation that will be given in \Erf eq. 
Here by the monodromy charge of $\lambdab$ \wrt $\J$ we mean the combination
  \be  Q_\J(\lambda):= \Delta_\lambdab+ \Delta_\J - \Delta_{\J\star\lambdab}
  \bmod\zet  \ee
of conformal weights; we write $\lambda$ rather than $\lambdab$ for the 
argument because this quantity is constant on $\Gs$-orbits.
The monodromy charges also satisfy 
  \be  Q_{\J^{-1}}(\mu) = - Q_\J(\mu) \,, \Labl14
and for every $\lambdab$ the map
  \be  \J \,\mapsto\, \exp(2\pi\ii Q_\J(\lambda))  \ee
furnishes a character of the simple current group $\Gs$.

To explain the meaning of the character $\psu_\lambda$, we first need to
introduce the stabilizer of $\lambdab$. This is the subgroup
  \be  \cals_\lambda := \{\J\iN\Gs\,|\, \J\,{\star}\,\lambdab\eq\lambdab \}
  \,\subseteq\Gs \,,  \ee
which is again constant on $\Gs$-orbits.
(The symbol `$\star$' stands for the fusion product, and for brevity
the basis elements of the fusion ring are just denoted by their labels
$\lambdab$.) When $\J\iN\cals_\lambda$, then one says that $\lambdab$ is a 
fixed point of $\J$; fixed points of an integer spin simple current $\J$ 
clearly have vanishing monodromy charge:
  \be  \J\iN\cals_\lambda \ \Rightarrow \
  Q_\J(\lambda)= \Delta_\lambdab+ \Delta_\J - \Delta_{\lambdab}
  \bmod\zet = 0 \,.  \Labl0Q
Now $\psu_\lambda$ is a character of a particular subgroup $\calu_\lambda$
of the full stabilizer $\cals_\lambda$, $\psu_\lambda\iN\calu_\lambda^*$.
This subgroup is called the \ustab\ of $\lambdab$; it is obtained as the subset
  \be  \calu_\lambda := \{\J\iN\cals_\lambda \,|\, F_\lambda(\JK,\J)\eq1\
  {\rm for\;all}\ \JK\iN\cals_\lambda \}  \ee
on which a certain \abihom\ 
  \be  F_\lambda:\quad \Gs\timeS\Gs\to U(1)  \ee
is trivial. The map $F_\lambda$, in turn, is determined by the matrices
$\SJ$ described in the next subsection through relation \Erf QF.

We remark that in \cite{fusS6} the notation $(\rhoB,\psu_\rho)$ was chosen 
in place of $\RhoB$. This is actually slightly misleading. Namely, in the 
equivalence relation that defines the classes $[\cdots]$, the simple currents
$\J$ act both on the primary label $\lambdab$ and on the character 
$\psu_\lambda$:
  \be  (\lambdab,\psu_\lambda) \,\sim\,
  \J\,(\lambdab,\psu_\lambda) = (\J{\star}\lambdab,\JJ\psu_\lambda)  \Labl eq
for all $\J\iN\Gs$, with
  \be  \JJ\psu_\lambda(\J') := F_\lambda(\J,\J')^*\,\psu_\lambda(\J')  \Labl JJ
for all $\J'\iN\calu_\lambda$; by the multiplicative property of the 
$F_\lambda$'s, the quantity $\JJ\psu_\lambda$ is again a character of 
$\calu_\lambda$. (The crucial point here is that we consider $F_\lambda$ also 
for currents $\J$ that are not in the stabilizer. When $\J$ is in the 
stabilizer, then $F_\lambda$ is equal to one by definition of the untwisted 
stabilizer.)

\subsection{The modular S-matrix}

The modular transformation matrix $S$ of the $\cala$-theory is given by
  \be  S_{\LambdaB,\MuB}
  := \Frac{|\Gs|}{\sqrt{\s\lambda\u\lambda\s\mu\u\mu}}
  \sum_{\J\in\calu_\lambda\cap\calu_\mu} \psu_\lambda(\J)\,
  \psu_\mu(\J)^*\, S^\J_{\lambdab,\mub} \,,  \labl S
where $\s\lambda\,{\equiv}\,|\cals_\lambda|$ and 
$\u\lambda\,{\equiv}\,|\calu_\lambda|$, and where $\{\SJ\,|\,\J\iN\Gs\}$ is 
a set of matrices which satisfy the following relations.
\nxt
$\SJ$ is non-vanishing only on fixed points:
  \be  S^\J_{\lambdab,\mub}\eq 0\eq S^\J_{\mub,\lambdab} \quad
  {\rm for}\ \J\,{\not\in}\,\cals_\mu \,.  \Labl00
\nxt
The restriction of $\SJ$ to the fixed points of $\J$ is unitary, and together 
with the restriction $T^\J$ of the T-matrix it obeys the usual relations
$(S^\J T^\J)^3\eq(S^\J)^2$ and $(S^\J)^4\eq\one$ of the two-fold cover \slz\
of the modular group. 
\nxt
For every element $\J'\iN\Gs$, $\SJ$ satisfies the simple current relations
  \be  
  \SJ_{\J'\lambdab,\mub} = F_\lambda(\J',\J)\,\eE^{2\pi\ii Q_{\J'}(\mu)}\,
  \SJ_{\lambdab,\mub} \,, \qquad 
  \SJ_{\lambdab,\J'\mub} = F_\mu(\J',\J)^*_{}\,\eE^{2\pi\ii Q_{\J'}(\lambda)}\,
  \SJ_{\lambdab,\mub} \,. 
  \Labl QF
\nxt
The matrices for inverse currents are transposed to each other:
  \be  S^{\J^{-1}}_{\mub,\rhob} = S^\J_{\rhob,\mub} \,.  \Labl-J
\nxt
The space of one-point \cblock s with insertion of the simple current $\J$ on 
the torus has a natural basis labelled by the fixed points $\lambdab$ of $\J$.
Upon a suitable canonical choice of normalization of the basis elements, $\SJ$ 
plays the role of the modular S-transformation matrix for those blocks 
\cite{bant6}. In particular, $S^\vacb$ is the ordinary modular S-matrix of 
the $\calap$-theory, $S^\vacb\eq\bS$. There is some freedom left in the 
canonical basis choice, which is irrelevant for the formula \erf S but does 
play a role for the definition of the matrix $\tS$ in \Erf tS.
We expect that the prescription given in \cite{bant6} can be refined
in such a manner that the remaining freedom in the normalization of
the blocks constitutes a character of the full stabilizer $\cals_\lambda$.
\nxt
In the case of \wzwts, $\SJ$ coincides, up to possibly a fourth root of unity,
with the ordinary S-matrix of another \wzwt\ that is determined by $\calap$
and $\J$ \cite{fusS3,fusS6}.
\nxt
The square of $\SJ$ obeys
  \be  (S^\J)^2_{} = \eta^\J\, C^\J = C^\J\,(\eta^\J)^*_{} \,,  \Labl eJ
where $C^\J$ is the restriction of the charge conjugation of the 
$\calap$-theory to the fixed points of $\J$ and $\eta^\J$ is a diagonal 
matrix, with properties to be specified below. 

When the extended theory has a surviving simple current $[\J']$ with
$\J'\,{\not\in}\,\Gs$, then the relations \Erf QF are 
still valid for that current. It follows that the matrix \erf S satisfies
  \be  S_{[\J']\star\LambdaB,\MuB}
  = \eE^{2\pi\ii Q_{\J'}(\mu)}_{} \cdot S_{\LambdaB,\MuB} \,.  \Labl QQ
In words, the monodromy charges \wrt $[\J']$ in the extended theory are
the same as those \wrt $\J'$ in the original theory.

The properties \Erf0Q of fixed points and \Erf QF of $\SJ$
can e.g.\ be employed to derive the alternating property of the
\bihom s $F_\lambda$. Indeed, for every $\J$ and every fixed point $\lambdab$ of
$\J$ there exists at least one $\mub$ such that $S^\J_{\lambdab,\mub}\nE0$. The
fact that $Q_J(\mub)\eq0$ then implies
  \be  S^\J_{\lambdab,\mub} = S^\J_{\J\lambdab,\mub}
  = S^\J_{\lambdab,\mub} \cdot \eE^{2\pi\ii Q_\J(\mu)}\,F_\lambda(\J,\J)
  = F_\lambda(\J,\J)\, S^\J_{\lambdab,\mub} \,,  \ee
from which it follows that
  \be  F_\lambda(\J,\J) = 1 \,.  \Labl0y
By the homomorphism property, one then concludes that the \bihom\ is trivial 
even within the whole cyclic group generated by $\J$, i.e.
  \be  F_\lambda(\J^m,\J^n) = 1  \ee
for all $m,n$.

\subsection{Properties of the matrices $\eta^\J$}

The entries $\eta^\J_\lambdab\delta_{\lambdab,\mub}$ of the matrix 
$\eta^\J$ which appears in \Erf eJ are constant on $\Gs$-orbits,
  \be  \eta^\J_{\JK\lambdab} = \eta^\J_\lambdab  \Labl ek
for all $\JK\iN\Gs$. Further, these matrices satisfy
  \be  \eta^{\J^{-1}}_\lambdab = (\eta^\J_\lambdab)^*  \Labl es
as well as
  \be  \eta^\J_\lambdabp = (\eta^\J_\lambdab)^*  \,.  \Labl ep
Moreover, for every fixed $\lambdab$,
$\eta_\lambdab$ constitutes a character of $\calu_\lambda$, i.e.\
  \be  \eta^\J_\lambdab\,\eta^{\JK}_\lambdab = \eta^{\J\JK}_\lambdab
  \Labl er
for all $\J,\JK\iN\calu_\lambda$; it is not necessarily a character of the
full stabilizer $\cals_\lambda$, though.

Combining \Erf-J, \Erf eJ and the fact that $C^\J$ is the restriction of a 
permutation matrix and hence has order two, it follows that
  \be  \eta^{\J^{-1}} = (\eta^\J)^*  \,.  \Labl et
Finally, based on results \cite{bave} about mapping class group \rep s,
for self-conjugate fixed points $\lambdab$ of $\J$ one can show that
  \be  \eta^\J_\lambdab = \fsi_\lambdab \sum_{\mub,\nub} Q_\J(\mu)\,
  \Sb\vac\mu\Sb\vac\nu\,\eE^{4\pi\ii(\Delta_\mub-\Delta_\nub)}\,
  \N\mu\nu\lambda\,, \ee
where $\fsi_\lambdab$ is the Frobenius\hy Schur indicator for $\lambdab$
\cite{bant6}.

\subsection{Conjugation properties}

The relations \Erf-J, \Erf eJ and \Erf es, \Erf ep imply the 
conjugation properties
  \be  S^\J_{\lambdabp,\mub} = \eta^{\J*}_\lambdab\,(S^\J_{\mub,\lambdab})^*_{}
  \,, \qquad
  S^\J_{\lambdab,\mubp} = \eta^{\J}_\mub\,(S^\J_{\mub,\lambdab})^*_{}  \Labl ec
of the matrices $\SJ$.
When combined with the simple current symmetry \Erf QF, the result \Erf ec
implies in particular that
  \be  \eE^{2\pi\ii Q_\JK(\mu)}\,F_\lambdap(\JK,\J)\, \SJ_{\lambdabp,\mub}
  = \SJ_{\JK\lambdabp,\mub} = \SJ_{(\JK^{-1}\lambdab)^+_{\phantom i},\mub}
  = \eta^{\J*}_\lambdab\,
  \Llb \eE^{2\pi\ii Q_{\JK^{-1}}(\mu)}\,F_\lambda(\JK^{-1},\J)^*_{}\,
  \SJ_{\mub,\lambdab} {\Lrb}^* \,,  \ee
from which with the help of the homomorphism property of $F$, the identity
\Erf14 for monodromy charges, and once more formula \Erf ec it follows that
the \bihom s for conjugate orbits are complex conjugate to each other,
  \be  F_\lambdap(\JK,\J) = F_\lambda^{}(\JK,\J)^*_{}  \Labl Fs
for all $\J,\JK\iN\Gs$.

The conjugation of the primary labels of the extended theory is defined by
  \be  \LambdaBP := [\lambdabp,\psu_\lambda^+]  \Labl 9m
with 
  \be  \psu_\lambda^+(\J) := \psu_\lambda^{}(\J)\, \eta^\J_\lambda \,.  \Labl9n
Because of the character property \Erf er of $\eta_\lambda$,
the quantity $\psu_\lambda^+$ introduced this way is again a character of
$\calu_\lambda$.
That the prescription \Erf9m is the right one can be checked by inserting
the relations \Erf9n and \Erf ec into the definition \erf S of $S$, which leads
to the correct conjugation property
  \be  S_{\LambdaB,\MuBP}^{} = S_{\LambdaB,\MuB}^*  \ee
of the modular S-matrix.

\subsection{Characters}

The irreducible characters $\X_\Lambdab$ of the $\cala$-theory are
expressed in terms of the irreducible characters $\chib_\lambdab$
of the $\calap$-theory as
  \be  \X_\Lambdab
  = d_\lambda \sum_{\J\in\Gs/\cals_\lambda} \chib_{(\J\lambdab,\psu)}
  = \Frac{d_\lambda}{|\cals_\lambda|} \sum_{\J\in\Gs} \chib_{(\J\lambdab,\psu)}
  = \Frac1{\sqrt{|\cals_\lambda|\,|\calu_\lambda|}}
  \sum_{\J\in\Gs} \chib_{(\J\lambdab,\psu)} \,.  \labl X
The factor 
  \be  d_\lambda = \sqrt{\s\lambda/\u\lambda} \,,  \ee
which is different from 1 when there is a genuine untwisted stabilizer,
accounts for the presence of the degeneracy space $\Vpsu$ in the
decomposition \erf{deco}.

We close this exposition with a remark on the $\calap$-theory.
One may always extend a summation over the untwisted sector to a summation
over all sectors by inserting the projection operator
  \be  \Frac1{|\Gs|}\sum_{\J\in\Gs} \eE^{2\pi\ii Q_\J(\lambda)} \,.  \labl P
For instance, we have
  \be  \bearll
  \dsty\sumbo\mu S^\J_{\mub,\rhob}\, (S^\J_{\mub,\sigmab})^*_{} \!\!
  &= \dsty\sum_\mub S^\J_{\mub,\rhob} (S^\J_{\mub,\sigmab})^*_{} \cdot
  \Frac1{|\Gs|}\sum_{\J'\in\Gs} \eE^{-2\pi\ii Q_{\J'}(\mu)}
  \\{}\\[-1.0em]
  &= \Frac1{|\Gs|}\dsty\sum_{\J'\in\Gs}
  \sum_\mub S^\J_{\mub,\rhob} (S^\J_{\mub,\J'\sigmab})^*_{}\,F_\sigma(\J',\J)^*
  = \Frac1{|\Gs|}\sum_{\J'\in\Gs} F_\sigma(\J',\J)^*\,
  \delta^{}_{\rhob,\J'\sigmab} \,.  \eear \Labl ip

\sect{Twisted group algebras and \bihom s}\label{s.b}
 
\subsection{Two-cocycles}
 
Let $\G$ be a finite group (not necessarily abelian), and 
$\F$ a two-cocycle on $\G$ with values in $\complex^\times$, i.e.\ a map
  \be  \F: \quad\G\times\G \to \complex^\times \ee
such that
  \be  \F(g_1,g_2)\,\F(g_1g_2,g_3) = \F(g_1,g_2g_3)\,\F(g_2,g_3)  \labl{prop1}
for all $g_1,g_2,g_3\iN\G$. The cohomologically trivial two-cocycles,
i.e.\ the coboundaries, are of the form
  \be  \F(g,h)=\epsilon(g)\,\epsilon(h)/\epsilon(gh)  \ee
with $\epsilon$ an arbitrary 
function $\epsilon{:}\ \G\to\complex^\times$. 
When $\G$ is cyclic, all two-cocycles are coboundaries.
 
By setting $g_1\eq e\eq g_2$, 
\resp\ $g_2\eq e\eq g_3$ (with $e$ the unit element of $\G$), we learn that
  \be  \F(e,g) = \F(g,e) = \F(e,e) =: \alfa  \labl2
for every $g\iN\G$. 
Let now $\F$ be some given two-cocycle; we change $\F$ by multiplying 
it with a coboundary obtained from any function $\epsilon$ such that 
$\epsilon(e)\eq\alfa^{-1}$. The so obtained cohomologous cocycle $\F'$ 
satisfies
  \be  \F'(g,e)= \F(g,e)\,\epsilon(g)\,\epsilon(e)\,\epsilon(ge)^{-1} =
  \F(g,e)\, \alfa^{-1} = 1 \, . \Labl8e
We will from now on assume this property, but for simplicity 
denote this cohomologous cocycle $\F'$ just
by $\F$; doing so, we have to keep in mind that we are now
only allowed to modify it by coboundaries coming from such $\epsilon$ 
which in addition fulfil
$\epsilon(e)\eq1$. (Sometimes this is also taken as part of the definition of
a cocycle.) Further, by looking at the triple $g,g^{-1},g$ we learn that
  \be  \F(g,g^{-1}) = \F(g^{-1},g)  \Labl3b
for every $g\iN\G$.

\subsection{Twisted group algebras}
 
To any finite group $\G$ one associates its {\em group algebra\/} $\complex\G$,
which is an associative unital algebra over the complex numbers.
The dimension of $\complex\G$ is $|\G|$, and it has a basis 
$\{b_g\}$ that is labelled by group elements $g\iN\G$ and has multiplication
  \be  b_g\,b_{g'} = b_{g g'} \,.  \Labl bb
The elements $b_g$ are units of $\complex\G$. The group algebra is
commutative if and only if $\G$ is abelian.
 
Given a two-cocycle $\F$ on the finite group $\G$, one can also
define the {\em $\F$-twisted group algebra\/}
$\complex_\F\G$ by modifying the multiplication \Erf bb to
  \be  b_g b_{g'} = \F(g,g')\, b_{g g'}\,.  \ee
Relation \erf{prop1} ensures that $\complex_\F\G$ is still associative. 
Twisted group algebras are unital; the unit element is given by 
$\F(e,e)^{-1}b_e$. In particular, for every cocycle with property \Erf8e,
$b_e$ is still a unit element. We will assume from now on that $\F$ is 
such a cocycle.
 
The isomorphism type of $\complex_\F\G$ as an algebra over $\complex$
depends only on the cohomology class of $\F$. The $\F$-twisted group algebra
$\complex_\F\G$ is isomorphic to the ordinary
group algebra $\complex\G$ if and only if 
$\F$ is a coboundary, and \cite{KArp4} if and only if there is a homomorphism 
of complex algebras from $\complex_\F\G$ to $\complex$ (i.e.\ if and only if 
$\complex_\F\G$ has a \onedim\ \rep). Further, 
a twisted group algebra is abelian if and only if $\G$ is abelian and the
cocycle is cohomologically trivial.

Every twisted group algebra is semisimple.
Let us list a few general properties of semisimple associative algebras $A$
over $\complex$:
\nxt Every  $A$-representation is fully reducible.
\nxt As an algebra over $\complex$, $A$ is isomorphic to a direct sum
  \be  A \,\cong\, \bigoplus_i M_{d_i}(\complex) \Labl1f
     of full matrix algebras, where the $d_i$ are the dimensions of the
     inequivalent \irrep s of $A$.
\nxt The number of (equivalence classes of) irreducible
     $A$-\rep s equals the dimension of the center of $A$.
\nxt The dimensions of the inequivalent \irrep s are the square roots $d_i$ of
     the dimensions of the simple summands $M_{d_i}(\complex)$ that appear 
     in the decomposition \Erf1f.

\subsection{Representation theory of twisted group algebras}

The representation theory of $\complex_\F\G$ is governed by the {\em center\/}
of  $\complex_\F\G$. 
For an explicit description of the center some additional concepts are required.
First, for a given cocycle $\F$, a group element $g\iN\G$
is called $\F$-{\em regular\/} if and only if
  \be  \F(g,h) = \F(h,g) \ee
for all $h$ in the centralizer $C_g(\G)$ of $g$. This is equivalent to saying
that the two elements $b_g$ and $b_h$ of $\complex_\F\G$ commute:
  \be  b_g b_h= b_h b_g \quad\mbox{for all}\quad h\iN\G \,\ {\rm with}\,\
  gh\eq hg \,.  \ee
The set of all $\F$-regular elements of $\complex_\F\G$ will be denoted by
$\GReg_{}\,{\equiv}\,\GReg_\F$.
With $g$ every conjugate element $hgh^{-1}\iN\G$ is $\F$-regular, too;
accordingly we also call a conjugacy class of $\F$-regular elements 
$\F$-regular. Further, if $\F'$ is a two-cocycle cohomologous to $\F$, then 
every $\F$-regular element is also $\F'$-regular.
 
Second, for many purposes it is convenient to choose special cocycles
within a cohomology class. 
A cocycle $\F$ is called {\em standard\/} if it satisfies both
  \be  \F(g,g^{-1}) =1 \qquad \mbox{for all } g\iN\G \Labl4s
and
  \be  \F(g,h)\,\F(gh,g^{-1}) = 1 \qquad \mbox{for all } h\iN\Greg,\,g\iN\G
  \,.  \Labl5s
In terms of the twisted group algebra this means that $b_{g^{-1}}$ is the 
inverse of $b_g$ and that conjugation of $\F$-regular elements works without 
additional factors, i.e.
  \be  \bearl
  b_{g^{-1}} = (b_g)^{-1} \quad \mbox{for all } g\iN\G \,, \\[.7em]
  b_g b_h (b_g)^{-1} = b_{ghg^{-1}}\quad \mbox{for all } h\iN\Greg,\,g\iN\G
  \,. \eear \Labl44
By a suitable diagonal change of the basis of the twisted group algebra, 
the validity of \Erf44 can always be achieved \cite{KArp4}. 
Finally, a {\em left transversal\/} for a subgroup $\HG$ of $\G$ is a set of
representatives for $\HG\backslash\G$, i.e.\ a subset of
$\G$ that contains precisely one element from each left coset $\HG x$.
 
The center of the twisted group algebra then
corresponds to $\F$-regular classes as follows \cite{KArp4}.
Let $\{g_1,g_2,\,..., g_\ell\}$ be a set of representatives for the $\F$-regular
classes, and, for each $i$, $T_i$ a left transversal for the conjugacy class 
$C_{g_i}$ of $g_i$. Then the $\ell$ elements
  \be  \sum_{h\in T_i} b_h b_{g_i} (b_h)^{-1} \ee
form a basis of the center. In particular, when $\F$ is standard, then the 
$\ell$ sums 
  \be  \sum_{h\in C_{g_i} } b_h  \ee
over the $\F$-regular conjugacy classes constitute a basis for the center.

The number of inequivalent \irrep s of a twisted group algebra is equal to the
dimension of the center (since the algebra is semisimple), and hence 
to the number of regular conjugacy classes. Notice, though, that there is 
no canonical correspondence between \irrep s and conjugacy classes.
 
\subsection{Commutator cocycles for abelian groups}
 
We now consider finite groups $\G$ that are {\em abelian\/}. 
Then the $\F$-regular elements are characterized by
  \be  \F(g,h) = \F(h,g) \quad \mbox{for all }\, h\iN\G \,.  \Labl gg
Further, in the abelian case
the subset $\Greg$ of $\F$-regular elements of $\G$ is actually a
sub{\em group\/}. Namely, if both $g_1$ and $g_2$ are regular, i.e.\ if $b_
{g_i}b_h\eq b_hb_{g_i}$ for $i\eq1,2$ and all $h\iN C_{g_i}\eq\G$, then one has
  \be  \F(g_1,g_2)\,b_{g_1g_2}b_h=b_{g_1}b_{g_2}b_h = b_{g_1}b_hb_{g_2}
  = b_hb_{g_1}b_{g_2} = b_h\cdot\F(g_1,g_2)\,b_{g_1g_2}  \ee
for all $h\iN\G$, and hence $b_{g_1g_2}$ is regular as well.

We will use the term {\em\bihom\/} for every function 
  \be  F: \quad \G\times\G \to \complex^\times \ee
that satisfies
  \be  F(g_1g_2,g_3) = F(g_1,g_3) F(g_2,g_3) \qquad\mbox{and}\qquad
  F(g_1,g_2g_3) = F(g_1,g_2) F(g_1,g_3)  \ee
for all $g_1,g_2,g_3\iN\G$.
It may be noted that every \bihom\ obeys \erf{prop1} and hence constitutes
a two-cocycle on the abelian group $\G$. But this property will not be 
important to us. Also, we are interested in definite \bihom s\ rather 
than their cohomology classes. In particular one may have to deal with 
\bihom s that are non-trivial even though cohomologically trivial.\,%
 \futnote{An example is provided by $G\eq\zet_2\timeS\zet_2$, with the \bihom\ 
$F$ obeying $F(1,\J)\eq1\eq F(\J,\J)$ for all $\J\iN\zet_2\timeS\zet_2$, while
all other values $F(\J,\J')$ are $-1$.
This situation is e.g.\ realized in the $D_4$ level 2 \wzwt.} 
 
A \bihom\ $F$ on $\G$ is called {\em alternating\/} if
  \be  F(g,g)= 1\quad \mbox{for all } g\iN\G \,,  \ee
which via the \bihom\ property implies (without using abelianness of $\G$) the
{\em antisymmetry\/} property\,%
 \futnote{The function $F$ that is present in the simple current relation 
\Erf QF constitutes an \abihom\ of the abelian simple current group $\Gs$. 
In the application to simple current extensions, the antisymmetry property holds
for the reason described around \Erf0y.}
  \be  F(h,g)=(F(g,h))^{-1} \quad\mbox{for all }\, g,h\iN\G  \,.  \Labl as

Every \abihom\ $F$ of an abelian group $\G$ can be written (see e.g.\
\cite[p.\,127]{BRow'} and \cite{hugh})
as the {\em commutator cocycle\/} of some two-cocycle $\F\,{\equiv}\,\F_F$ of 
$\G$, i.e.
  \be  F(g,h) = \F(g,h) / \F(h,g)  \Labl FF
for all $g,h\iN\G$. Note that $F$ determines only the cohomology class of $\F$;
put differently, $F$ depends on $\F$ only through the cohomology class of $\F$.
Moreover, since only for cohomologically trivial cocycles the twisted group
\alg\ is isomorphic to the ordinary group \alg, only for such cocycles the 
commutator cocycle is trivial (i.e.\ ${\equiv}\,1$). It follows that distinct 
cohomology classes of cocycles also possess distinct commutator cocycles.
 
The subgroup of $\F$-regular elements can be
expressed in terms of the commutator cocycle as
  \be  \GReg_\F = \GF \,,  \Labl gF
where
  \be  \GF:=\{h\iN\G\,|\, F(h,g)\eq1 \quad\mbox{for all }\, g\iN\G \}
  \,.  \Labl GF
This is indeed a subgroup of $\G$, and its definition is symmetric in the two 
arguments of $F$. (In our application in the main text, $\GF\eq\calu$ 
is the \ustab\ for the stabilizer subgroup $\cals\,{\subseteq}\,\Gs$.) 

\subsection{Representation theory and traces}

The \irrep s of $\complex_\F\G$ are labelled by the characters $\psit$ of 
the center $\GReg_\F\eq\GF$.
Now let $R$ be a $d$-dimensional \irrep\ of the twisted group algebra 
$\complex_\F\G$ of an abelian group $\G$, and let $R_\circ$ be an \irrep\ of 
$\G$ (in particular $R_\circ$ is \onedim). Then $R\,{\otimes}\,R_\circ$, which
maps $b_g\iN\complex_\F\G$ to
  \be  (R\,{\otimes}\,R_\circ)(b_g) := R(b_g)\,R_\circ(g) \,,  \ee
is again an \irrep\ of $\complex_\F\G$ of the same dimension
$d$. It can be shown that all \irrep s of $\complex_\F\G$ are
related this way, and thus in particular they all have the same dimension. 
By the Artin\hy Wedderburn theorem we then have
  \be  |\G| = {\rm dim}(\complex_\F\G) = d^2 \cdot N_{\rm irr.rep.} \,. \ee
Now the number $N_{\rm irr.rep.}$ of inequivalent \irrep s equals the dimension
of the center. We thus learn that when $G$ is abelian, then for all $d_i$ in 
\Erf1f one has 
  \be  d_i=d= \sqrt{[\G\,{:}\,\GReg_\F]} \,.  \ee
In particular, the index of $\GReg_\F$ in $\G$ is a complete square.

The elements of the center $\Greg$ act as multiples of the identity 
in every \irrep\ $R_{\psit}$:
  \be  R_{\psit}(b_g) = \psit(g)\,\one_d \quad\mbox{for all }\, h\iN\Greg
  \,.  \Labl1p
Next we show that if $g\iN\G$ is not regular, then it has vanishing trace
in every \irrep\ $R_{\psit}$ of $\complex_\F\G$, i.e.
  \be  \tr_{R_{\psit}}\, b_g = 0 \quad \mbox{for all }\, g\,{\not\in}\,\Greg \,.
  \Labl0t
We first consider the regular representation $R$ of $\complex_\F\G$.
In this representation $g\iN\G$ acts by mapping $b_h$ to a multiple of
$b_{gh}$. Hence the only group element that has a trace in the regular
representation is the identity element $e$. We thus have
  \be  \tr_R\,b_g = |\G|\,\delta_{g,e} \,.  \ee
Now since $\Greg$ is a subgroup of $\G$, for $g\,{\not\in}\,\Greg$ also $gh$ 
is not in $\Greg$ for all $h\iN\Greg$; in particular, $gh$ is not the identity
element, and therefore $\tr_R\, b_{gh}\eq0$. Now for a semisimple \alg\
the regular representation is a direct sum of all inequivalent
\irrep s, each appearing with multiplicity 1. Therefore the identity
  \be  R_{\psit}(b_{gh}) = R_{\psit}(b_g)\,R_{\psit}(b_h)
  = \psit(b_h)\, R_{\psit}(b_g) \,,  \ee
which follows with the help of \Erf1p, allows us to compute
  \be  0 = \tr_R\, b_{gh}= \sum_{\psit\iN\Gregs} \tr_{R_{\psit}}\,b_{gh} =
  \sum_{\psit\iN\Gregs} \psit(h)\, \tr_{R_{\psit}}\,b_g  \ee
for all $h\iN\Greg$. Fourier-transforming this relation over $\Greg$ then
finally yields $\tr_{R_{\psit}}\,b_g\eq0$
for all $\psit\iN\Gregs$ and all $g\,{\not\in}\,\Greg$, thus proving \Erf0t.

We also have
  \be  \sum_{\ssty\psi\in\G^*\atop\psi|_\Greg=\psu} \psi(g) = 
  d\,\delta_{g\in\Greg}\, \psu(g)  \labl{lema}
for every $g\iN\G$ and every $\psu\iN\Gregs$. This is derived as follows.
For $g\iN\Greg$ the result follows immediately from the fact that in any
\irrep\ central elements are represented by multiples of the unit matrix.
Suppose then that $g\,{\not\in}\,\Greg$; 
then the completeness of the $\G$-characters implies that the sums
  \be  \sum_{\psi\in\G^*} \psi(gh) = 0  \Labl29
over {\em all\/} $\G$-characters vanish for every $h\iN\Greg$. This means that
  \be  \sum_{\psu\in\Gregs} \psu(h) \sum_{\ssty\psi\in\G^*\atop\psi|_\Greg=\psu}
  \psi(g) = 0 \,. \ee
We now multiply the relation \Erf29 with $\phu^*(h)$ for some
$\phu\iN\Gregs$ and sum over $h\iN\Greg$; by the properties of 
$\Greg$-characters we then arrive at
  \be  0 = \sum_{\psu\in\Gregs} \sum_{h\in\Greg} \phu^*(h)\,
  \psu(h) \sum_{\ssty\psi\in\G^*\atop\psi|_\Greg=\psu} \psi(g)
  = \sum_{\ssty\psi\in\G^*\atop\psi|_\Greg=\phu} \psi(g) \,.  \ee
This finishes the proof of \erf{lema}.

When considering the tensor product of two (\findim)
projective \rep s $R_1$ and $R_2$
of a finite abelian group $\G$, one should allow for the possibility that the
cohomology classes of the two relevant two-cocycles $\F_i$ are different. The
tensor product \rep\ $R_1\otimeS R_2$ is again endowed with the structure 
of a projective $\G$-\rep\ via
  \be  (R_1{\otimes}R_2)(b_g) := R_1(b_g)\otimeS R_2(b_g) \,,  \ee
and one immediately checks that the cocycle relevant to the tensor product 
\rep\ ist the product $\F_1\F_2$. The most interesting case is the one where
the cohomology classes of $\F_1$ and $\F_2$ are complex 
conjugate (i.e.\ when they contain
representatives that are each others' complex conjugates). Then the product
$\F_1\F_2$ is cohomologically trivial, so that the tensor product is a
{\em honest\/} \rep\ of $\G$ and hence fully reducible into a direct sum of
\onedim\ irreducible $\G$-\rep s.

One easily verifies that for complex conjugate cocycles the set of regular 
elements coincide, $\GReg_1\eq\GReg_2\,{=:}\,\Greg$. It follows that in the
case of irreducible projective modules $V_1\,{\equiv}\,V_{\psu_1}$ and 
$V_2\,{\equiv}\,V_{\psu_2}$, both 
$V_1$ and $V_2$ have dimension $d\,{:=}\,\sqrt{|\G|/|\Greg|}$, so that the 
tensor product module $V_1\otimeS V_2$ has dimension $d^2$. Further, for
every $g\iN\Greg$ one has
  \be  (R_1{\otimes}R_2)(b_g) = \psu_1\psu_2\,\one  \,,  \ee
which means in particular that only those irreducible $\G$-\rep s appear
in the decomposition whose restriction to $\Greg$ is the irreducible
$\Greg$-\rep\ with character $\psu\eq\psu_1\psu_2\iN\Gregs$. There are $d^2$ 
inequivalent $\G$-\rep s $V_\psi$ with this property. To determine their
multiplicity in the tensor product, we use the results \Erf1p and \Erf0t to
compute
  \be  {\rm mult}_{\psu_1\psu_2}(\psi) = \sum_{g\in\G} \psi^*(g)\,\tR
  (R_1{\otimes}R_2)(b_g) = \sum_{g\in\Greg} \psi^*(g)\,\tR(R_1{\otimes}R_2)(b_g)
  \,,  \ee
which only depends on the restriction of $\psi$ to $\Greg$. Thus all of the
$d^2$ $\G$-\rep s $V_\psi$ with the same restriction $\psi|_\Greg\eq\psu_1\psu
_2$ have the same multiplicity, and by comparing dimensions we learn that this
multiplicity is equal to one, i.e.\ each of these \rep s appears precisely once
in the decomposition of $R_1{\otimes}R_2$: 
  \be  R_{\psu_1}\otimeS R_{\psu_2} \cong \bigoplus_{\psi\gt\psu_1\psu_2}
  R_\psi \,.  \Labl41
We conclude in particular that
the trivial \irrep\ of $\G$ appears in the tensor product of two projective
irreducible $\G$-\rep s  if and only if the two cocycles are complex
conjugate and the two characters are complex conjugate as well, and then
it appears precisely once.

\subsection{Branching rules}

Consider now the situation where we are given a two-cocycle $\F$ on
the abelian group $\G$ and in addition a subgroup $\Gp\,{\subset}\,\G$.
Manifestly, the restriction 
  \be  \Fp := \F|_{\Gp\times\Gp}   \Labl Fp
of $\F$ to the subgroup constitutes a two-cocycle on $\Gp$. Likewise, the
commutator cocycles are related by restriction as well,
  \be  F' \equiv F_\Fp = F_\F|_{\Gp\times\Gp} \,.  \ee
The twisted group algebra $\complex_\Fp\Gp$ is a semisimple
subalgebra of $\complex_\F\G$. Hence every \irrep\ $R_\psu$ of
$\complex_\F\G$ is fully 
reducible into \irrep s $R_\psup$ of $\complex_\Fp\Gp$. Our goal in this
subsection is to determine the corresponding branching rules
  \be  R_\psu \cong \bigoplus_{\psup\in\Gregps} \br_\psu^{\;\psup}\, R_\psup
  \,, \ee
where the symbol `$\cong$' stands for isomorphy of $\complex_\Fp\Gp$-\rep s,
explicitly for all $\psu\iN\Gregs$.
For ease of notation from now on we will use the abbreviations
  \be  \U := \Greg \equiv \GReg_\F \qquad{\rm and}\qquad
  \Up := \Gregp \equiv \Gregp_{\!\!\Fp}  \ee
for the subgroups of regular elements of $\G$ and $\Gp$, \resp.
We remark that in general $\Up$ is not a subgroup of $\U$.
On the other hand, $\Uc$ is a subgroup of $\Up$.

We first observe that $\Uc$ is contained both in $\U$ and in $\Up$, so that the
\rep\ matrices $R_\psu(b_g)$ and $R_\psup(b'_g)$ are diagonal and act with the 
same eigenvalue (compare \Erf1p). Thus a necessary condition for $R_\psup$ to
appear in the branching of $R_\psu$ is that the restrictions of $\psu$ and
$\psup$ to $\Uc$ coincide:
  \be  \br_\psu^{\;\psup}\ne0 \ \Rightarrow\ \psu|_\Uc^{} = \psup|_\Uc^{} \,.
  \Labl1c
For every $\psu\iN\Us$ there are $|\Up|/|\Uc|$ many 
$\complex_\Fp\Gp$-characters $\psup\iN\Ups$ that satisfy the criterion \Erf1c. 
We claim that each of the corresponding \irrep s $R_\psup$ appears
with the same multiplicity in the branching of $R_\psu$ and that this
multiplicity is the same for all irreducible $\complex_\F\G$-\rep s, i.e.\ that
  \be  R_\psu \cong \br \cdot\!\!\bigoplus_{\ssty
  \psup\in\Gregps \atop \ssty \psup\gt\psu|_\Uc^{}} R_\psup  \Labl2c
with $\br$ independent of $\psu$ and $\psup$.

To verify this assertion we consider an arbitrary $\complex_\Fp\Gp$-\rep\
$R'$ and define
  \be  P_\psup := \Frac1{|\Up|} \sum_{g'\in\Gp} \psup(g')^*\, R'(b'_{g'})  \ee
for every $\psup\iN\Ups$. Direct computation shows that the operators $P_\psup$
are projectors,
  \be  P_\psup P_\phup = \delta_{\psup,\phup}^{}\, P_\psup  \ee
for all $\psup,\phup\iN\Ups$.
Moreover, combining the orthogonality relation for $\Up$-characters
with the result \Erf0t about traces one finds that
  \be \tr_{R_\phup}\, P_\psup = d'\,\delta_{\psup,\phup}^{}  \ee
for every irreducible $\complex_\Fp\Gp$-\rep\ $R_\phup$ and every 
$\psup\iN\Ups$, where, as usual, $d'\eq\sqrt{|\Gp|/|\Up|}$.
Together it follows that the multiplicity of $R_\phup$ in $R'$ is given by
$\tr_{R'}^{} P_\phup/d'$. Applying now this result to the irreducible
$\complex_\F\G$-\rep\ $R_\psu$ (considered as a, generically reducible,
$\complex_\Fp\Gp$-\rep) we find
  \be  \bearll
  \br_\psu^{\;\phup} \!\! 
  &= \dsty\Frac1{d'}\,\tr_{R_\psu}\, P_\phup
   = \Frac1{|\Up|}\,\Frac d{d'}\, \sum_{g'\in\Uc} \phup(g')^*\, \psu(g')
   = \Frac{|\Uc|}{|\Up|}\,\Frac d{d'}\, \delta_{\phup|_\Uc,\psu|_\Uc} 
  \,.  \eear \Labl3c
This means that only the restriction of $\phup$ to $\Uc$ matters, which 
finally proves our claim \Erf2c. It also shows that the multiplicity $\beta$
in the branching rule \Erf2c is given by
  \be  \br \equiv \br_{}^{(\Gp\subseteq\G)}
  = \Frac{|\Uc|}{|\Up|}\,\Frac d{d'} \,.  \Labl4c

\section{The homothety property of $\n$}\label{s.c}

In this appendix we derive the identity \Erf y1 that is equivalent to the 
homothety property of the mapping \Erf35 and hence enters crucially in the 
calculation of inner products of the boundary blocks. To this end we define 
  \be  \tkappa(v\ot p,v'\ot p')
  := \kappab(p,p')\, \sum_{j=1}^{d_\lambda}\n(v\ot w_j)^*_{}\n(v'\ot w_j)
  \Labl y2
for every pair $p,p'\iN\calhbl$, where $\{w_j\}$ is an orthonormal basis of 
$\Vpsup$. Note that by definition the form $\tkappa$ is sesquilinear and
independent of the choice of orthonormal basis $\{w_j\}$.

The relation \Erf y1 is equivalent to the assertion that 
  \be  \tkappa(v\ot p,v'\ot p') = \xi\, \Kappa(v\ot p,v'\ot p')  \Labl y3
for all $p,p'\iN\calhbl$.
This relation, in turn, is proven once we have shown that $\tkappa$
is a non-degenerate and invariant scalar product on $\calhl$,
since such scalar products are unique up to a scalar.
Non-degeneracy is immediate. Indeed, since the scalar product $\kappab$ is 
non-degenerate, for every $v\ot p\iN\Vpsu{\otimes}\calhbl$
one can find a $p'\iN\calhbl$ such that $\kappab(p,p')\nE0$.
Furthermore, for $v'\eq v$ the expression $\sum_j\n(v\ot w_j)^*_{}\n(v'\ot w_j)$
is non-vanishing.

To establish invariance,
we first choose a basis $\{y_\vphi\}$ of $\End(\Vpsu)$ that consists
of unitary elements, which is always possible.
Then we expand any $Y{\in}\,\cala$ \wrt this basis, i.e.\ we write
  \be  Y = \Sumphipsu\lambda y_\vphi \oT \bar Y^{(\vphi)}  \,,  \Labl y4
where the first tensor factor acts on $\Vpsu$ and the second on
$\calhbl$. To be precise, we also have to account
for the fact that $\bar Y$ is in general not an endomorphism of
$\calhbl$, but rather a map
  \be  \bar Y:\quad  \calhbl \to \plujl \calhb_{\J\lambdab} \,.  \Labl y5

By linearity we can concentrate on the individual summands in the expression
\Erf y4, i.e.\ on elements of $\cala$ of the form
  $ Y\eq y_\vphi \oT \bar Y$.
For the second tensor factor we just invoke invariance of the
scalar product $\kappab$ and are done.\,%
 \futnote{Strictly speaking, because of \Erf y5
we cannot directly work with $\kappab$, but must consider a scalar product
$\kappabt\eq\plujl\bar\kappa_{\J\lambdab}$, where $\bar\kappa_{\J\lambdab}$ is
a scalar product on $\calhb_{\J\lambdab}$. Note that $\kappabt$ is defined
for the {\em reducible\/} module $\plujl
\calhb_{\J\lambdab}$ and hence is not unique up to multiplication. But this
is irrelevant, because in our considerations always at least one of its
arguments is a vector in $\calhbl$ and because the two operations of restricting
to a submodule and taking the hermitian conjugate commute, so that effectively 
we still only work with $\kappab$. Accordingly we will refrain from using the 
notation $\kappabt$ below, but rather just assume that $\bar Y$ only has
a component in the endomorphisms of $\calhbl$.}
Concerning the first factor we note that the grade of $y_\vphi$ is zero, just 
because the grade of the degeneracy space $\Vpsu$ is zero.
Further we use the Ward identity \Erf WI
for $\BB$ and write $(-1)^{\Delta_Y-1}\,{=:}\,\zetay$ so as to obtain
  \be  |\BBB(\po,\qo)|^2\,
  \tkappa(Y(v\ot p),v'\ot p')
  = - \zetay \sum_{j=1}^{d_\lambda} \BB(v\ot\po\oT (y_\vphi w_j)\ot\qo)^*\,
  \BB(v'\ot\po\oT w_j\ot\qo) \, \kappab(\bar Y p,p')  \ee
as well as
  \be  \bearll  |\BBB(\po,\qo)|^2\,
  \tkappa(v\ot p,Y^\dagger(v'\ot p')) \!\!\!
  &= - \zetay \dsty\sum_{j=1}^{d_\lambda} \BB(v\ot\po\oT w_j\ot\qo)^* \,
  \BB(v'\ot\po\oT (y_\vphi^\dagger w_j)\ot\qo)\,
  \kappab(p,\bar Y^\dagger p')
  \,,  \eear \ee
where $\po$ and $\qo$ are the elements of $\calhbl$ and $\calhblp$ used in the
definition \erf n of $\n$. Next we implement the
fact that $\kappab(p,\bar Y^\dagger p')\eq\kappab(\bar Y p,p')$. The final step 
in establishing the invariance relation then consists in using the identity
  \be  \dsty\sum_{j=1}^{d_\lambda} \BB(v\ot\po\oT (y_\vphi w_j)\ot\qo)^*\,
  \BB(v'\ot\po\oT w_j\ot\qo)
  = \dsty\sum_{j=1}^{d_\lambda} \BB(v\ot\po\oT w_j\ot\qo)^* \,
  \BB(v'\ot\po\oT (y_\vphi^\dagger w_j)\ot\qo) \,,  \ee
which holds because $w_j\,{\mapsto}\,y_\vphi w_j$ is a basis transformation 
between two orthonormal bases of $\Vpsup$ (recall that
$y_\vphi$ was chosen to be unitary).

\newpage
\small
 \newcommand\wb{\,\linebreak[0]} \def\wB {$\,$\wb}
 \newcommand\Bi[1]    {\bibitem{#1}}
 \renewcommand\J[5]   {{\sl #5}, {#1} {#2} ({#3}) {#4} }
 \newcommand\PhD[2]   {{\sl #2}, Ph.D.\ thesis (#1)}
 \newcommand\Prep[2]  {{\sl #2}, preprint {#1}}
 \newcommand\BOOK[4]  {{\em #1\/} ({#2}, {#3} {#4})}
 \def\jf    {J.\ Fuchs}
 \def\comp  {Com\-mun.\wb Math.\wb Phys.}
 \def\duke  {Duke\wB Math.\wb J.}
 \def\foph  {Fortschr.\wb Phys.}
 \def\ijmp  {Int.\wb J.\wb Mod.\wb Phys.\ A}
 \def\imrn  {Int.\wb Math.\wb Res.\wb Notices}
 \def\jams  {J.\wb Amer.\wb Math.\wb Soc.}
 \def\maan  {Math.\wb Annal.}
 \def\mpla  {Mod.\wb Phys.\wb Lett.\ A}
 \def\nuci  {Nuovo\wB Cim.}
 \def\nupb  {Nucl.\wb Phys.\ B}
 \def\pams  {Proc.\wb Amer.\wb Math.\wb Soc.}
 \def\phlb  {Phys.\wb Lett.\ B}
 \def\phrl  {Phys.\wb Rev.\wb Lett.}
 \def\NH     {{North Holland Publishing Company}}
 \def\SV     {{Sprin\-ger Ver\-lag}}
 \def\Ad     {{Amsterdam}}
 \def\Be     {{Berlin}}

\end{document}